\RequirePackage{fix-cm}
\documentclass[smallextended,authoryear]{svjour3}       
\smartqed  
\usepackage{graphicx}
\usepackage{amssymb}

\usepackage{color}
\definecolor{gr}{cmyk}{1,0,1,0.09}

\begin{document}

\title{Natural Highways for End-of-Life Solutions in the LEO Region}



\author{Elisa Maria Alessi \and Giulia Schettino \and Alessandro Rossi \and Giovanni B. Valsecchi
}


\institute{E. M. Alessi  \at
              IFAC-CNR, via Madonna del Piano 10, 50019 Sesto Fiorentino (FI), Italy \\
              \email{em.alessi@ifac.cnr.it}           
            \and G. Schettino, A. Rossi \at
              IFAC-CNR, via Madonna del Piano 10, 50019 Sesto Fiorentino (FI), Italy 
           \and
           G. B. Valsecchi \at
              IAPS-INAF, via Fosso del Cavaliere 100, 00133 Rome, Italy\\ 
              IFAC-CNR, via Madonna del Piano 10, 50019 Sesto Fiorentino (FI), Italy 
}

\date{Received: date / Accepted: date}

\maketitle

\begin{abstract}
We present the main findings of a dynamical mapping
performed in the Low Earth Orbit region. The results were obtained by
propagating an extended grid of initial conditions, considering two
different epochs and area-to-mass ratios, by means of a
singly-averaged numerical propagator. It turns out that dynamical
resonances associated with high-degree geopotential harmonics,
lunisolar perturbations and solar radiation pressure can open natural
deorbiting highways. For area-to-mass ratios typical
of the orbiting intact objects, these corridors can be exploited only
in combination with the action exerted by the atmospheric drag. For
satellites equipped with an area augmentation device, we show the boundary of application
of the drag, and where the solar radiation pressure can be exploited.

\end{abstract}

\section{Introduction}

Thanks to the effort made by scientists, public organizations and
industries worldwide, there now exists a strong awareness on the value of
preserving the circumterrestrial environment for future
generations. The need of controlling the debris environment and
limiting the growth of its population is addressed at several
levels, which include, among others, long-term modeling,
operational activities like collision avoidance operations and reentry
campaigns, and mitigation actions.

For its strategic importance and its well-attested unstable evolution
\cite{liou}, the Low Earth Orbit (LEO) region is, in
particular, subject to a thorough assessment by the community. The
consequences of possible fragmentations in the long-term, active and
passive debris removal campaigns, but also the new scenarios driven by
the launch of mega constellations are the aspects where there is the
strongest commitment \cite{rossi_frag}, \cite{lewis}. Besides the technological challenges, a
renovated focus is put on the theoretical features which cannot be
ignored.

We are here interested in passive disposal solutions, to be adopted at
the end-of-life of a given mission.  As very well known, the LEO
region is one of the ``Protected Regions" defined by international
agreements \cite{ESA}. It is formally defined as ``the spherical shell
that extends from the Earth's surface up to an altitude of 2000 km",
and the mitigation guidelines require that ``a space system operating
in LEO shall be disposed of by reentry into the Earth's atmosphere
within 25 years after the end of the operational phase."

\begin{figure}[ht!]
  \begin{center}
 \includegraphics[width=90mm]{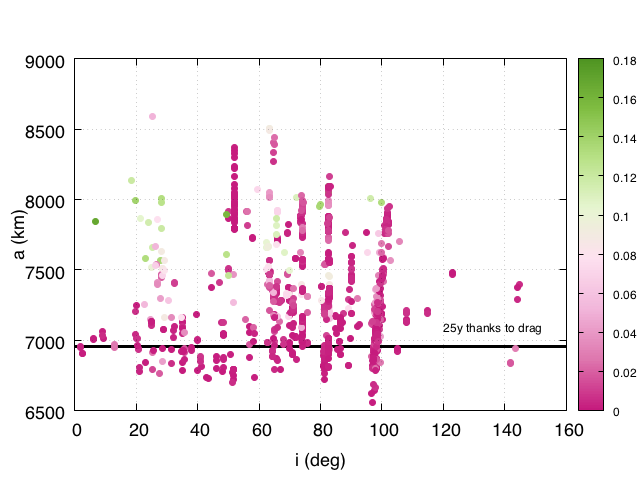}
\caption{Orbital distribution in terms of inclination $i$ and semi-major axis $a$ of the objects in LEO with a mass greater than 100 kg, according to MASTER \cite{MASTER}. The color bar reports the eccentricity of a given object.}\label{fig:MASTER_LEO}
\end{center}
\end{figure}

Figure \ref{fig:MASTER_LEO} shows the orbital
distribution in terms of inclination $i$, semi-major axis $a$ and
eccentricity $e$ of the objects in LEO with a mass greater than 100
kg, according to the ESA MASTER model \cite{MASTER}. The most critical
regions in LEO are located in the altitude ranges $[700:1000]$
km and $[1300:1600]$ km, where the orbits are mostly circular
(with eccentricities usually lower than 0.02). In the same figure, we have
indicated the mean altitude corresponding to a reentry in 25 years for
circular orbits, thanks to the perturbation exerted by the atmospheric
drag. It is clear from the figure that for most
  of the objects an impulsive de-orbiting
strategy must be applied to comply with the guidelines. From the
operational point of view, from a certain altitude onward (typically 1400
km), it is more convenient, in terms of propellant budget, to send a
spacecraft to a graveyard orbit in a higher LEO. As a matter of fact, if
we consider the payloads for which a maneuver must be applied, the
percentage of compliance with the 25-year rule is actually very low,
at less than a 15\% \cite{FL17}-\cite{F17}. In order to avoid the
accumulation of spent, uncontrolled spacecraft in a restricted region
of space, which would lead inevitably to a significant number of
collisions, it is mandatory to look for affordable solutions which
ensure reentry within the Earth's atmosphere in a reasonable time
frame.

The first step towards this direction is to understand if the
dynamical perturbations in the region can drive the spacecraft towards
natural reentry corridors.  In this perspective, we present the
simulations performed within the ReDSHIFT H2020 project \cite{aR17} on
the dynamical behavior in LEO. The results corresponding to Medium Earth Orbits (MEO) and Geostationary Earth Orbits, performed within the ReDSHIFT H2020 project, can be found in \cite{Grecia_SD},  and \cite{Colombo_SD}, \cite{GC17}, respectively. General results on the whole circumterrestrial space, also obtained within the ReDSHIFT H2020 project, can be found in \cite{Grecia_Grecia}.

To our knowledge, such an extensive and
systematic exploration was never performed. The main objective is to
detect ``deorbiting highways'', associated with a significant change in
eccentricity due to natural perturbations, apart from the atmospheric
drag. The results are collected in a series of color maps, which
describe at once whether a reentry is feasible, and, in such case, the
required time and initial dynamical configuration.

This study provides a new insight on the dynamical effects caused by
three different sources: the lunisolar gravitational attraction, the
solar radiation pressure (SRP), and high-degree geopotential. At
specific values of inclinations, these perturbations can foster the
orbital decay, if a well-defined resonance condition is
satisfied. Notice that resonances associated with these effects and the theoretical possibility to exploit them to lower the lifetime of spacecraft at the end-of-life were
already theorized in the past (see, e.g., \cite{B1999}-\cite{B2001a}). Nonetheless, so far space agencies have conceived their application mainly for MEO (see, e.g., \cite{A16}) and Highly Elliptical Orbits (see, e.g., \cite{C2014}). The only work, as far as we know, which considered operational aspects of these natural leverages also for LEO is \cite{LetAl12}, in which however a very narrow range of initial conditions is analyzed.

In the work, two values of area-to-mass
ratios are considered in order to show how the exploitation of a
drag- or SRP-enhancing device might change the whole picture, but also
to get a deeper understanding of the boundary between the realms of
application of these two effects. A full analysis on the case of the sail is not considered here, but only the main features will be described. As a matter of fact, this topic deserves a specific focus, given also the copious literature which exists on the subject and the crucial interaction between atmospheric drag and solar radiation pressure.

A full understanding of the dynamics at stake in the most populated
and operational region is the key starting point for many of the
actions the community will or must take. The results can be exploited
to design passive end-of-life solutions, to optimize the impulsive
strategies aimed at reentering, and to support the roadmap for
eventually taking advantage of a drag or a solar sail, but they could
help also, e.g., in choosing the nominal orbits during the
operational phase of a spacecraft, or for observational campaigns.

\section{Numerical Simulation}

The numerical simulation performed consists in propagating a series
of initial conditions, which are representative of the whole LEO
region, for 120 years, starting from two different initial epochs,
namely, December 22, 2018 at 17:50:21 (JD 2458475.2433), and June 21,
2020 at 06:43:12 (JD 2459021.7800). They correspond to a December full
Moon and a June new Moon solstice, where the longitude of the Sun with
respect to the ecliptic plane is $\lambda_S\approx 270^{\circ}$ and
$\lambda_S\approx 90^{\circ}$, respectively. On the second epoch a
solar eclipse will occur. The choice of the initial epochs was suggested by the will of comparing, in a second step, the results of the numerical simulations with simpler, averaged, analytical models where the relative configuration of the Sun-Earth-Moon system was expected to play a role. In particular, the constraints set consisted in  configurations
\begin{itemize}
\item close to solstices, so that the Earth rotation axis points towards the Sun;
\item
at a new or a full Moon, so that Sun, Earth and Moon are aligned;
\item
with the line of nodes of the lunar orbit possibly close to $0^\circ$, $90^\circ$, $180^\circ$ or 
$270^\circ$ with respect to the line containing the three bodies.
\end{itemize}

Similar to what done in \cite{A16}, during the 120 years time
interval, the orbital evolution of semi-major axis, eccentricity,
inclination, right ascension of the ascending node, argument of
perigee $(a,e,i,\Omega,\omega)$ was recorded at a step of 1 day.
Moreover, maximum and minimum values attained by semi-major axis,
eccentricity and inclination were stored. The orbit was considered to
have reentered into the atmosphere, whenever the altitude of perigee,
say $h_p$, had decreased down to 300 km.

In the following, the details on the orbit propagator and on the initial conditions are given.

\subsection{FOP}

The numerical method used is the Fast Orbit Propagator (FOP)
\cite{lAesa}-\cite{aR09}, which is an accurate, long-term orbit predictor,
based on the Long-term Orbit Predictor (LOP) \cite{jK86}. This
singly-averaged formulation integrates the Lagrange or Gauss planetary equations for a set of
orbital elements, which are non-singular for circular orbits
and singular only for equatorial orbits. The numerical integrator is a multi-step, variable step-size and order
integrator.  The dynamical
model accounts for the gravitational and non-gravitational
perturbations stemming from the Earth, the Moon, and the Sun, namely,
\begin{itemize}
\item geopotential harmonics (up to degree and order 5):
\item lunisolar perturbations;
\item atmospheric drag;
\item solar radiation pressure, including
shadows. 
\end{itemize}
 To expedite the computations, the perturbations are analytically
averaged over the mean anomaly of the satellite. For tesseral resonant effects (located at specific values of semi-major
axis, where there exists a commensurability between the satellite's
mean motion and the Earth's rotation rate), a partial averaging
procedure is applied to retain only the long-periodic perturbations
associated with these harmonics. The positions of the Moon and the
Sun, which are held constant during the averaging process, are
determined by means of accurate analytical ephemerides. The SRP
effect is represented by the cannonball model, accounting also for
shadowing intervals. The shadows are modeled as solar occultations, with a cylindrical model. The algorithm is based on the assumption that the Sun is a point at infinity and that the spacecraft and the Sun are frozen during the occultation. The atmospheric drag is applied for altitudes
below 1500 km, adapting the Jacchia-Roberts density model assuming an
exospheric temperature of 1000 K and a variable solar flux at 2800 MHz
(obtained by means of a Fourier analysis of data corresponding to the
interval 1961-1992). To average the disturbing accelerations
associated with solar radiation pressure and atmospheric drag, a standard
8th-order Gaussian quadrature method is used. FOP was widely
validated in the past \cite{lAesa}, and thus no additional analysis was needed for the
present work. Notice that in LEO none of the effects considered in the dynamical model can be neglected. 

\subsection{Grid Definition}

The grid of initial orbital elements explored is shown in
Table \ref{tab:grid}. Note that the LEO region is usually defined for
circular orbits up to $2000$ km in altitude, which is lower than the
maximum value of semi-major axis adopted here. On the one hand, this
choice is justified by the fact that orbits with larger semi-major
axes but also higher eccentricities have perigee altitudes falling
below that limit. On the other hand, a precise understanding of the dynamics just beyond $2000$ km is important for the long-term stability of graveyard orbits.

The range of eccentricity allowed is such that
the initial perigee is higher than 300 km. Moreover, to depict
accurately the situation of the most critical regions in LEO, besides
the refinement in semi-major axis displayed in the table, a finer step
in eccentricity, namely, $\Delta e = 10^{-3}$, was adopted for $a<
R_{E} +1600 $ km and $e<0.01$, being $R_E=6378.1363$ km the mean radius of
the Earth.

Concerning the step-size used for the longitude of the ascending node
and the argument of perigee, the standard definition adopted was $90^{\circ}$,
but for some specific values of $(a,e,i)$ a higher-resolution
exploration was performed, taking $\Delta \Omega=10^{\circ}$ and
$\Delta \omega=10^{\circ}$, both $\Omega$ and $\omega$ in the range
$[0:360)$ deg.

\begin{table}
\caption{Grid of initial conditions in semi-major axis $a$, eccentricity $e$ and inclination $i$ for the exploration performed. $R_E=6378.1363$ km is the radius of the Earth. }
\begin{tabular}{llllll}
\hline
$a$ (km) & $\Delta a$ (km) & $e$ & $\Delta e$ & $i$ (deg) & $\Delta i$ (deg) \\
\hline
$[500:700]$ + $R_E$ & $50$ & $[0:0.28]$ & $0.01$ & $(0:120]$ & $2$  \\
$[700:1000]$  + $R_E$ & $20$ & $[0:0.28]$ & $0.01$ & $(0:120]$ & $2$  \\
$[1000:1300]$  + $R_E$ & $50$ & $[0:0.28]$ & $0.01$ & $(0:120]$ & $2$  \\
$[1300:1600]$  + $R_E$ & $20$ & $[0:0.28]$ & $0.01$ & $(0:120]$ & $2$  \\
$[1600:2000]$  + $R_E$ & $50$ & $[0:0.28]$ & $0.01$ & $(0:120]$ & $2$  \\
$[2000:3000]$  + $R_E$ & $100$ & $[0:0.28]$ & $0.01$ & $(0:120]$ & $2$  \\
\hline
\end{tabular}
\label{tab:grid}
\end{table}

For all the initial conditions, the drag coefficient was set to
$C_D=2.1$, while the reflectivity coefficient to $C_R=1$ or
$C_R=2$. For the area-to-mass ratio, the standard value used was
$A/m=0.012$ m$^2/$kg, which reflects the average value of the orbiting
intact population. This value was the result of a thorough analysis carried out on the MASTER catalogue \cite{Colombo_SD}. For the 2020 epoch, the simulations
were repeated assuming $C_R=1$ and $A/m=1$ m$^2/$kg, which can be
considered as a realistic value achievable for small satellites equipped with a drag- or
SRP-enhancing device.

The total number of initial conditions simulated, not accounting for the specific analysis described in Sec. \ref{sec:BomPom}, is 3.6 millions.

\begin{figure}[ht!]
  \begin{center}
  \includegraphics[width=0.7\textwidth]{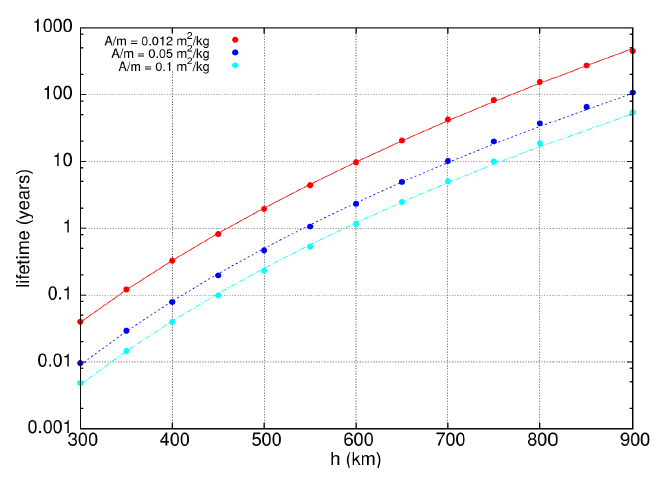}
\caption{Lifetime corresponding to different altitude values for
  circular LEO and different area-to-mass ratios, obtained by
  propagation with DROMO (see, e.g., \cite{hU16}).}\label{fig:lifetime_circular_LEO}
\end{center}
\end{figure}

\section{Numerical Results}
\label{num_res}

The analysis presented is based on the study of the lifetime and the
maximum eccentricity computed for each initial condition in the
grid. The
  goal is to build, for the first time, a detailed map of the
  dynamical behavior in the whole LEO region. As a matter of fact, at
altitudes where the effect of the drag is not dominant, the lifetime
of an orbit can vary significantly only {on account of eccentricity excitations.} Speaking of unstable orbits, we will
always refer to this behavior. In general, focusing for the moment our
analysis on the lower value of area-to-mass ratio,
beyond the drag realm, the maximum variation computed in semi-major
axis is of the order of tens of km, and it is due to tesseral
effects \cite{EK97}-\cite{GC15}.  The maximum change in inclination is
generally up to 1 degree, apart from specific retrograde orbits for
which variations can reach up to 4 degrees.

\subsection{Lifetime}

For low values of area-to-mass ratio, the lifetime estimates corresponding to
circular LEO (within the limits of the solar flux predictability) are
generally known, but , as far as we know, complete and detailed analyses on elliptic orbits are not
available.  As a reference, in
Figure \ref{fig:lifetime_circular_LEO}, we show the lifetime of
circular LEO corresponding to different values of altitude and
area-to-mass ratio of the satellite, obtained from propagations with
DROMO (see, e.g., \cite{hU16}) assuming the NRLMSISE--00 atmospheric
model with the solar flux defined by ESA's model \cite{ESA}.
Note that, as seen in the figure, beyond $a = R_E+800$ km a spacecraft
with $A/m=0.012$ m$^2/$kg on a circular orbit takes more than
100 years to reenter.

\begin{figure}[hbpt!]
  \begin{center}
  \includegraphics[width=0.6\textwidth]{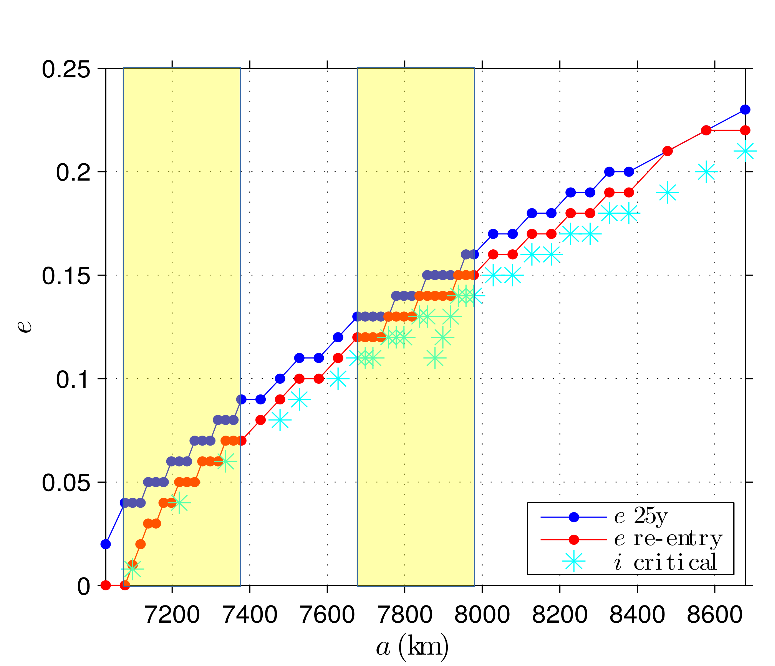}
\caption{As a function of the semi-major axis, the minimum
  value of eccentricity required to reenter in 25 years (blue curve),
  and in less than 120 years (red curve), assuming $A/m=0.012$
  m$^2/$kg, and $\Omega=180^{\circ}$ and $\omega=0^{\circ}$ and the initial epoch 2020. The cyan stars represent
  cases where perturbations different from drag facilitate the reentry at 
  specific values of inclination (see text for details).
  In yellow,  the most populated altitudes.}\label{fig:lifetime_elliptic_LEO}
\end{center}
\end{figure}

Our first step in the analysis of the results is to discriminate
  between the cases which can naturally comply with the 25-year rule,
  thanks to some perturbations, and those which cannot. The general
  behavior computed for $A/m=0.012$ m$^2/$kg (2020 epoch) is
  shown in Figure \ref{fig:lifetime_elliptic_LEO}. For each initial
  condition considered, we look for the minimum value of eccentricity,
  as a function of the semi-major axis, which guarantees a lifetime of
  at most 25 years (blue line in the figure). At the same time, we
  identify the minimum value of eccentricity, for any value of
  inclination in the range explored and for each given semi-major
  axis, which ensures to reenter within the assumed 120 years window
  of propagation (red line in the figure).

We stress that, for the standard value of area-to-mass ratio, in LEO
the reentry cannot be achieved without the action exerted by the
atmospheric drag, whose effect, as known, is to decrease both the
semi-major axis and the eccentricity. To this end, the initial
pericenter altitude must be lower than a given threshold, which
depends on the semi-major axis. For the whole range of eccentricity
explored, considering initial values of semi-major axis such that
$a>R_E+800$ km, the minimum pericenter altitude required to reenter
spans from about $580$ km down to about $350$ km, for increasing
semi-major axis values.  The cyan stars in Figure
\ref{fig:lifetime_elliptic_LEO} represent specific cases where
perturbations different from drag facilitate the reentry (but not on
their own).  In other words, depending on the initial configuration,
i.e., on the initial values of $(i, \Omega, \omega)$, given $a$, a
reentry could be achieved at a lower value of eccentricity. In the
following, we will refer to these specific inclinations as {\it
  resonant inclinations}; the reason for this will be clarified in
Section \ref{sec:res}.

\begin{figure}[htbp!]
  \begin{center}
         \includegraphics[width=0.4\textwidth]{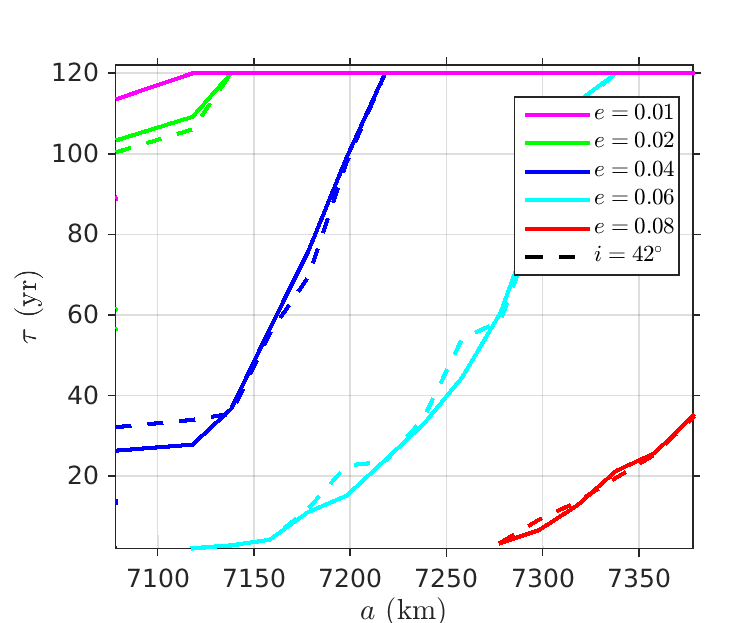}  \includegraphics[width=0.4\textwidth]{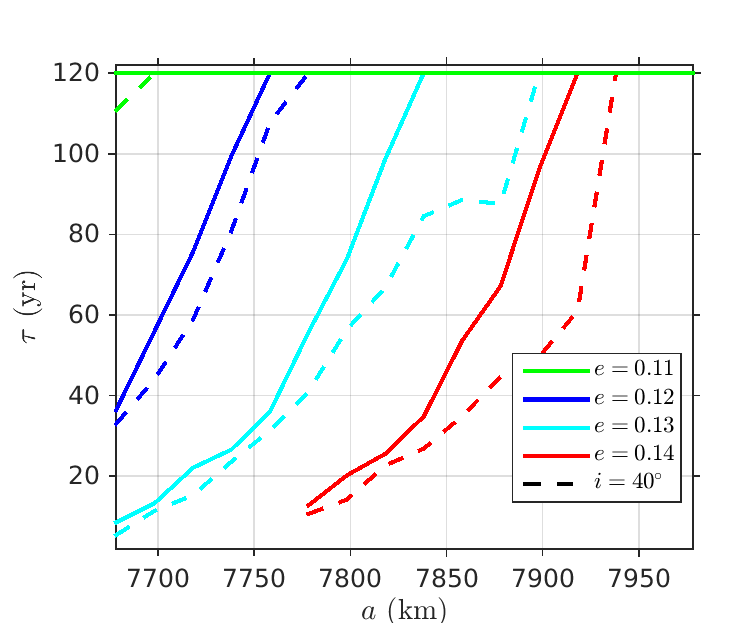}
 \includegraphics[width=0.4\textwidth]{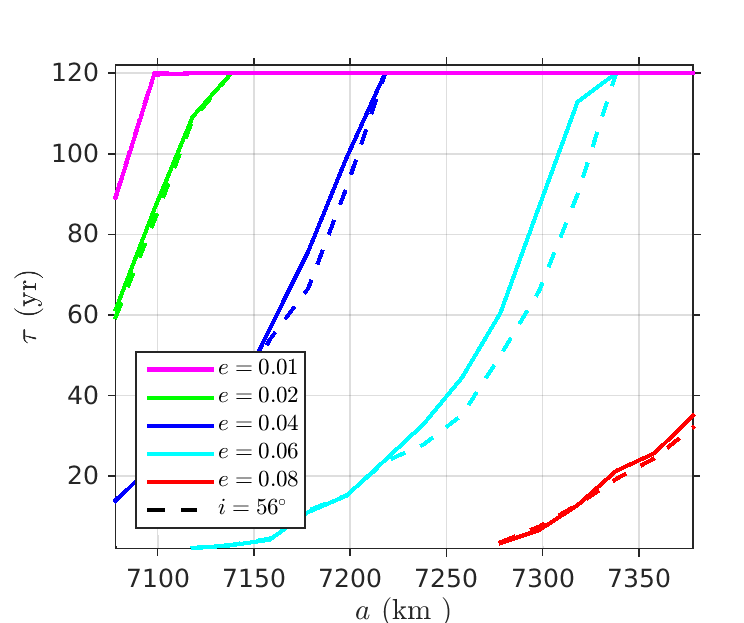}\includegraphics[width=0.4\textwidth]{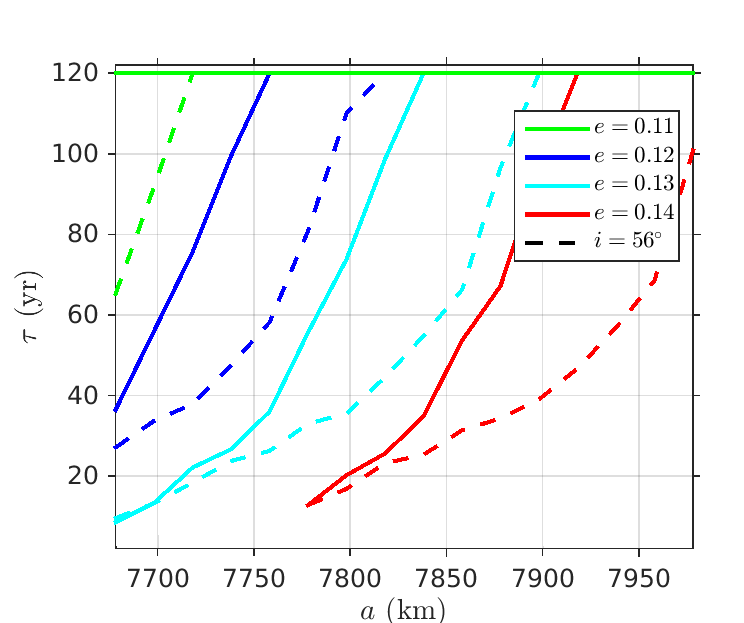}
\includegraphics[width=0.4\textwidth]{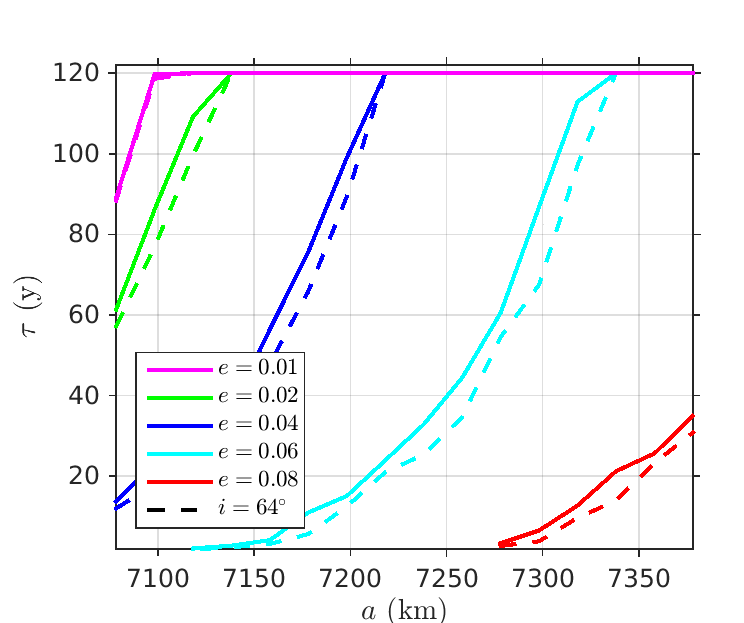}  \includegraphics[width=0.4\textwidth]{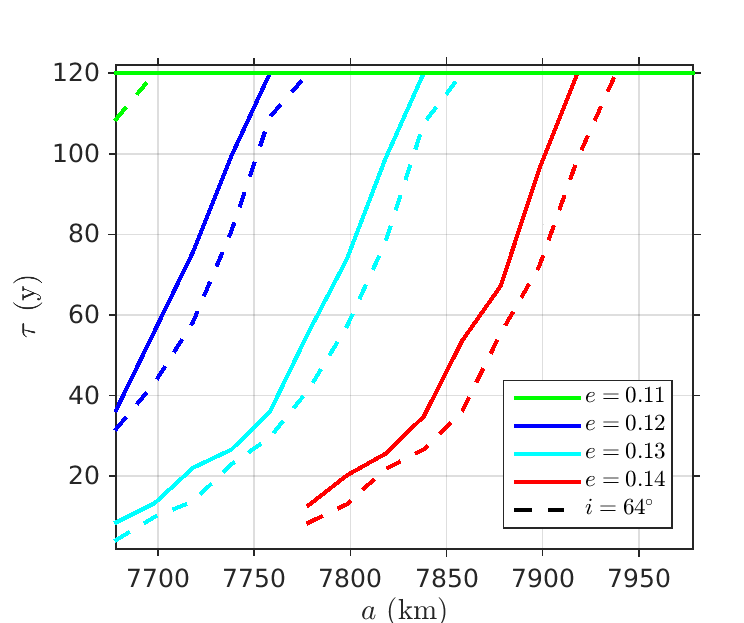}
\caption{Comparison of the lifetime computed in correspondence of a resonant inclination (dashed curve) or non-resonant inclination (solid curves). Left: $a\in[7078.14:7378.14]$ km. Right: $a\in[7678.14:7978.14]$ km. Top:
  $i=40^{\circ}$. Middle: $i=56^{\circ}$. Bottom:
  $i=64^{\circ}$. Initial
  epoch 2020, $C_R (A/m)=0.024$ m$^2/$kg, $\Omega=0^{\circ}$, $\omega=90^{\circ}$.}  \label{fig:lifetime_7001000_13001600}
\end{center}
\end{figure}

In Figure \ref{fig:lifetime_7001000_13001600}, we show the standard
lifetime curves computed for two different ranges of semi-major
axis and high-enough values of eccentricity, together with the
lifetime curves computed in the same regions for three of the resonant
inclination values detected, namely, $40^{\circ}$, $56^{\circ}$ and
$64^{\circ}$. The improvement in the reentry times in the cases
where the resonant inclinations are exploited (dashed lines) is noticeable, when the eccentricity is sufficiently high for the drag to be effective. From the figure, it can be seen that for  $a\in[7078.14:7378.14]$ km, the  effect of  the atmospheric drag combined with lunisolar perturbations, geopotential or solar radiation pressure can be exploited starting from $e > 0.02$, while for $a\in[7678.14:7978.14]$ km, the eccentricity must be as high as $e>0.11$. Notice also that in the latter case the reduction in lifetime is much larger, mainly because the eccentricity and the semi-major axis values correspond to a lower pericenter altitude and thus a stronger atmospheric effect. 

\begin{figure}[htbp!]
  \begin{center}
         \includegraphics[width=0.3\textwidth]{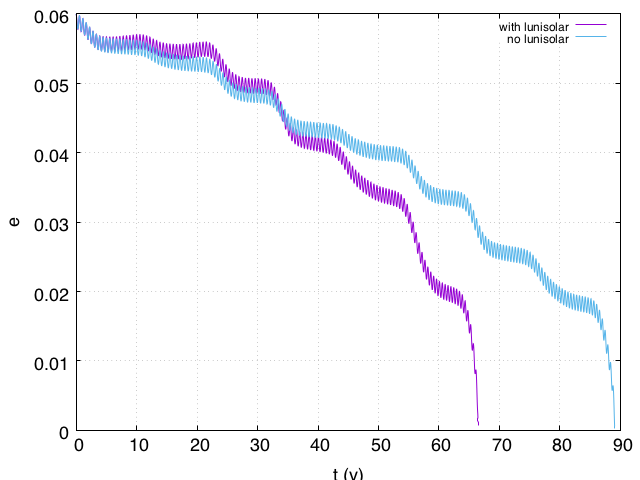} \includegraphics[width=0.3\textwidth]{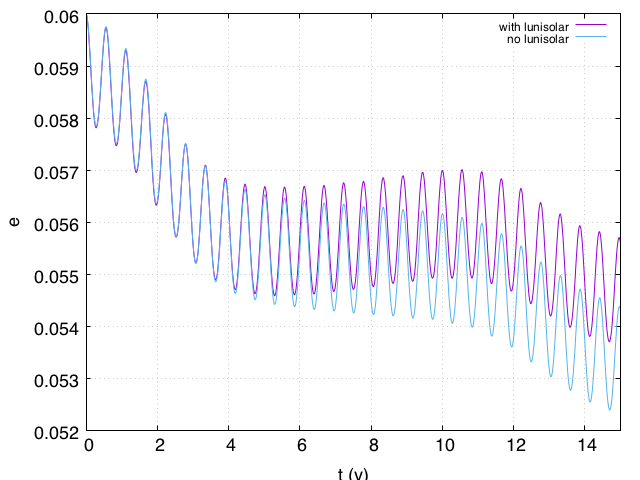}  \includegraphics[width=0.3\textwidth]{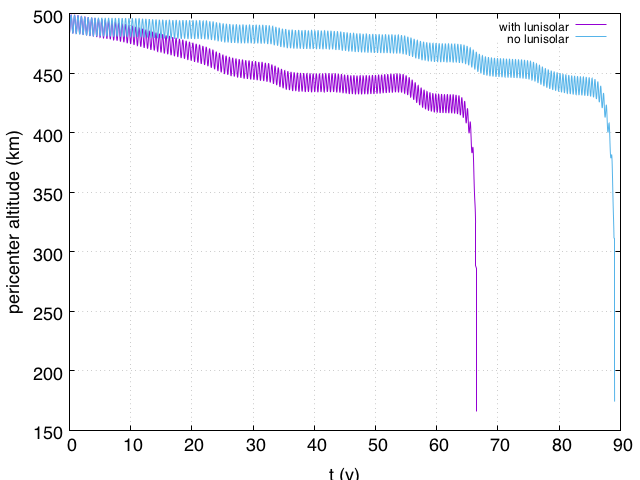}
\caption{Comparison of the lifetime computed when the effect of the drag is combined with a lunisolar gravitational resonance (purple), and when lunisolar perturbations do not act (blue). Left: eccentricity behavior in time. Middle: detailed behavior at the time when the relative eccentricity increase takes place. Right: pericenter altitude behavior in time. Initial condition: $a=7300.14$ km, $e=0.06$, $i=56^{\circ}$, $\Omega=0^{\circ}$, $\omega=90^{\circ}$,  epoch 2020, $C_R (A/m)=0.012$ m$^2/$kg.}  \label{fig:example1_drag_lunisolar}
\end{center}
\end{figure}

\begin{figure}[htbp!]
  \begin{center}
         \includegraphics[width=0.3\textwidth]{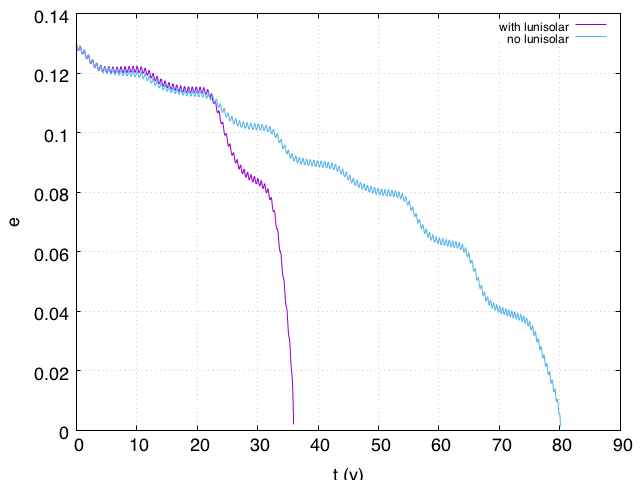}  \includegraphics[width=0.3\textwidth]{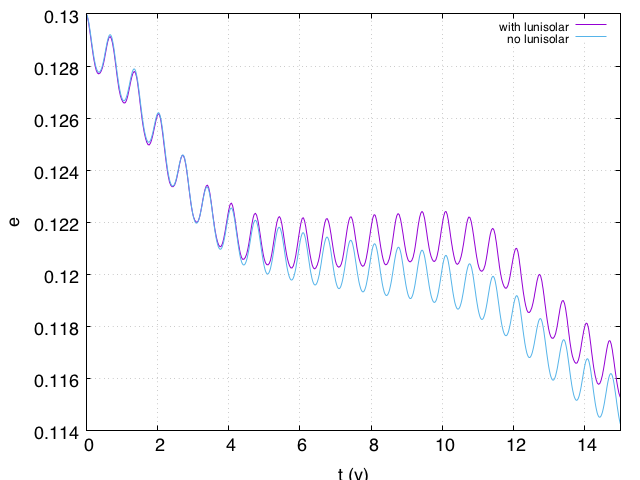}  \includegraphics[width=0.3\textwidth]{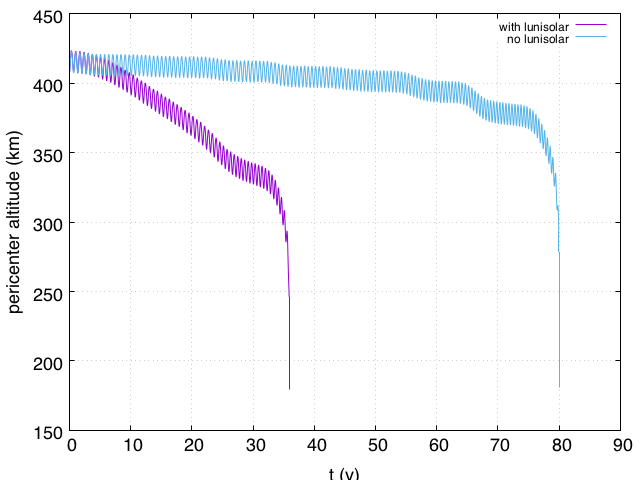}
\caption{Comparison of the lifetime computed when the effect of the drag is combined with a lunisolar gravitational resonance (purple), and when lunisolar perturbations do not act (blue). Left: eccentricity behavior in time.  Middle: detailed behavior at the time when the relative eccentricity increase takes place. Right: pericenter altitude behavior in time. Initial condition: $a=7800.14$ km, $e=0.13$, $i=56^{\circ}$, $\Omega=0^{\circ}$, $\omega=90^{\circ}$,  epoch 2020, $C_R (A/m)=0.012$ m$^2/$kg.}  \label{fig:example2_drag_lunisolar}
\end{center}
\end{figure}

As a further illustration of the dynamical interplay between the drag and one of the other perturbations considered, in Figures  \ref{fig:example1_drag_lunisolar}--\ref{fig:example2_drag_lunisolar}, we compare the evolution in time of the eccentricity and the pericenter altitude for $i=56^{\circ}$, $\Omega=0^{\circ}$, $\omega=90^{\circ}$,  epoch 2020, $C_R (A/m)=0.012$ m$^2/$kg, for two values of semi-major axis and eccentricity  ($a=7300.14$ km, $e=0.06$ and $a=7800.14$ km, $e=0.13$), when the lunisolar perturbation is included and when it is not in the numerical propagation. In both examples, the reduction in lifetime is due to a slight, which could appear negligible,  increase in eccentricity (see middle panels in both figures). This variation takes place after several years from the initial epoch and yields a value which is lower than the initial one. What matters, however, is the effect on the pericenter altitude, taking into account also the fact that the drag has, in the meantime, reduced the semi-major axis by an amount of 50-100 km in both examples. Recall that the atmospheric drag depends on the atmospheric density, which behaves exponentially. The lower the altitude, the stronger the effect.
  
  \subsection{Cartography}
\label{Cartography}
  
  To visualize more in detail the dynamical mechanisms, we build
  contour maps of both the lifetime and the maximum eccentricity
  attained in 120 years, as a function of the initial inclination and
  altitude of perigee or of the initial inclination and eccentricity,
  for each value of initial semi-major axis and for each of the 16
  initial $(\Omega, \omega)$ configurations. They represent what we
  call the {\it cartography} of the LEO region. Note that, at the end of the
  ReDSHIFT project, all the maps will be available on the ReDSHIFT
  website (http://redshift-h2020.eu/).
	
\begin{figure}[th!]
  \begin{center}
  \includegraphics[width=0.25\textwidth]{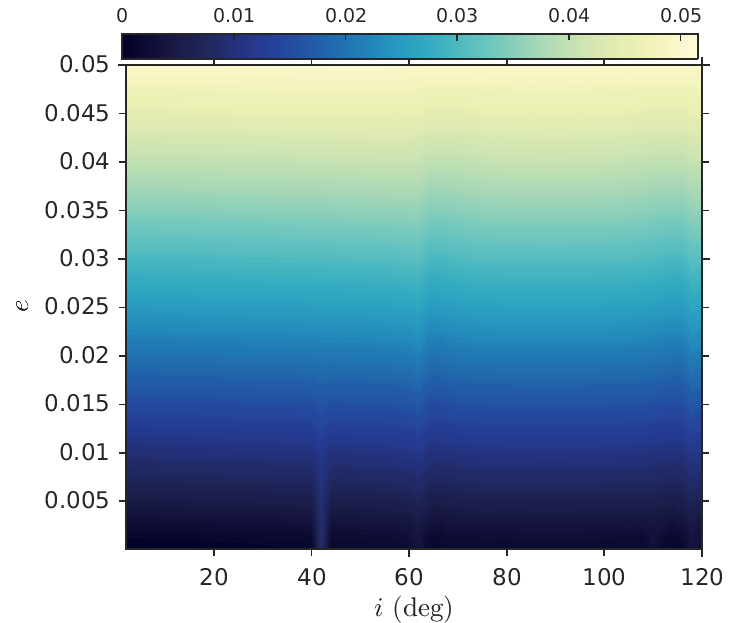}  \hspace{-0.3cm}  \includegraphics[width=0.25\textwidth]{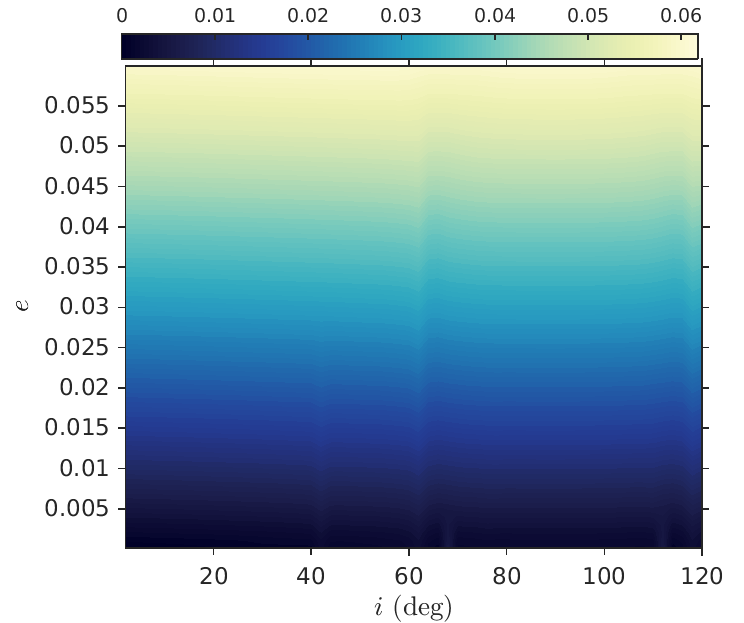}   \hspace{-0.5cm}   \includegraphics[width=0.25\textwidth]{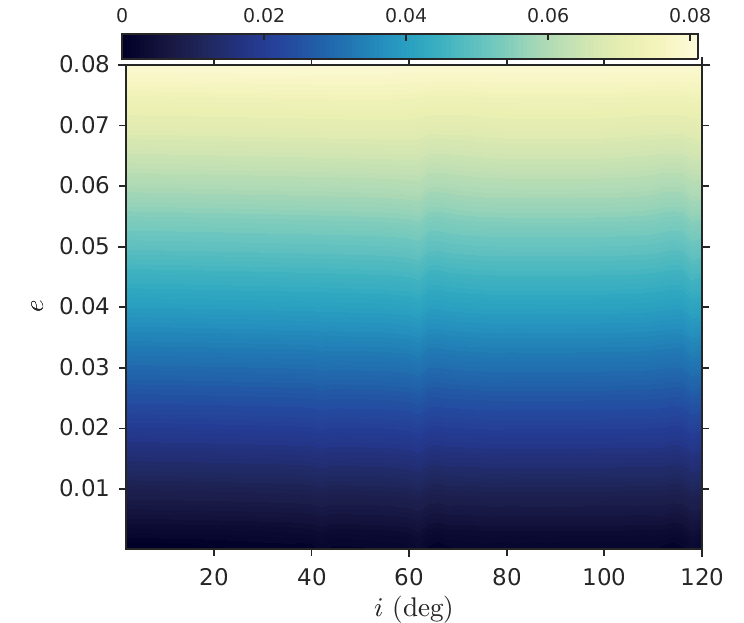}   \hspace{-0.5cm}   \includegraphics[width=0.25\textwidth]{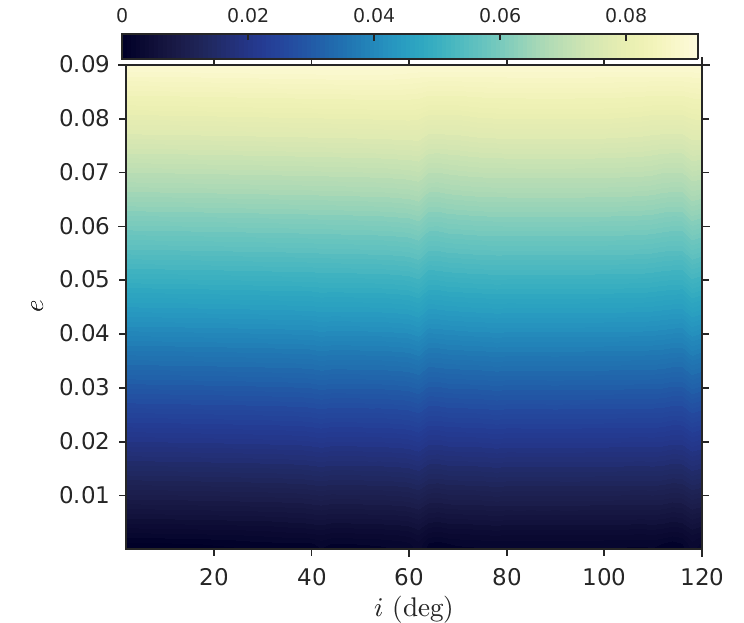} 
    \includegraphics[width=0.25\textwidth]{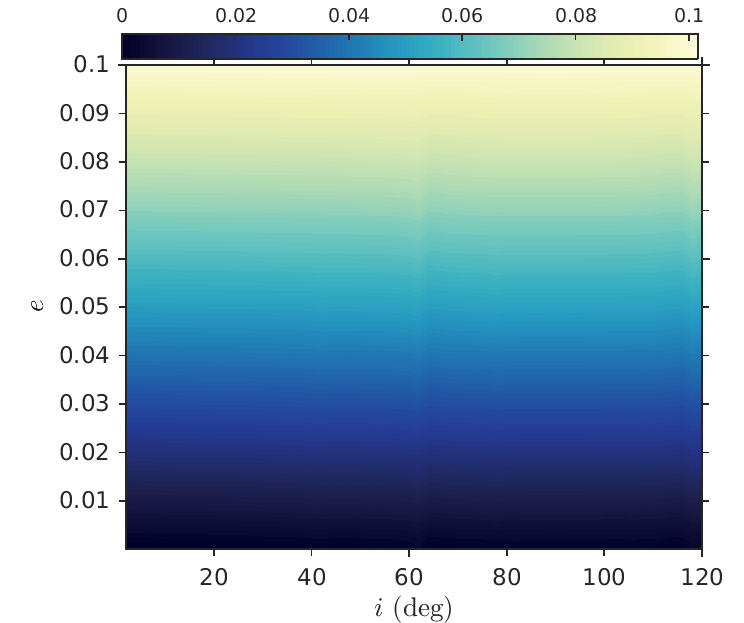}  \hspace{-0.3cm}  \includegraphics[width=0.25\textwidth]{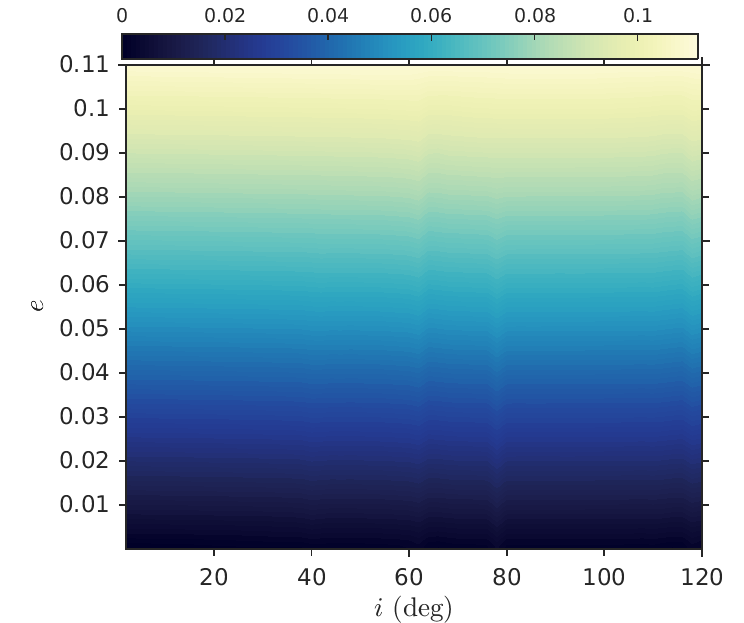}   \hspace{-0.5cm}   \includegraphics[width=0.25\textwidth]{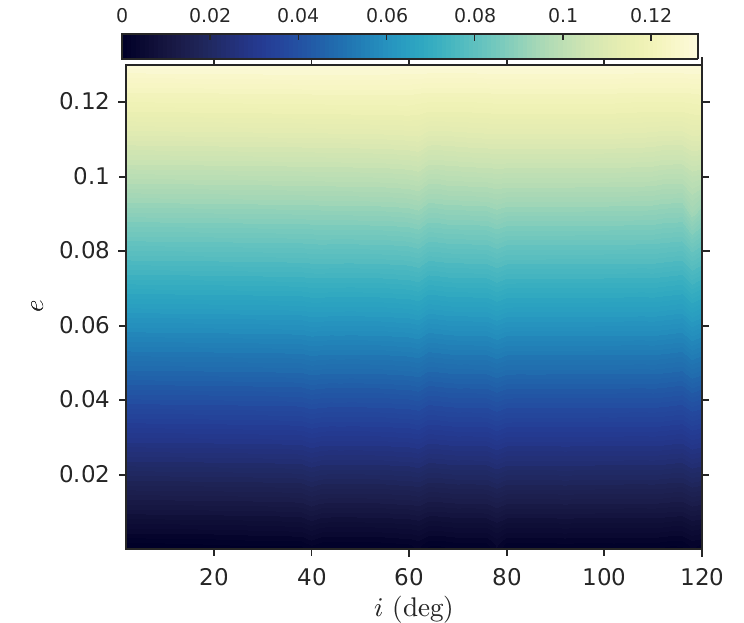}   \hspace{-0.5cm}   \includegraphics[width=0.25\textwidth]{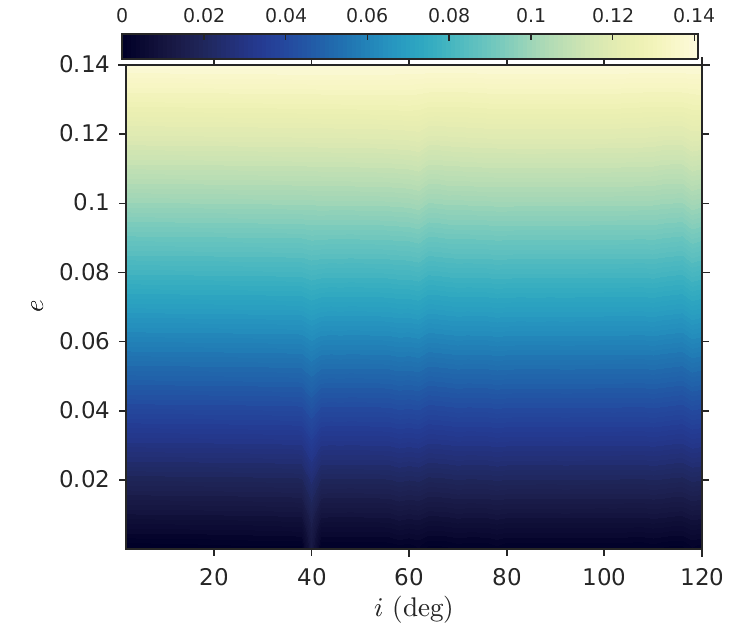} 
      \includegraphics[width=0.25\textwidth]{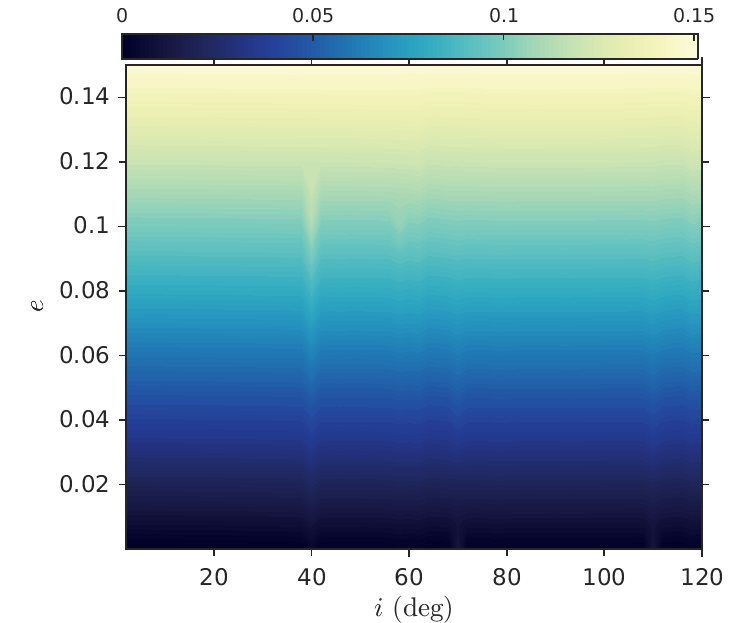}  \hspace{-0.3cm}  \includegraphics[width=0.25\textwidth]{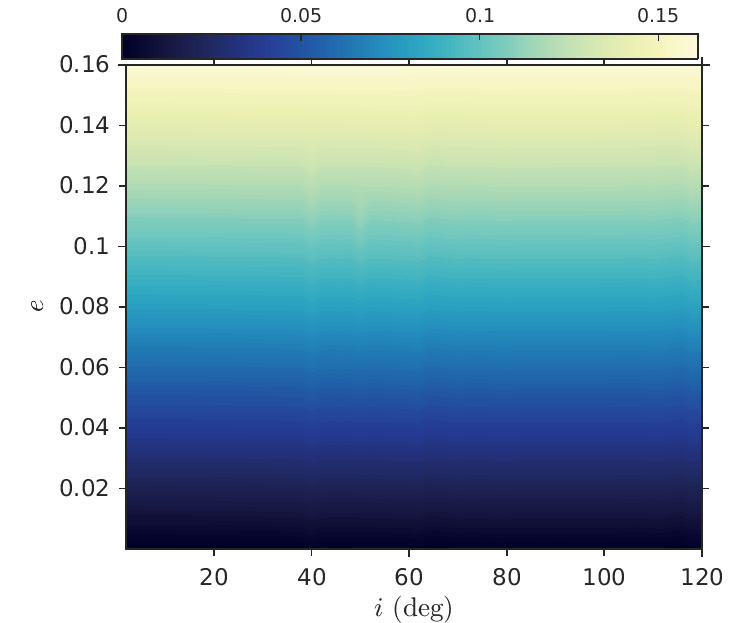}   \hspace{-0.5cm}   \includegraphics[width=0.25\textwidth]{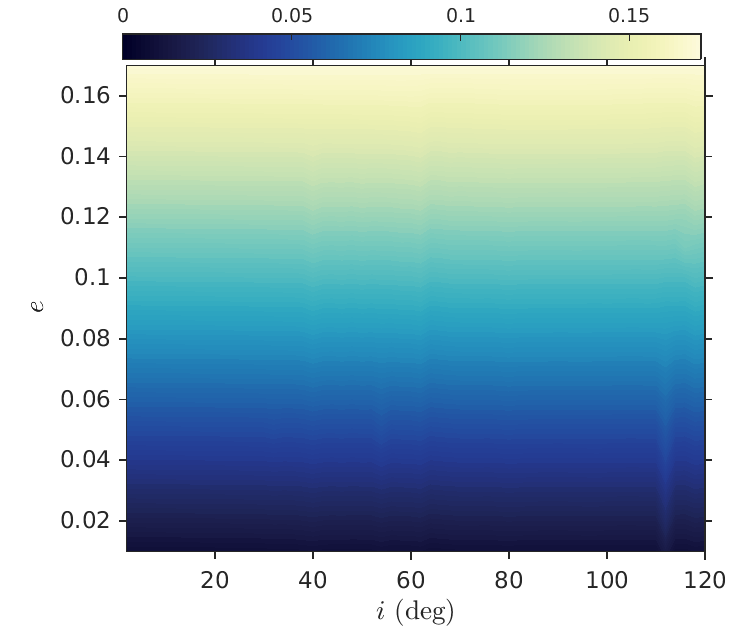}   \hspace{-0.5cm}   \includegraphics[width=0.25\textwidth]{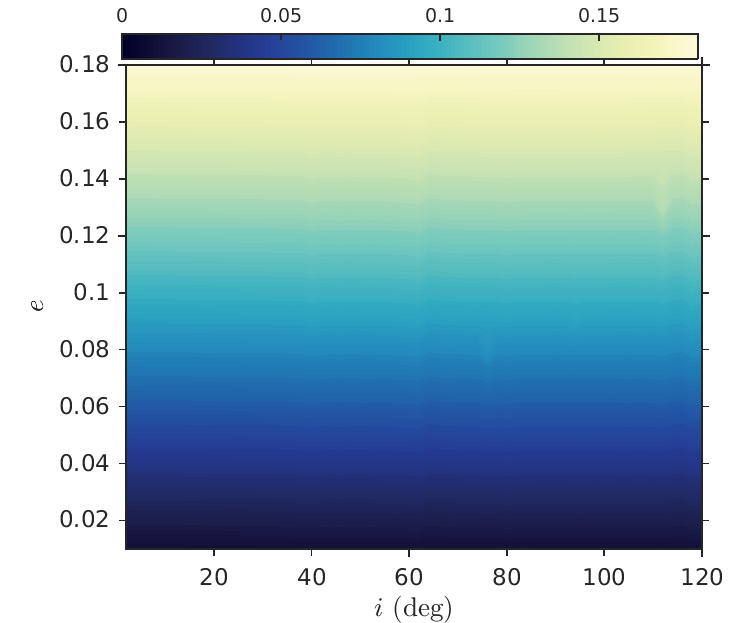} 
        \includegraphics[width=0.25\textwidth]{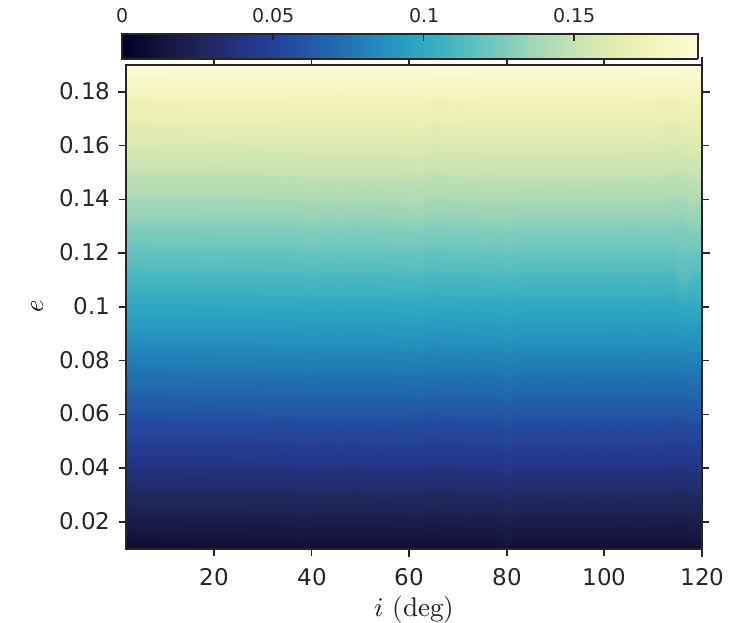}  \hspace{-0.3cm}  \includegraphics[width=0.25\textwidth]{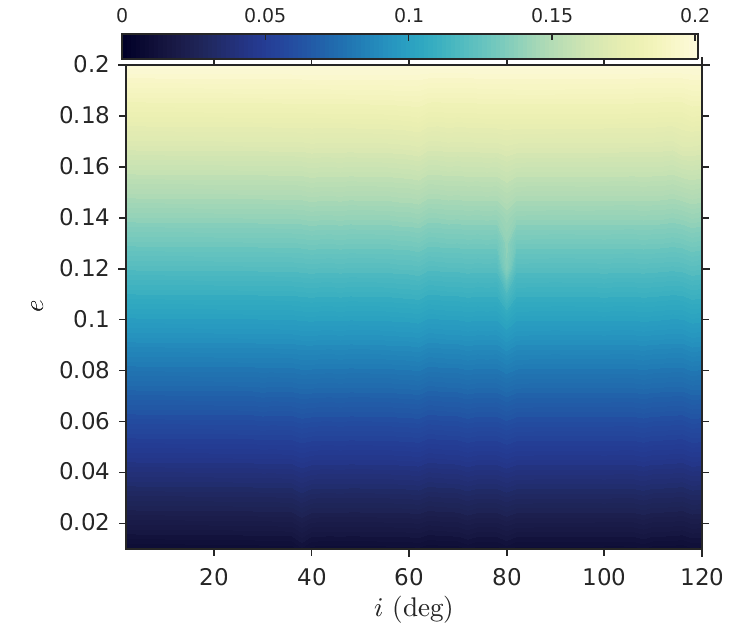}   \hspace{-0.5cm}   \includegraphics[width=0.25\textwidth]{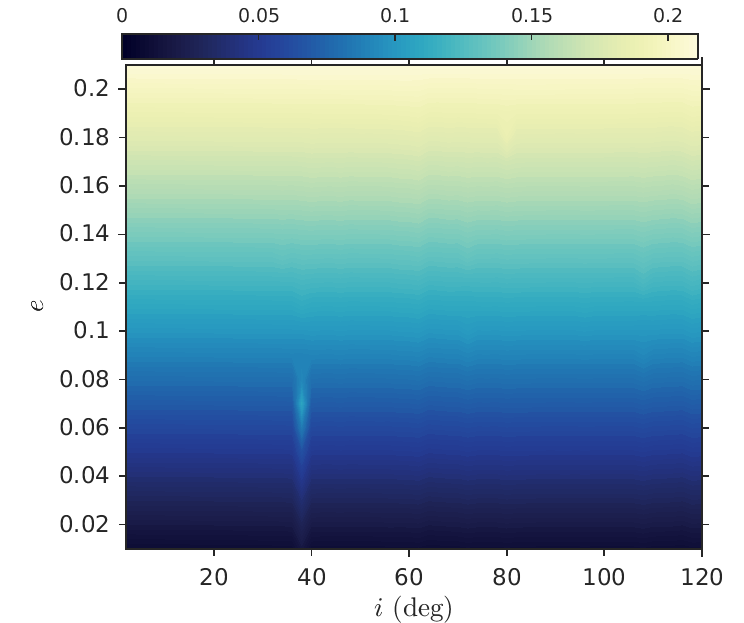}   \hspace{-0.5cm}   \includegraphics[width=0.25\textwidth]{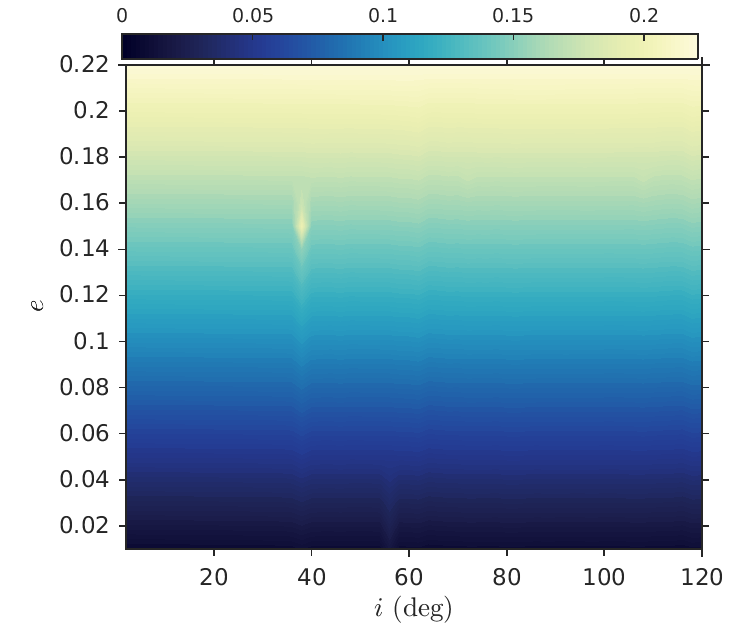} 
          \includegraphics[width=0.25\textwidth]{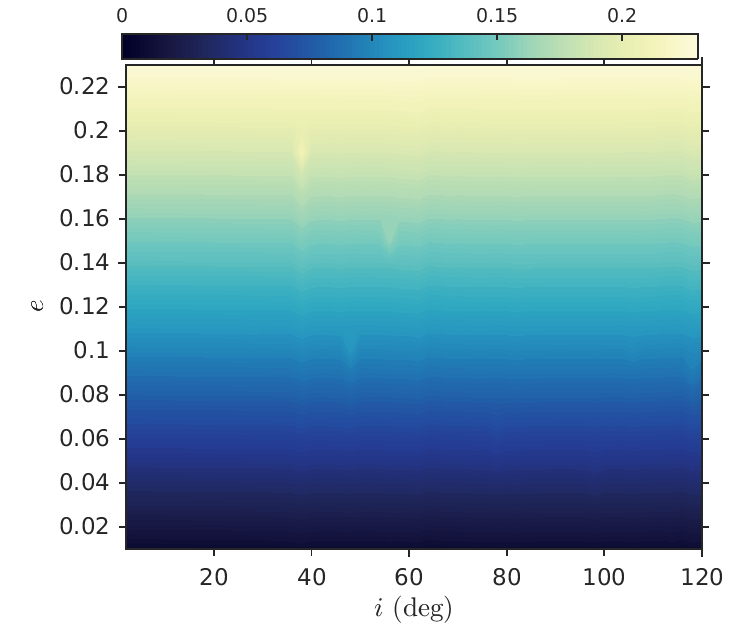}  \hspace{-0.3cm}  \includegraphics[width=0.25\textwidth]{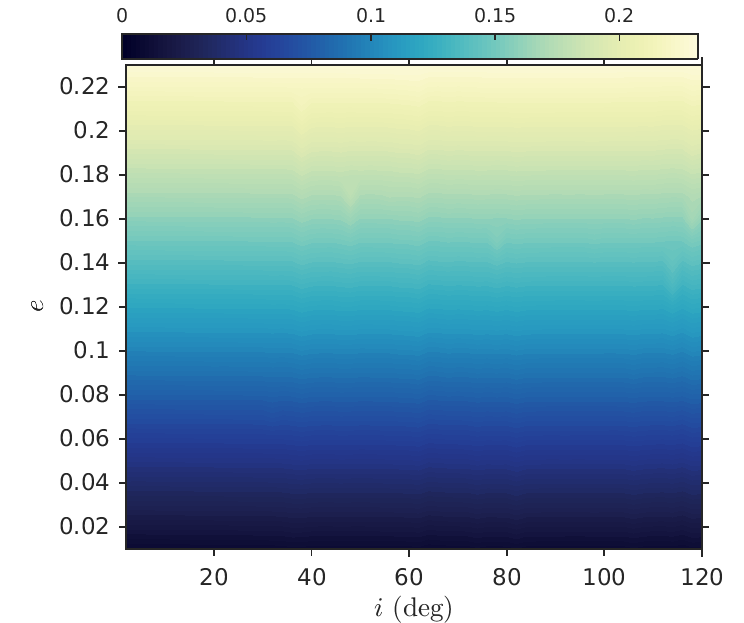}   \hspace{-0.5cm}   \includegraphics[width=0.25\textwidth]{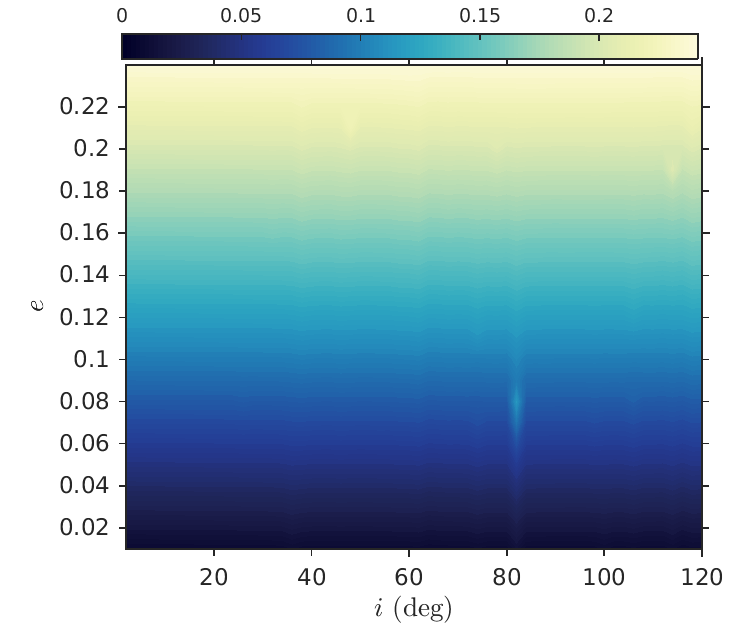}   \hspace{-0.5cm}   \includegraphics[width=0.25\textwidth]{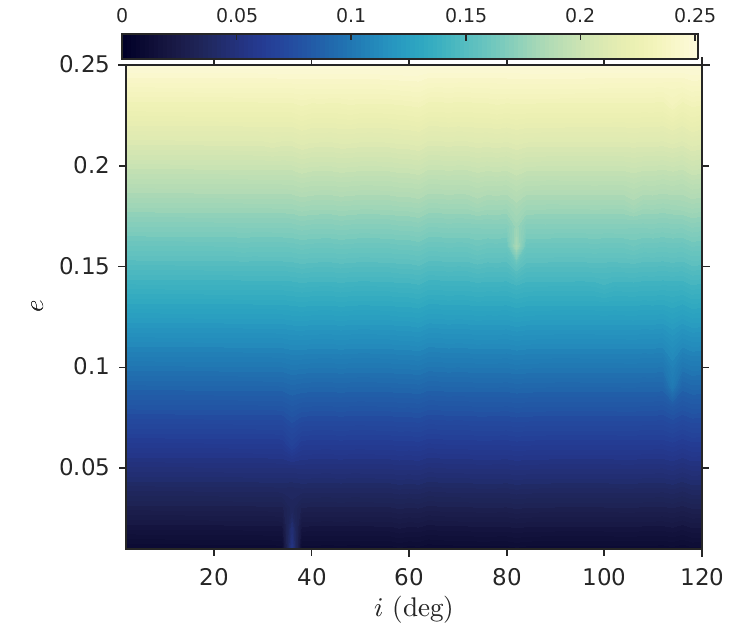} 
            \includegraphics[width=0.25\textwidth]{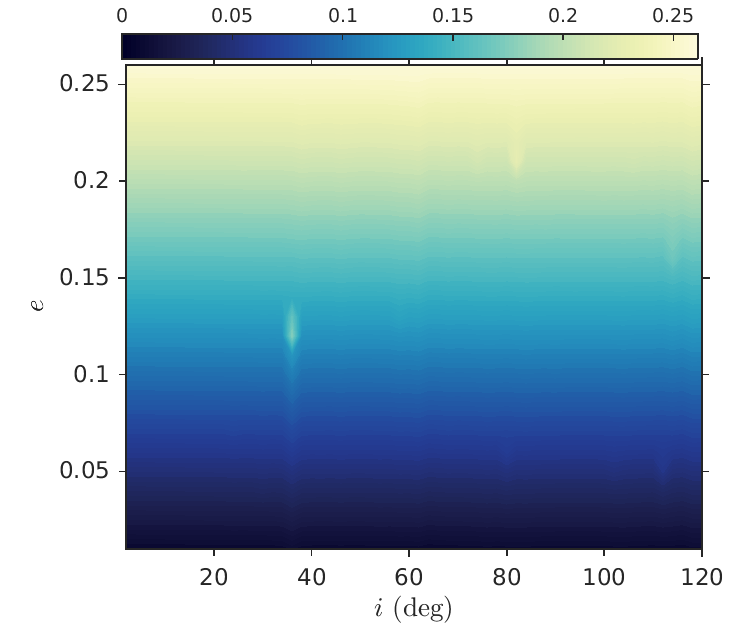}  \hspace{-0.25cm}  \includegraphics[width=0.25\textwidth]{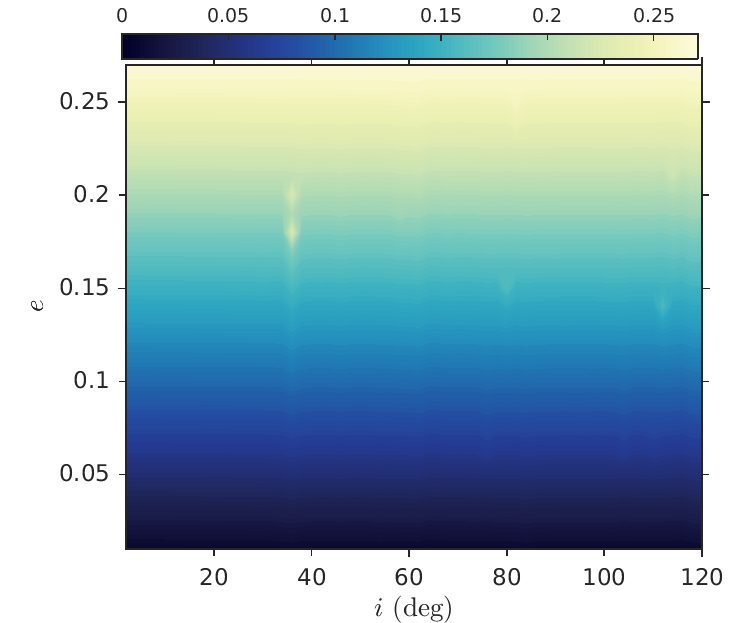}   \hspace{-0.5cm}   \includegraphics[width=0.25\textwidth]{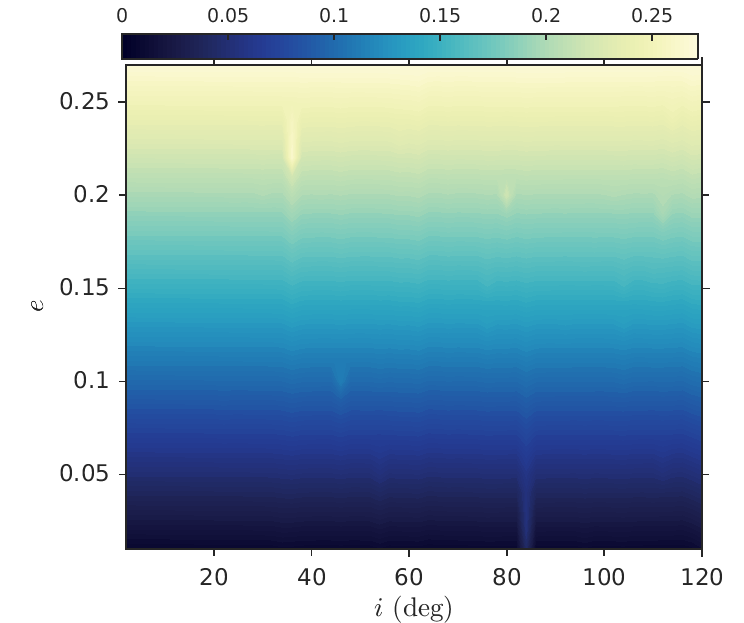}   \hspace{-0.5cm}   \includegraphics[width=0.25\textwidth]{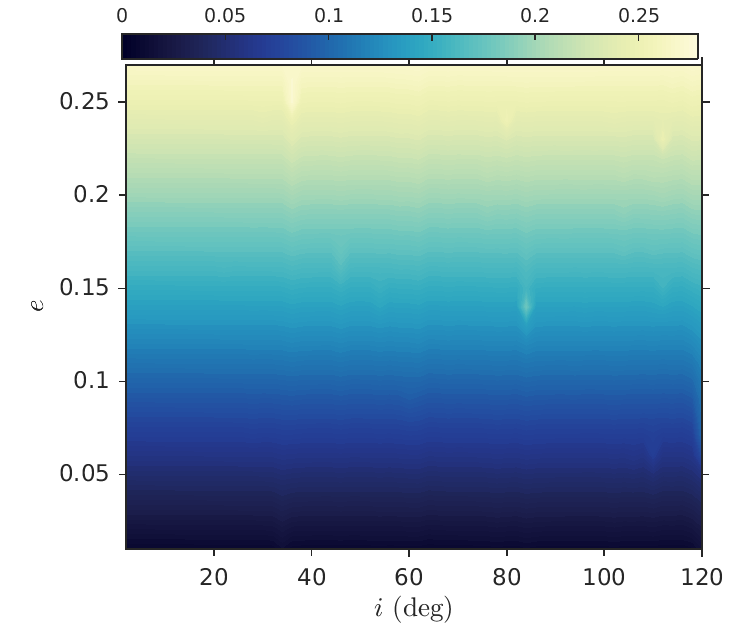} 
            \end{center}
        \caption{Maximum eccentricity computed (color bar) as a function of initial inclination and eccentricity for the initial epoch 2020 and $C_R (A/m)=0.024$
  m$^2/$kg, assuming $\Omega=0^{\circ}$ and $\omega=0^{\circ}$ at the initial epoch. Each plot depicts the behavior computed starting from a different value of initial semi-major axis. From the top left to the bottom right: $a=R_E+ 700$ km to $a=R_E+3000$ km at a step of 100 km.}\label{fig:ie_emax_2020_low_span_a}
  \end{figure}

\begin{figure}[th!]
  \begin{center}
  \includegraphics[width=0.32\textwidth]{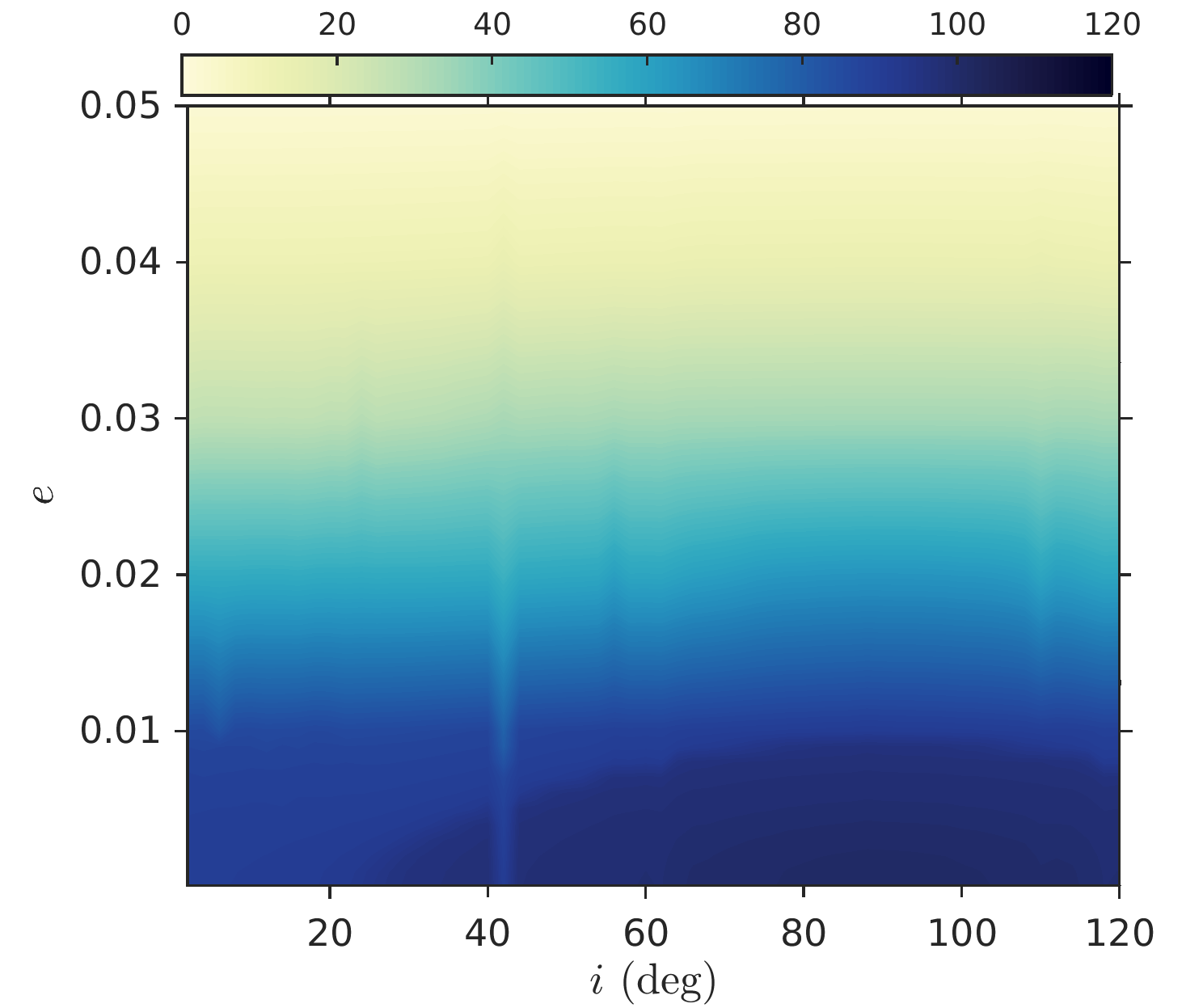}  \hspace{0cm}  \includegraphics[width=0.32\textwidth]{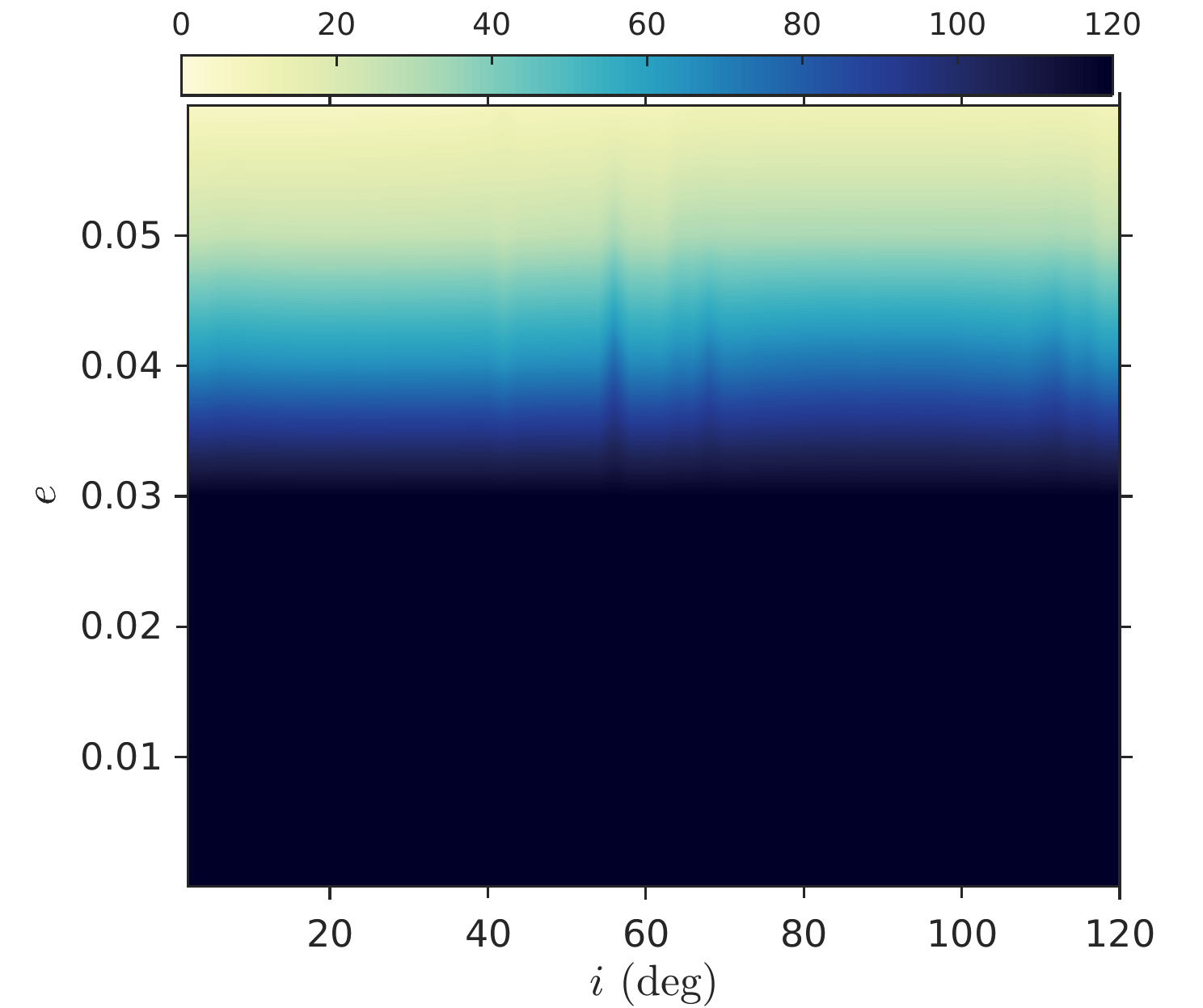}   \hspace{0cm}   \includegraphics[width=0.32\textwidth]{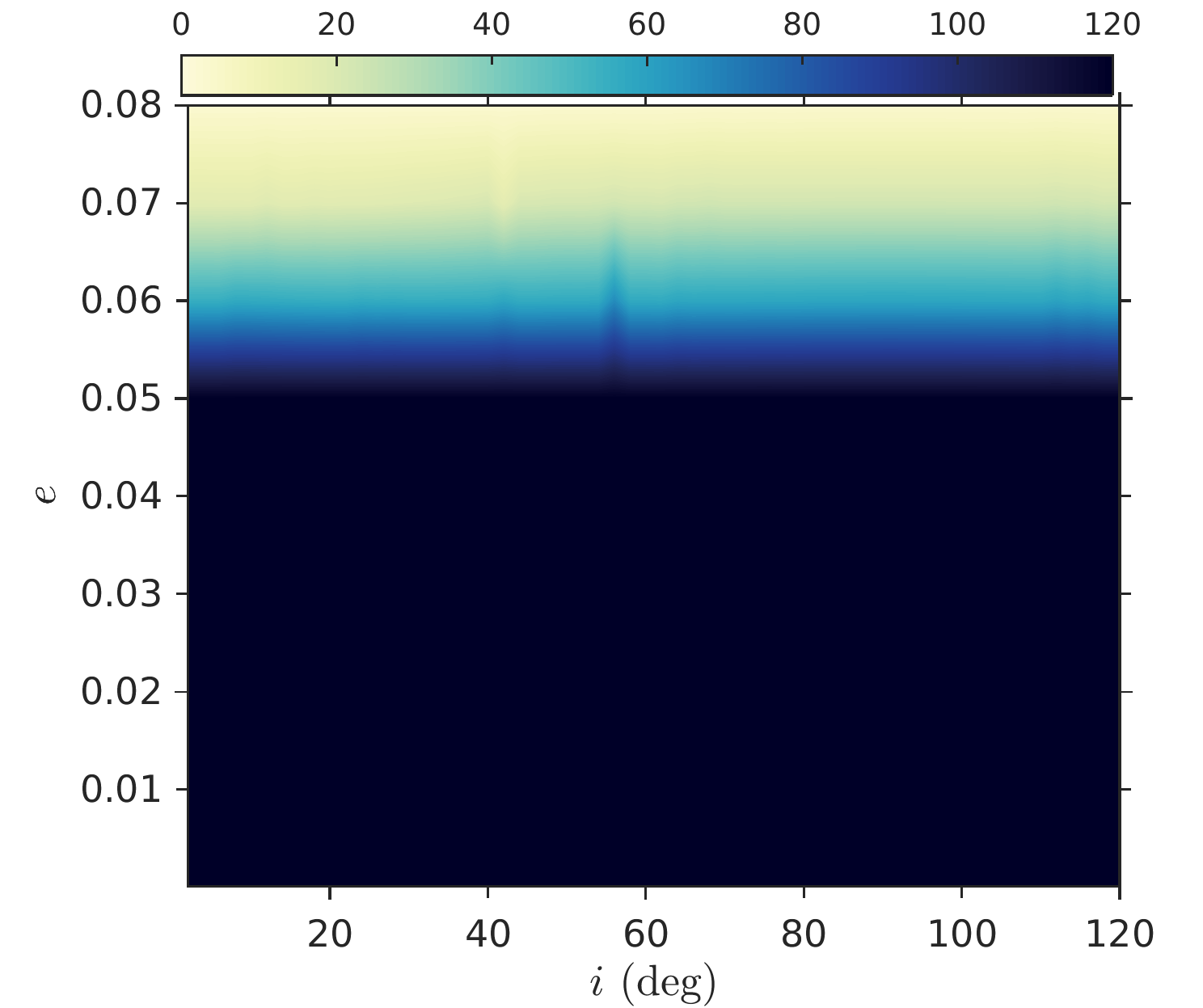}   \hspace{0cm}   \includegraphics[width=0.32\textwidth]{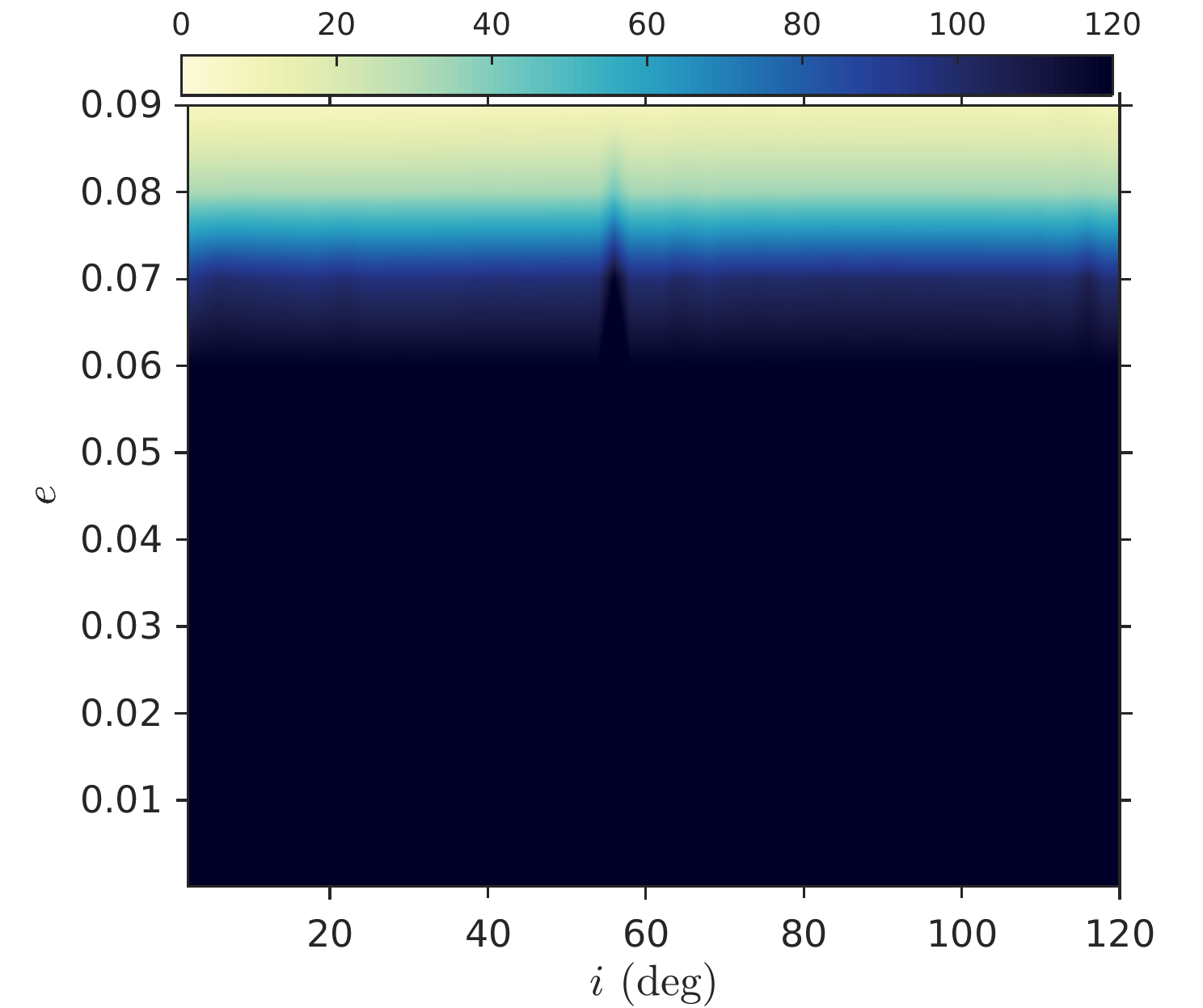} 
    \includegraphics[width=0.32\textwidth]{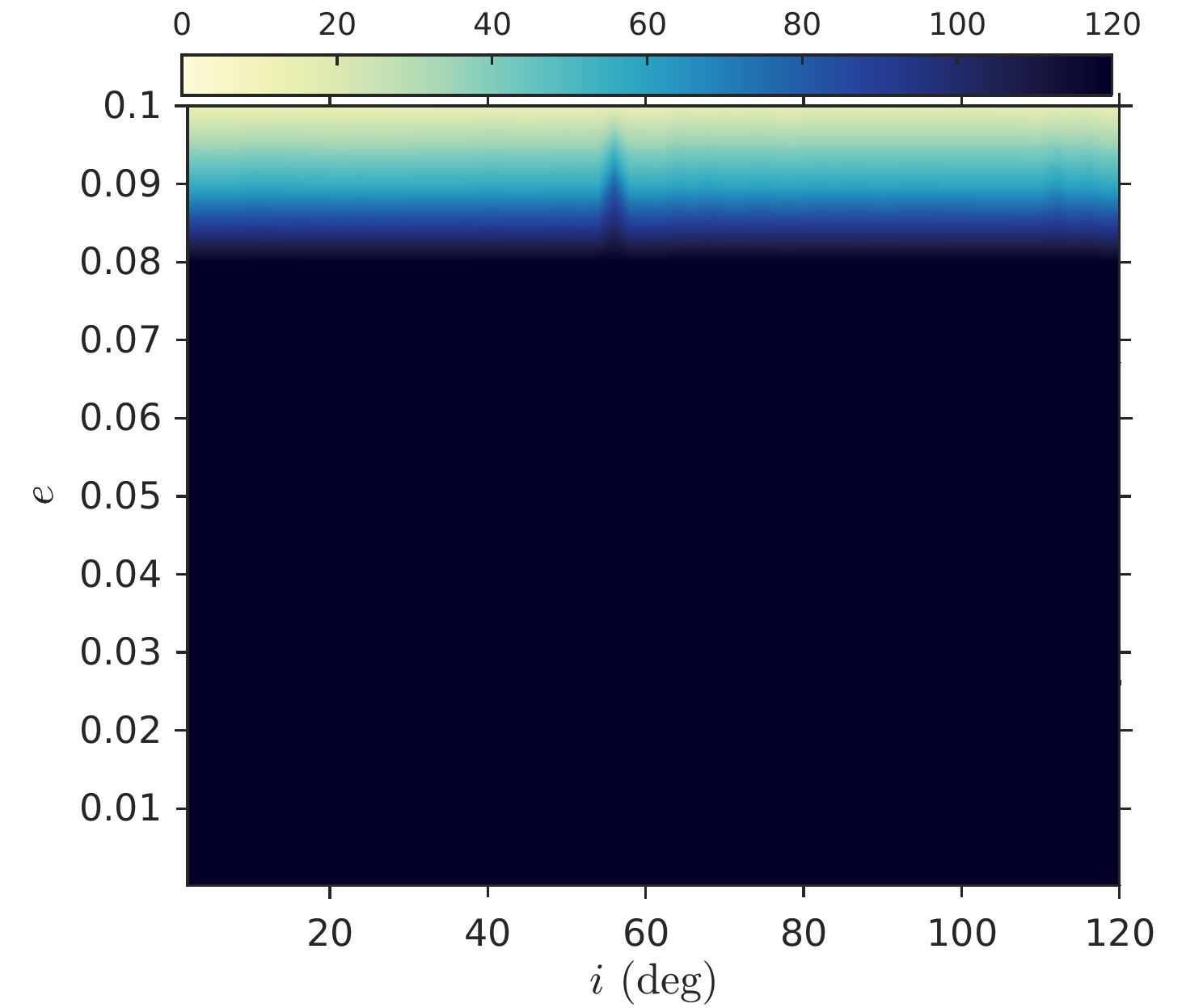}  \hspace{0cm}  \includegraphics[width=0.32\textwidth]{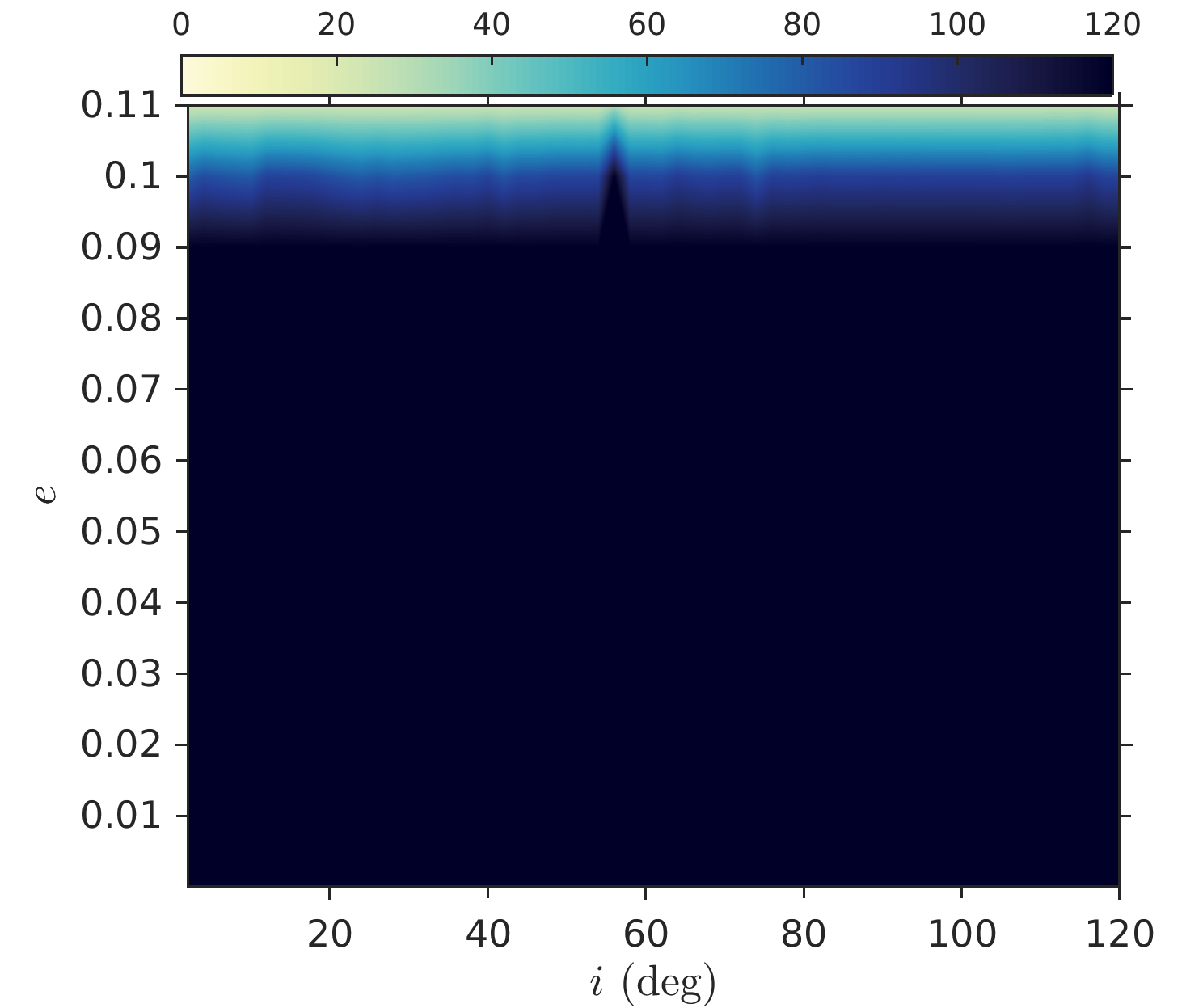}   \hspace{0cm}   \includegraphics[width=0.32\textwidth]{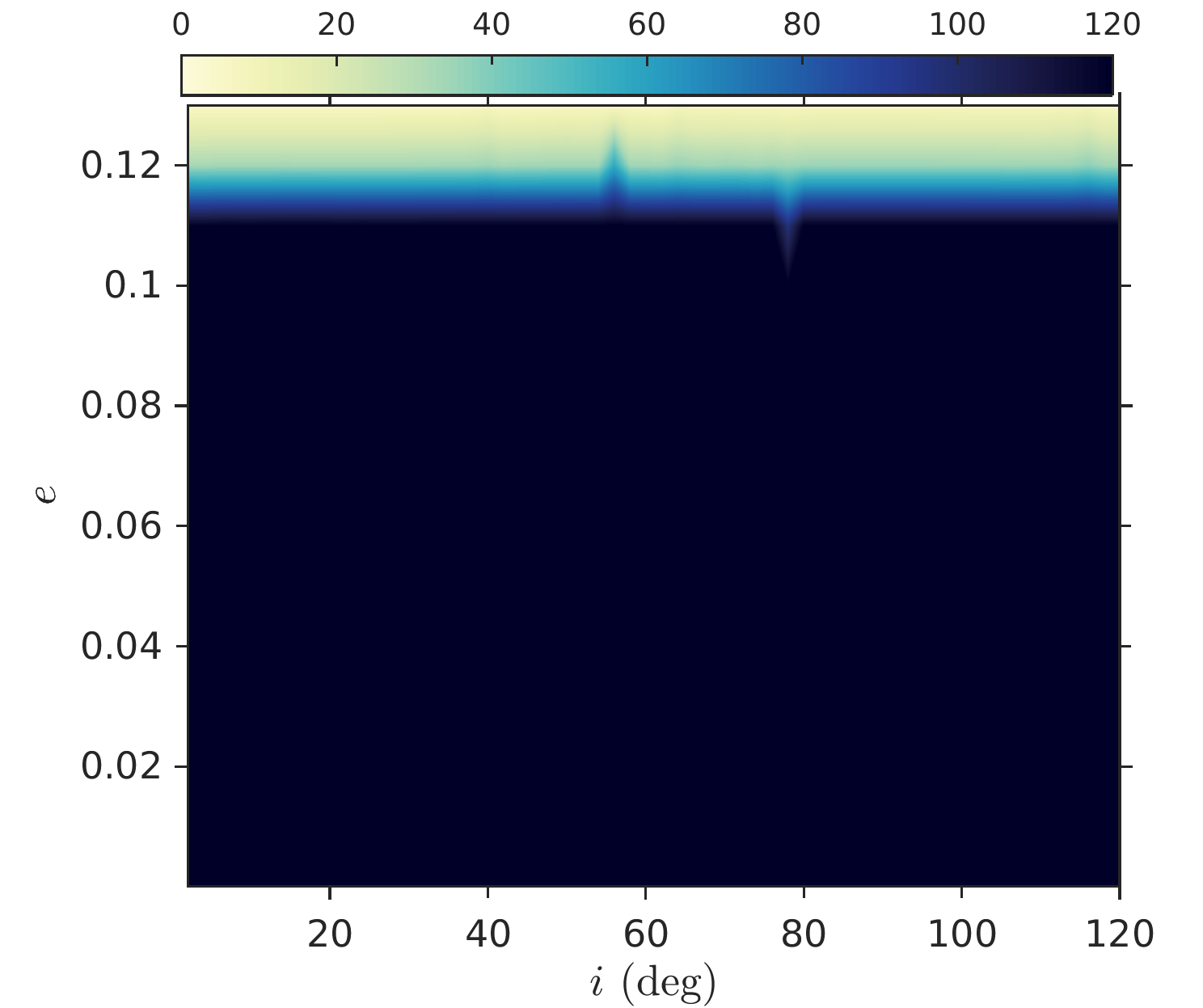}   \hspace{0cm}   \includegraphics[width=0.32\textwidth]{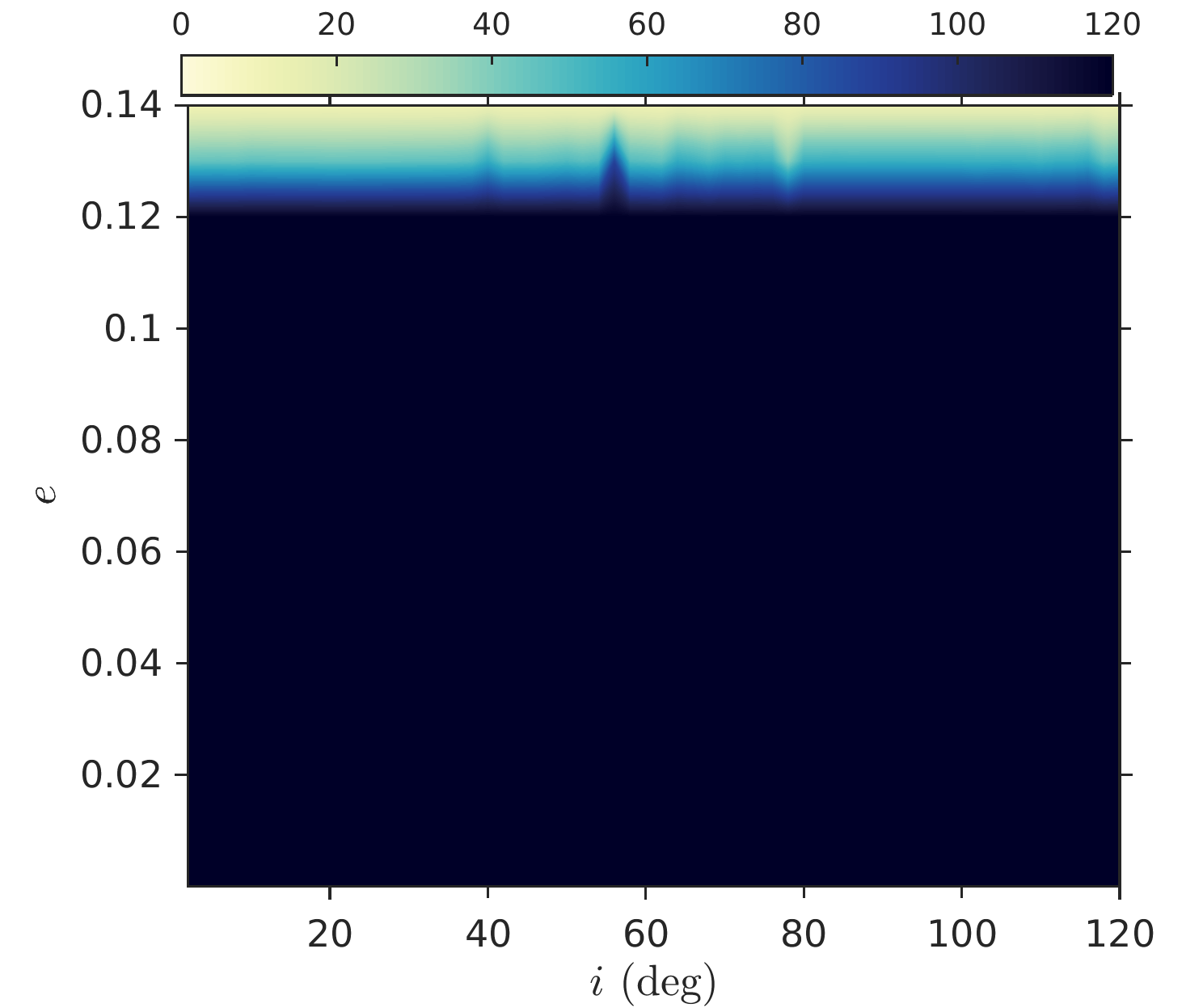} 
      \includegraphics[width=0.32\textwidth]{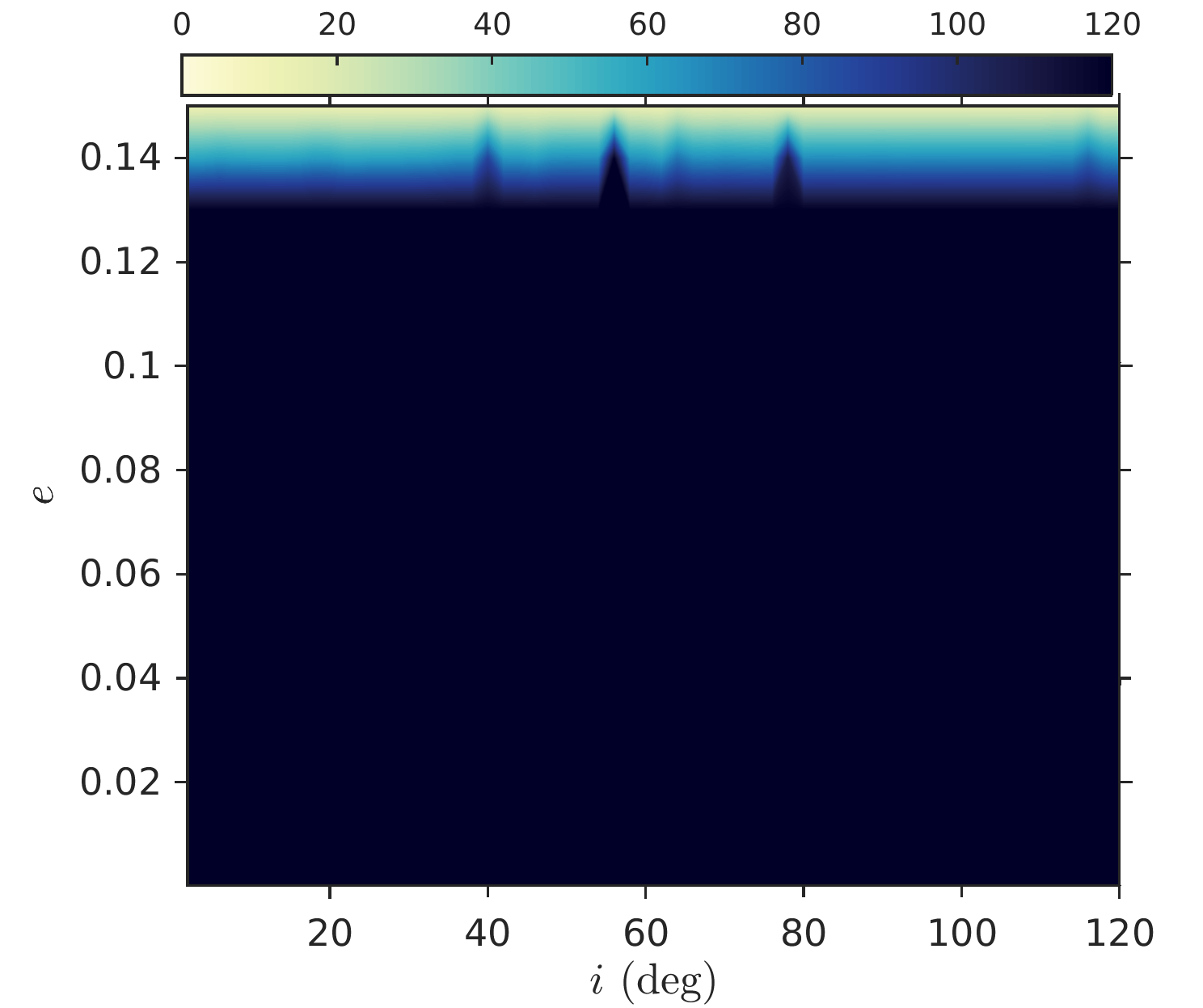}  \hspace{0cm}  \includegraphics[width=0.32\textwidth]{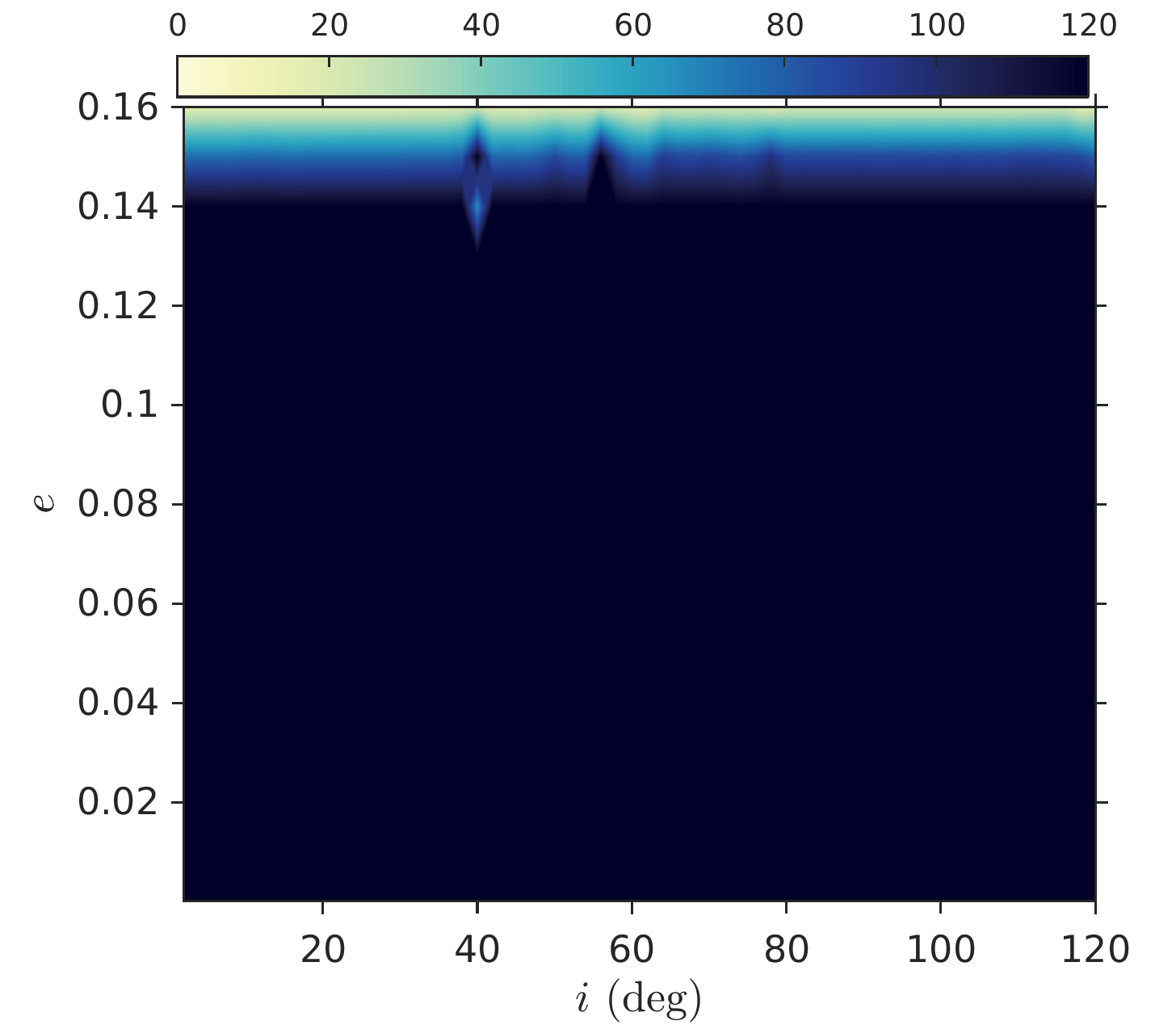}   \hspace{0cm}   \includegraphics[width=0.32\textwidth]{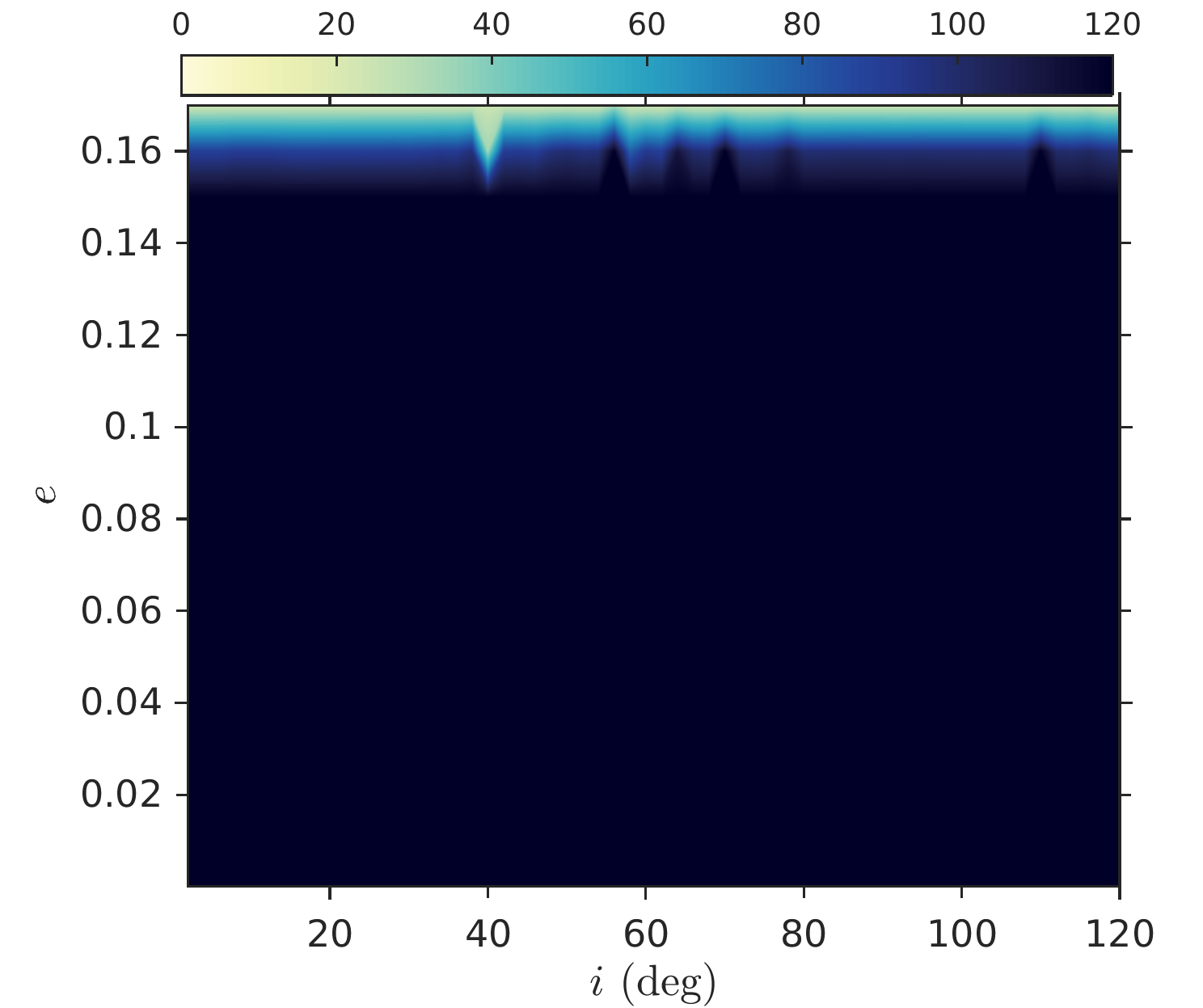}   \hspace{0cm}   \includegraphics[width=0.32\textwidth]{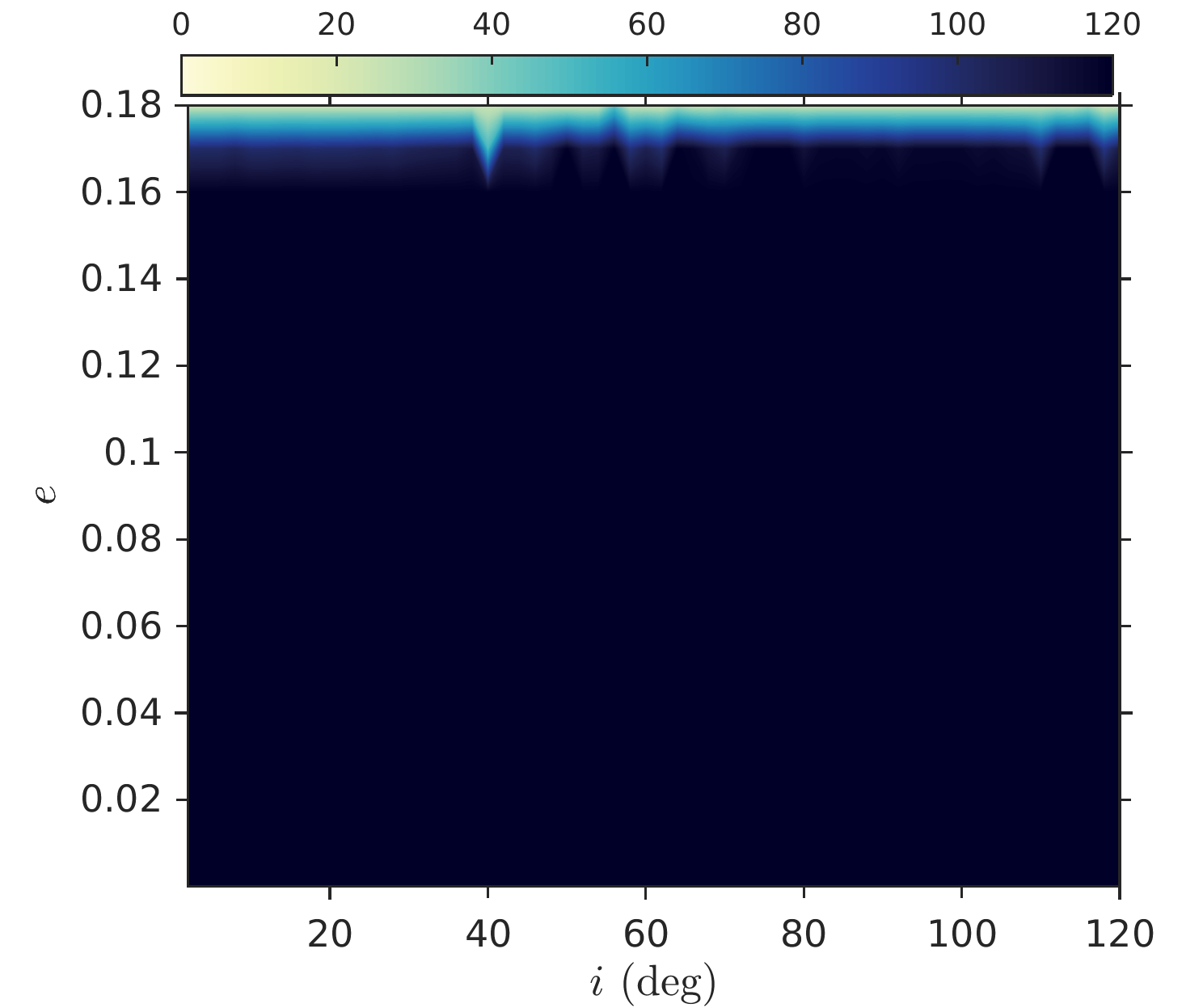} 
            \end{center}
        \caption{Lifetime computed (color bar) as a function of initial inclination and eccentricity for the initial epoch 2020 and $C_R (A/m)=0.024$
  m$^2/$kg, assuming $\Omega=0^{\circ}$ and $\omega=0^{\circ}$ at the initial epoch. Each plot depicts the behavior computed starting from a different value of initial semi-major axis. From the top left to the bottom right: $a=R_E+ 700$ km to $a=R_E+1800$ km at a step of 100 km. For higher values of semi-major axis, the range of eccentricity allowing for a reentry is as narrow as the last plot shown, or even less.}\label{fig:ie_lifetime_2020_low_span_a}
  \end{figure}

\begin{figure}[th!]
  \begin{center}
  \includegraphics[width=0.25\textwidth]{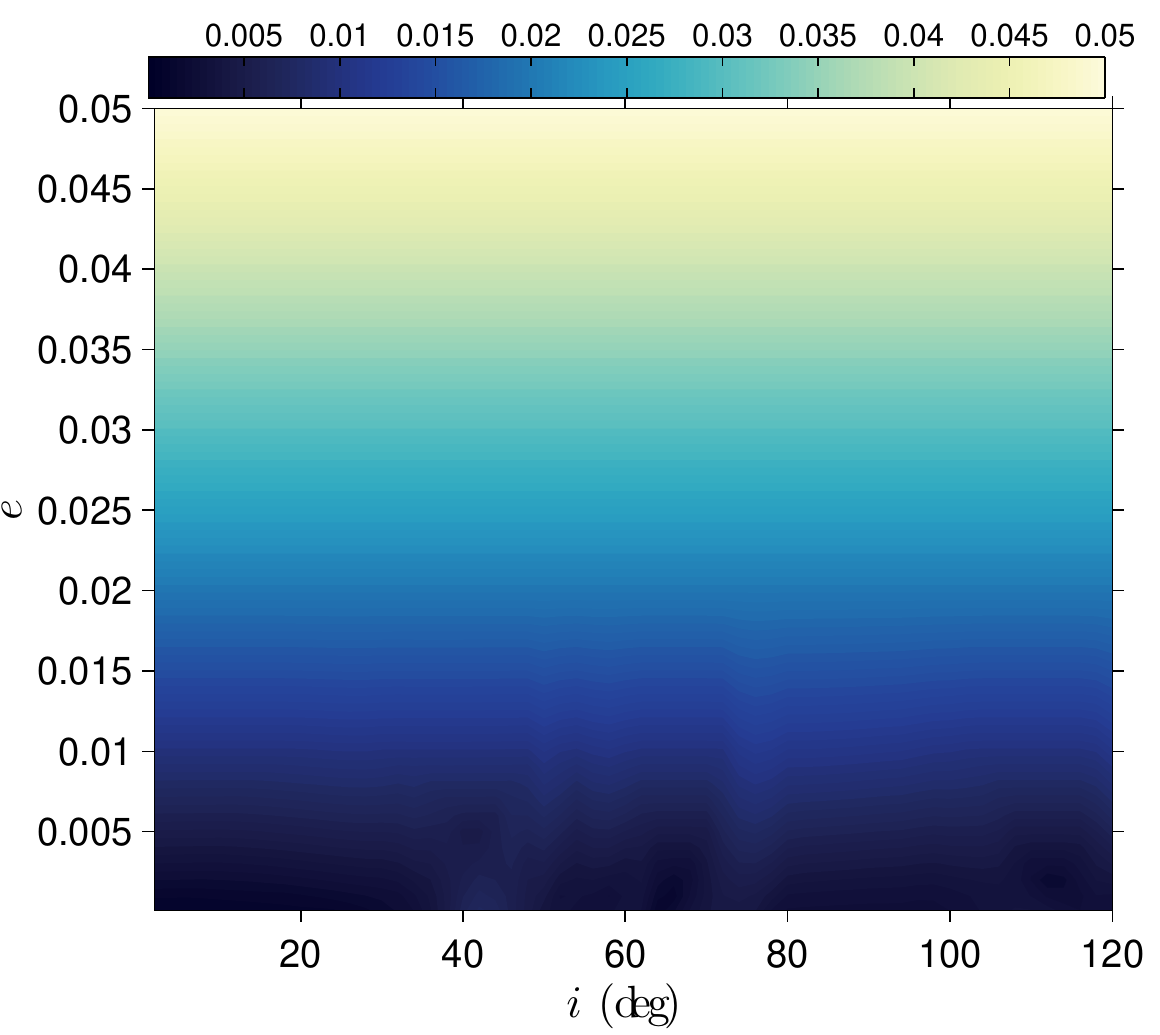}  \hspace{-0.25cm}  \includegraphics[width=0.25\textwidth]{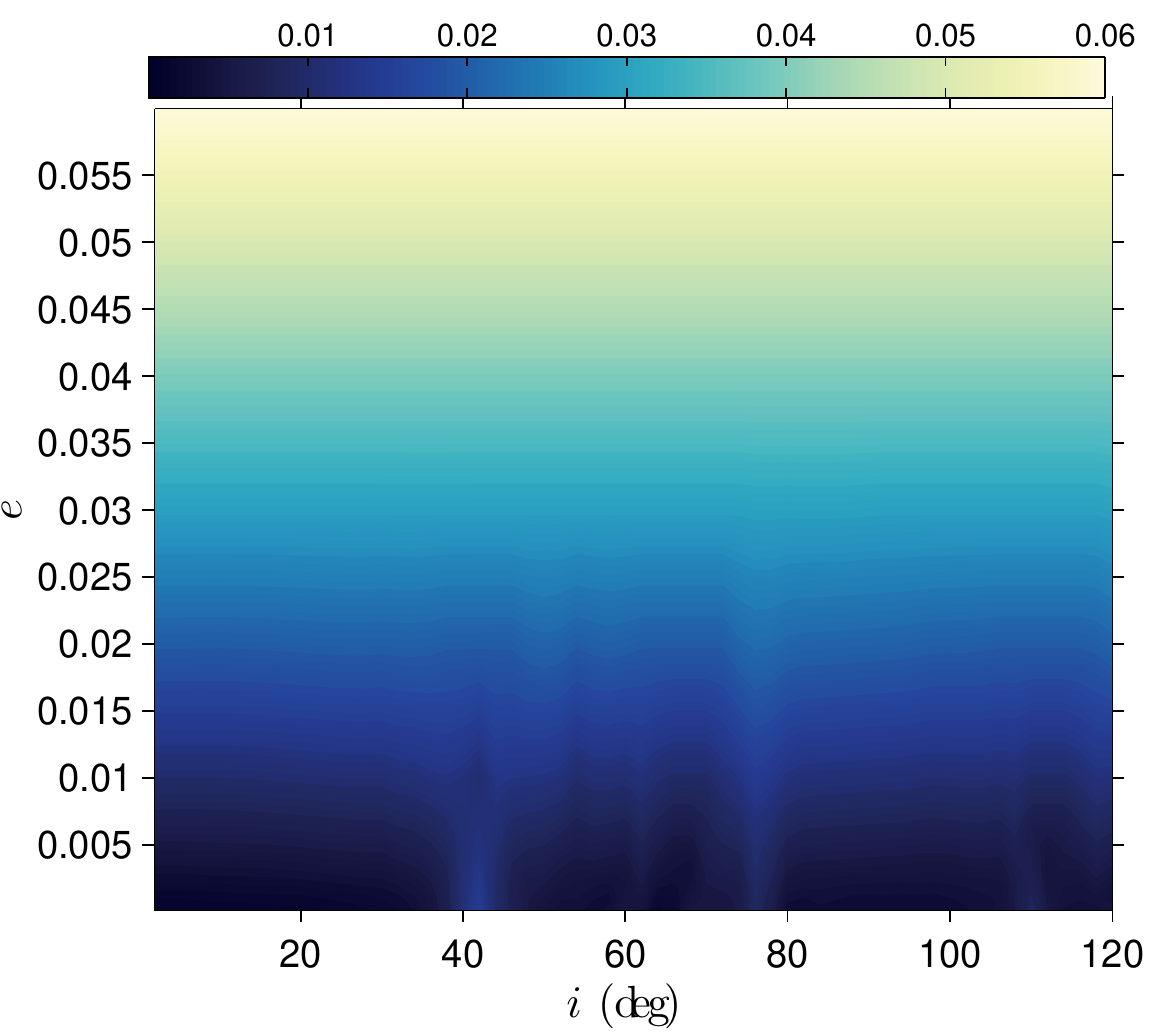}   \hspace{-0.3cm}   \includegraphics[width=0.25\textwidth]{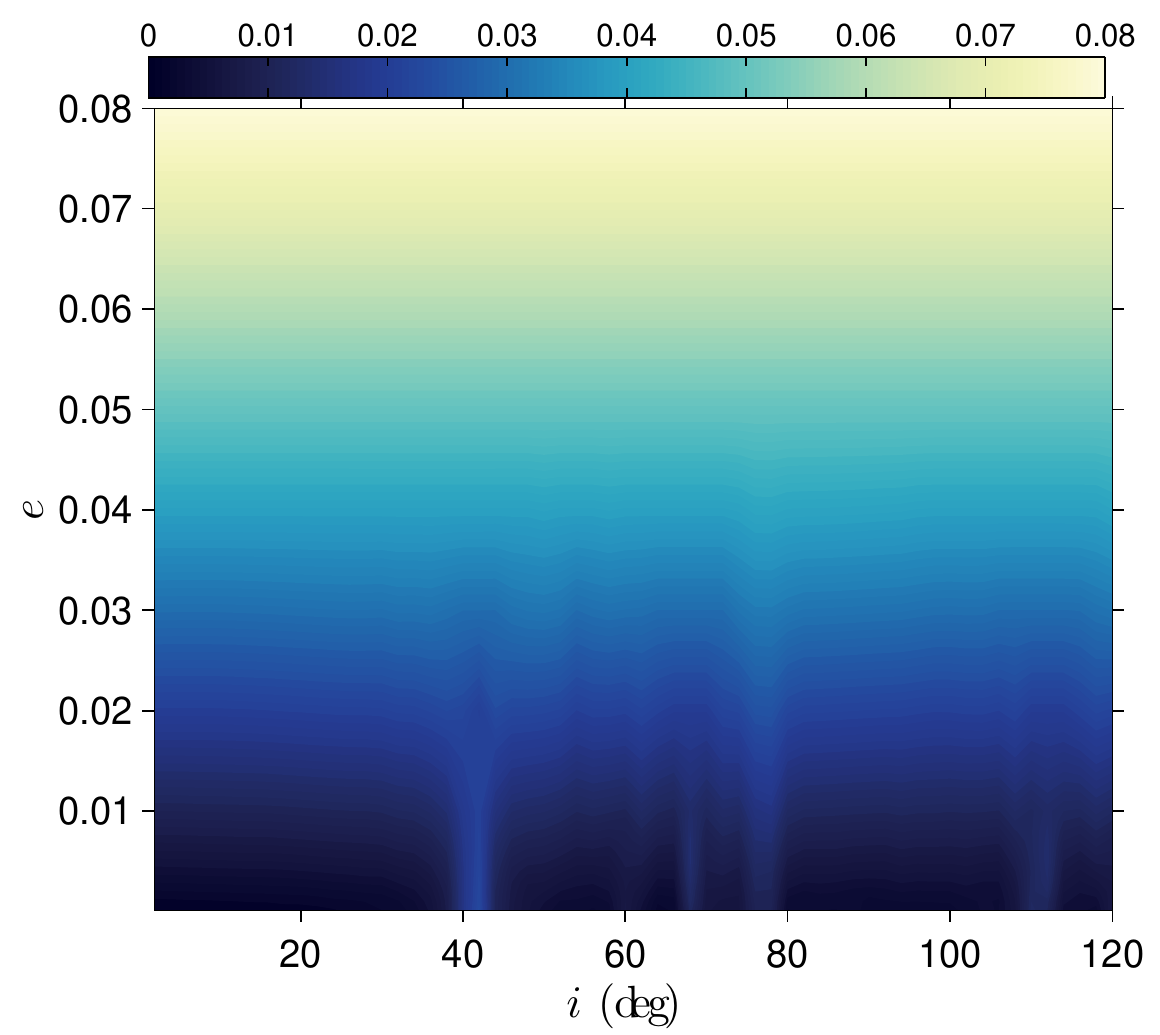}   \hspace{-0.3cm}   \includegraphics[width=0.25\textwidth]{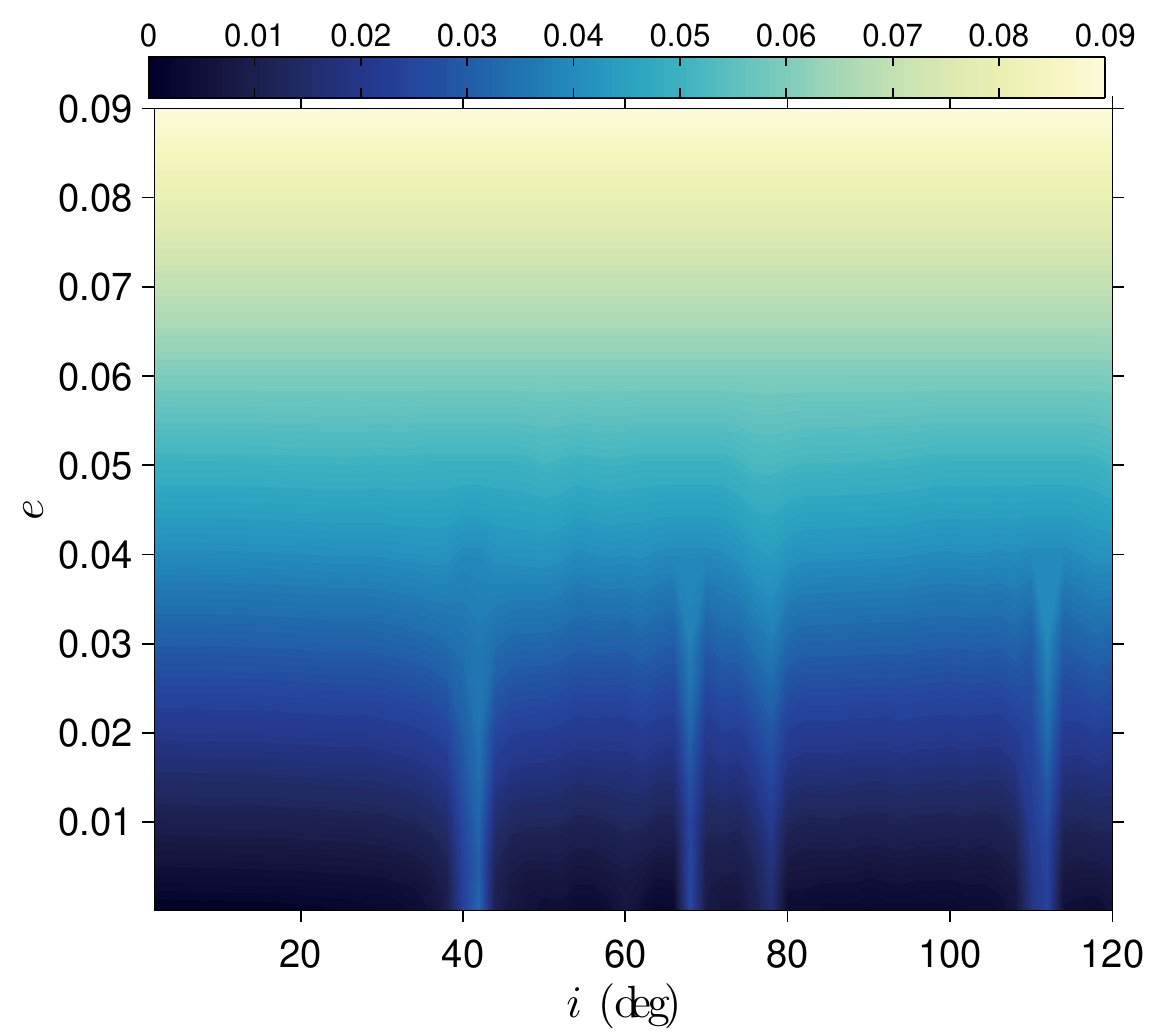} 
    \includegraphics[width=0.25\textwidth]{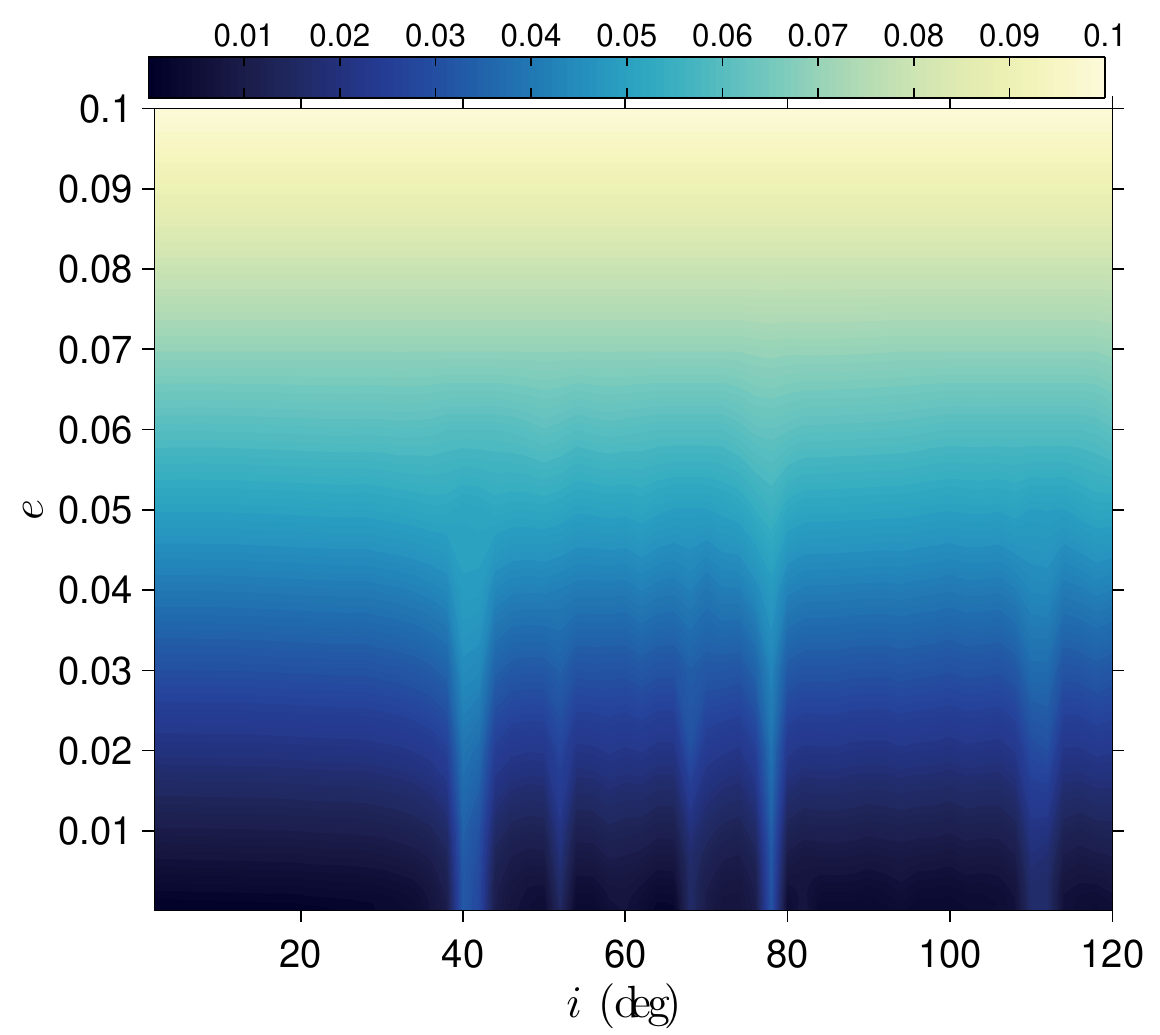}  \hspace{-0.25cm}  \includegraphics[width=0.25\textwidth]{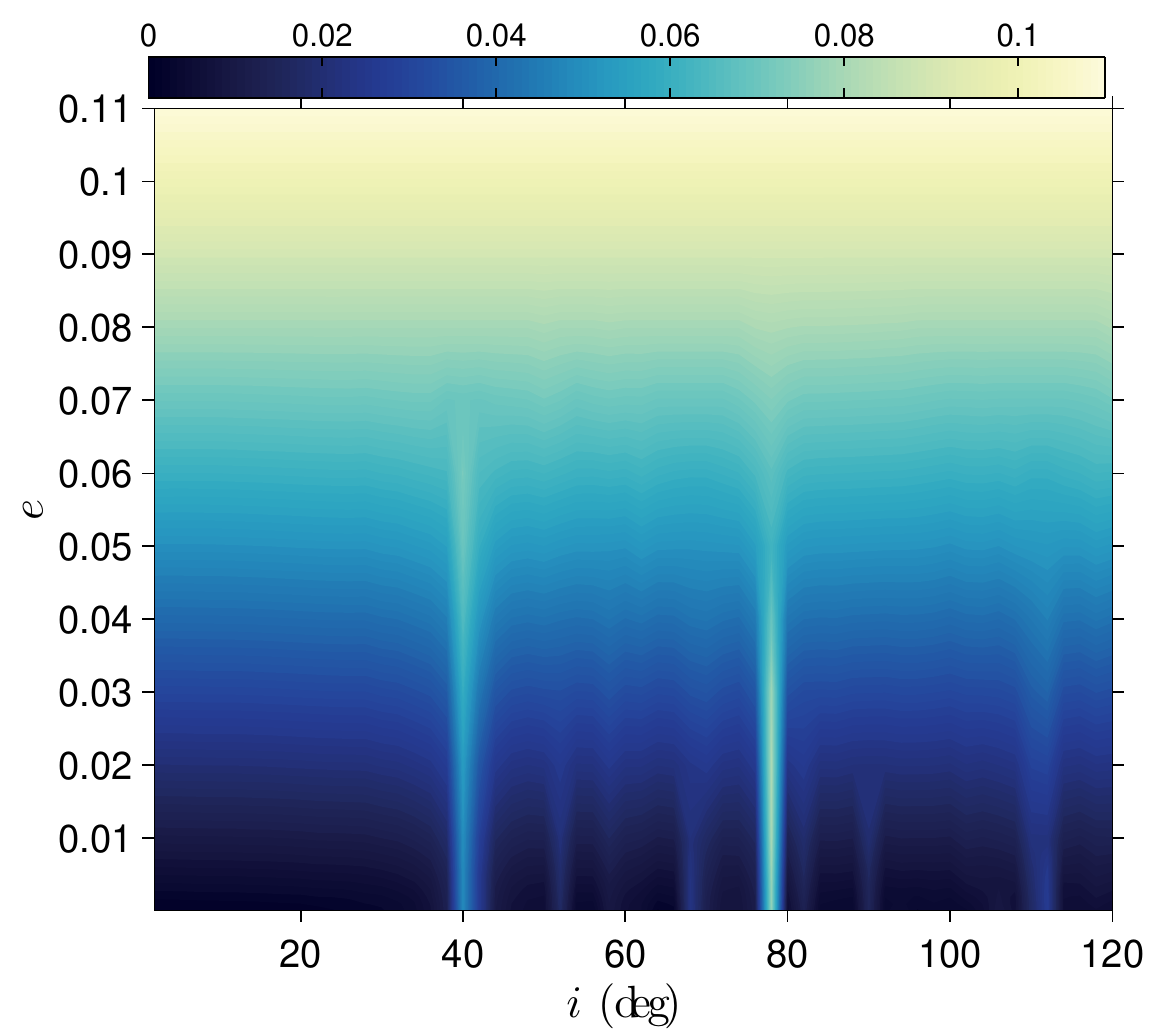}   \hspace{-0.3cm}   \includegraphics[width=0.25\textwidth]{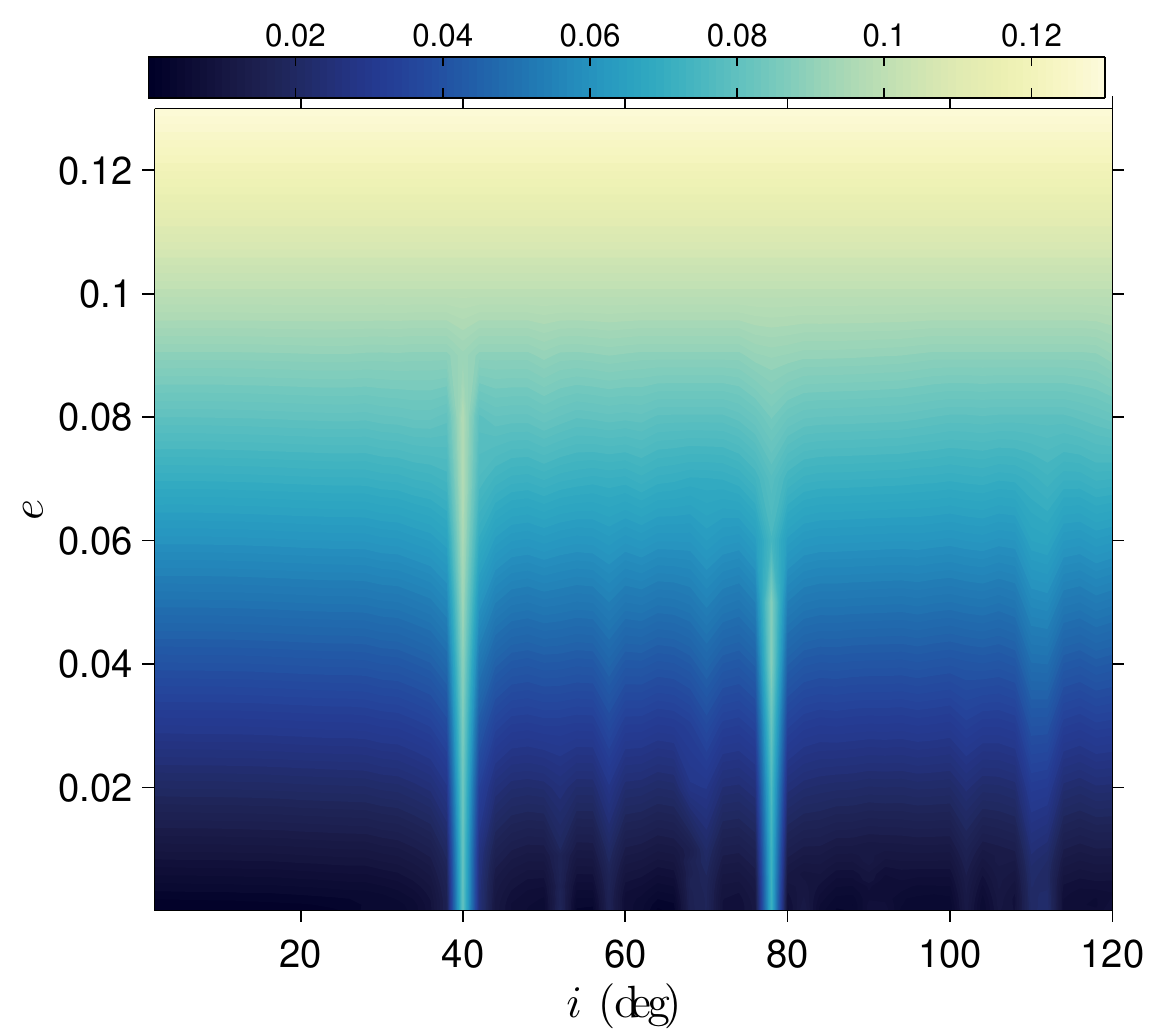}   \hspace{-0.3cm}   \includegraphics[width=0.25\textwidth]{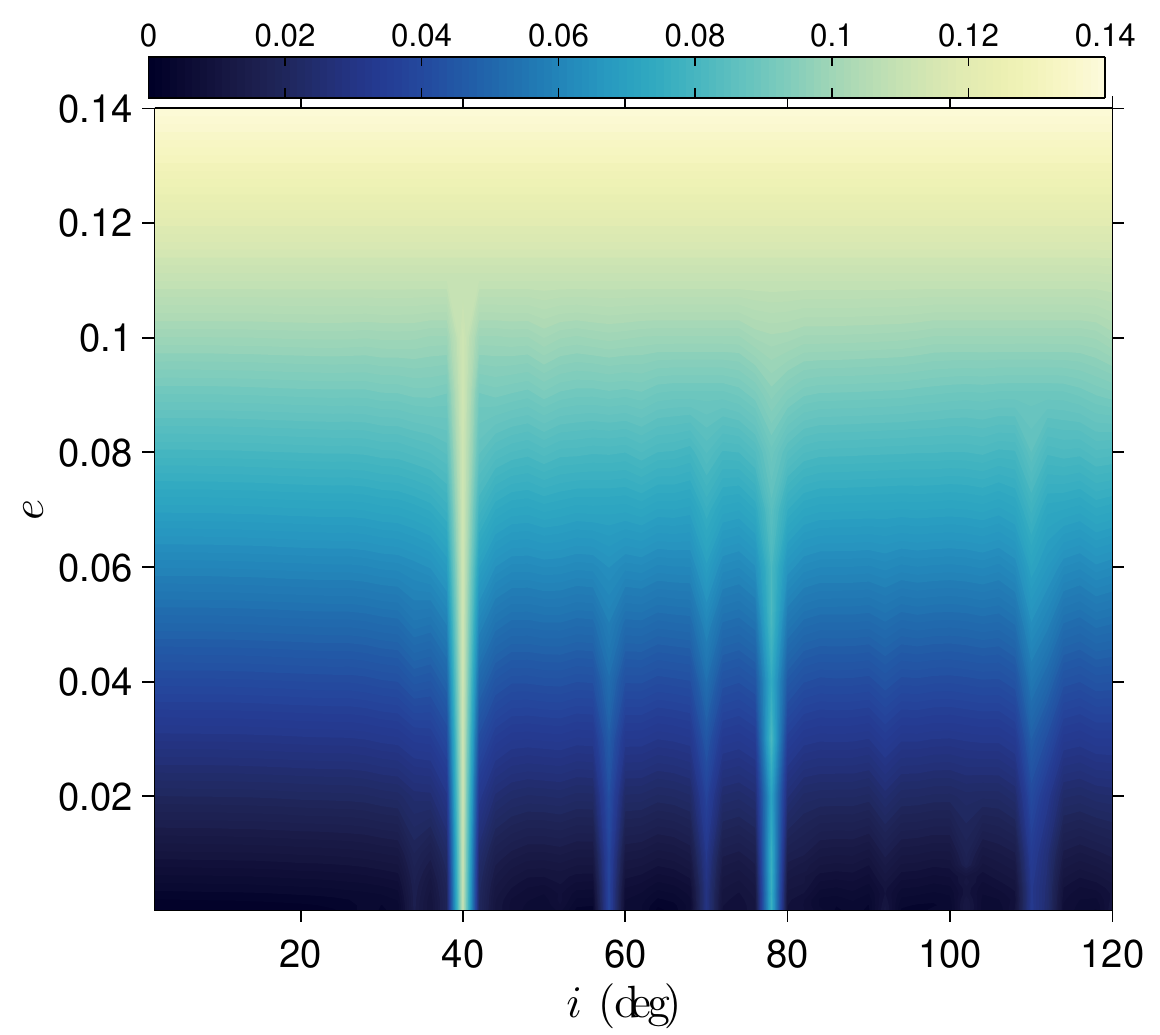} 
      \includegraphics[width=0.25\textwidth]{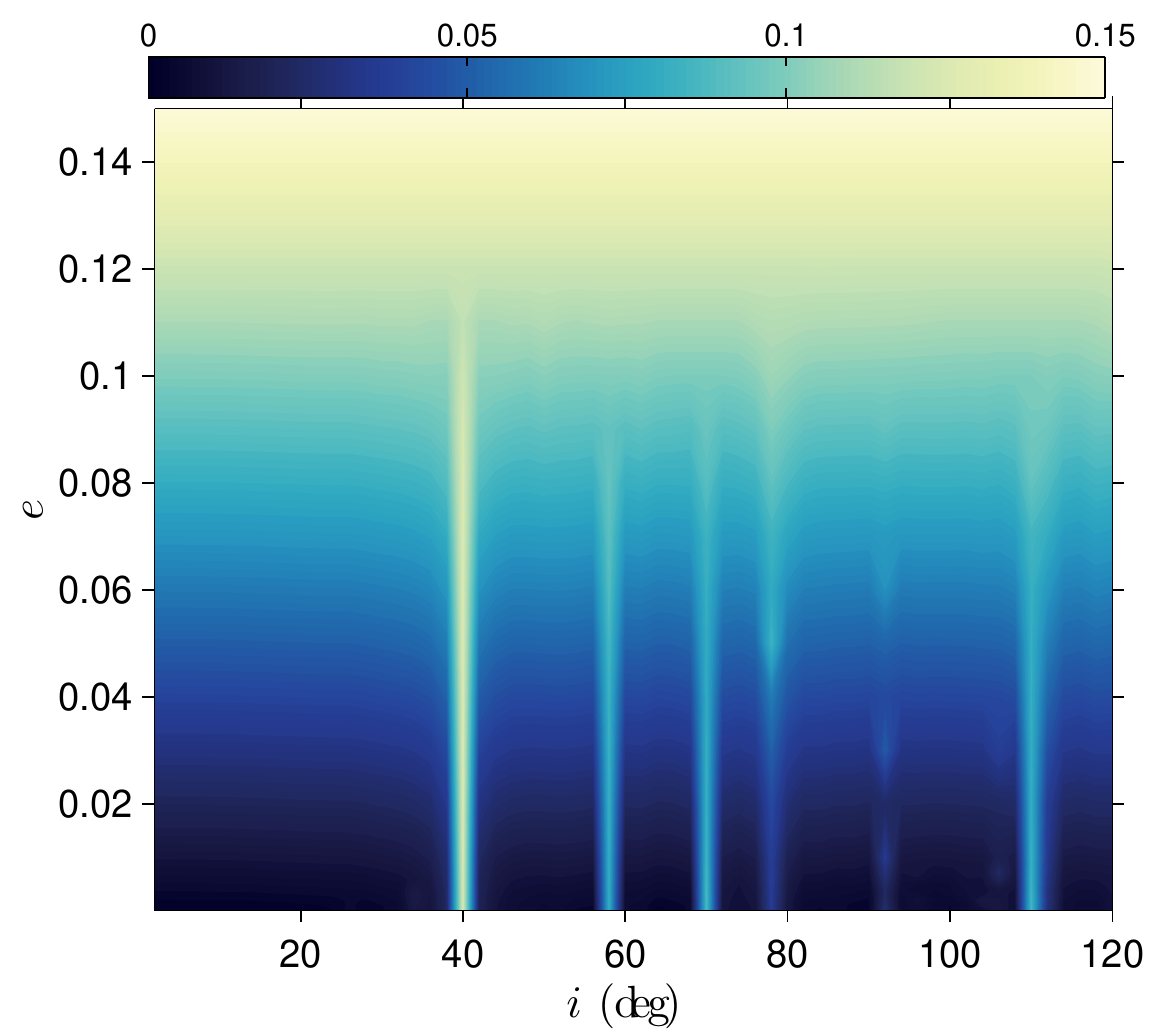}  \hspace{-0.25cm}  \includegraphics[width=0.25\textwidth]{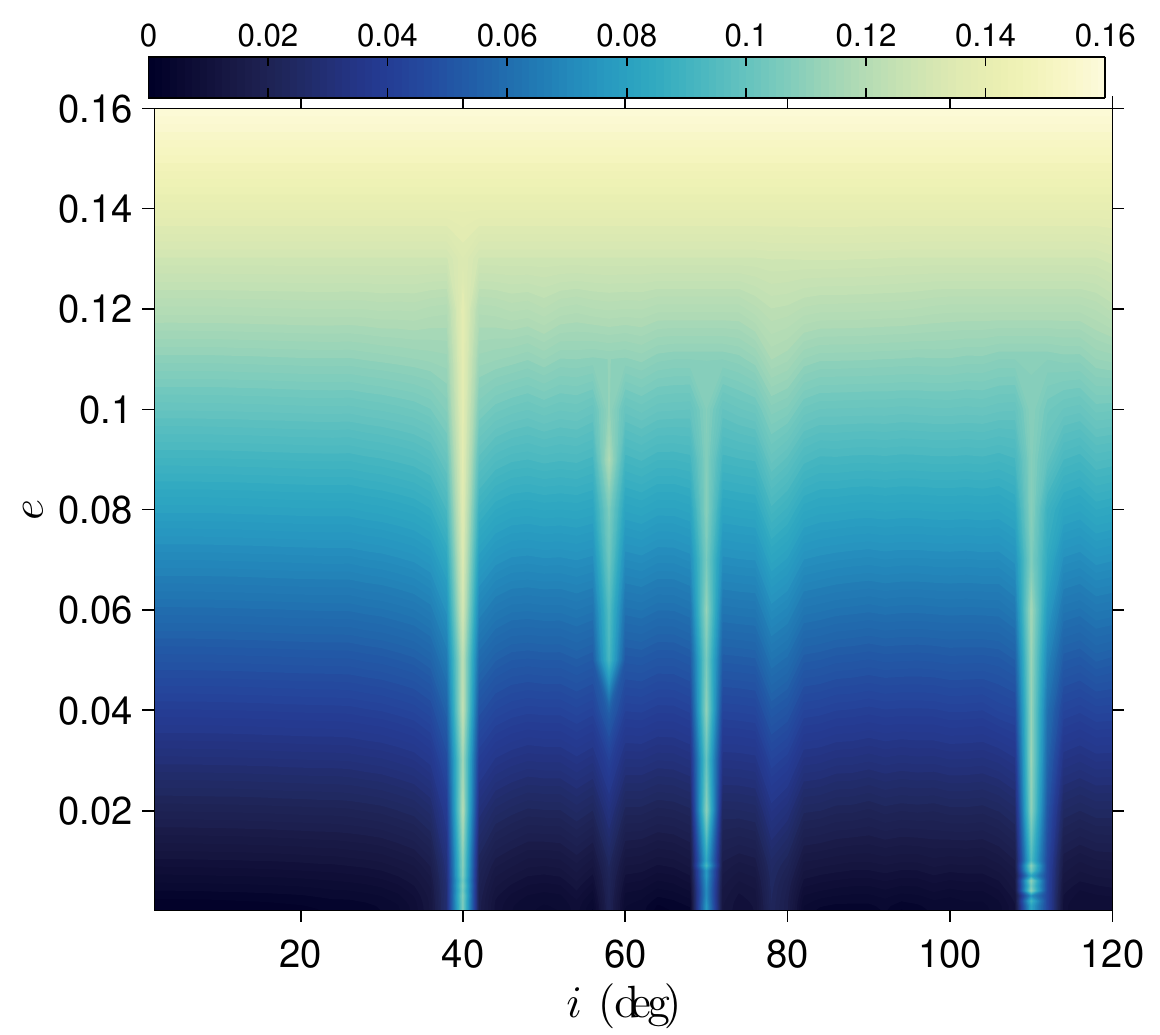}   \hspace{-0.3cm}   \includegraphics[width=0.25\textwidth]{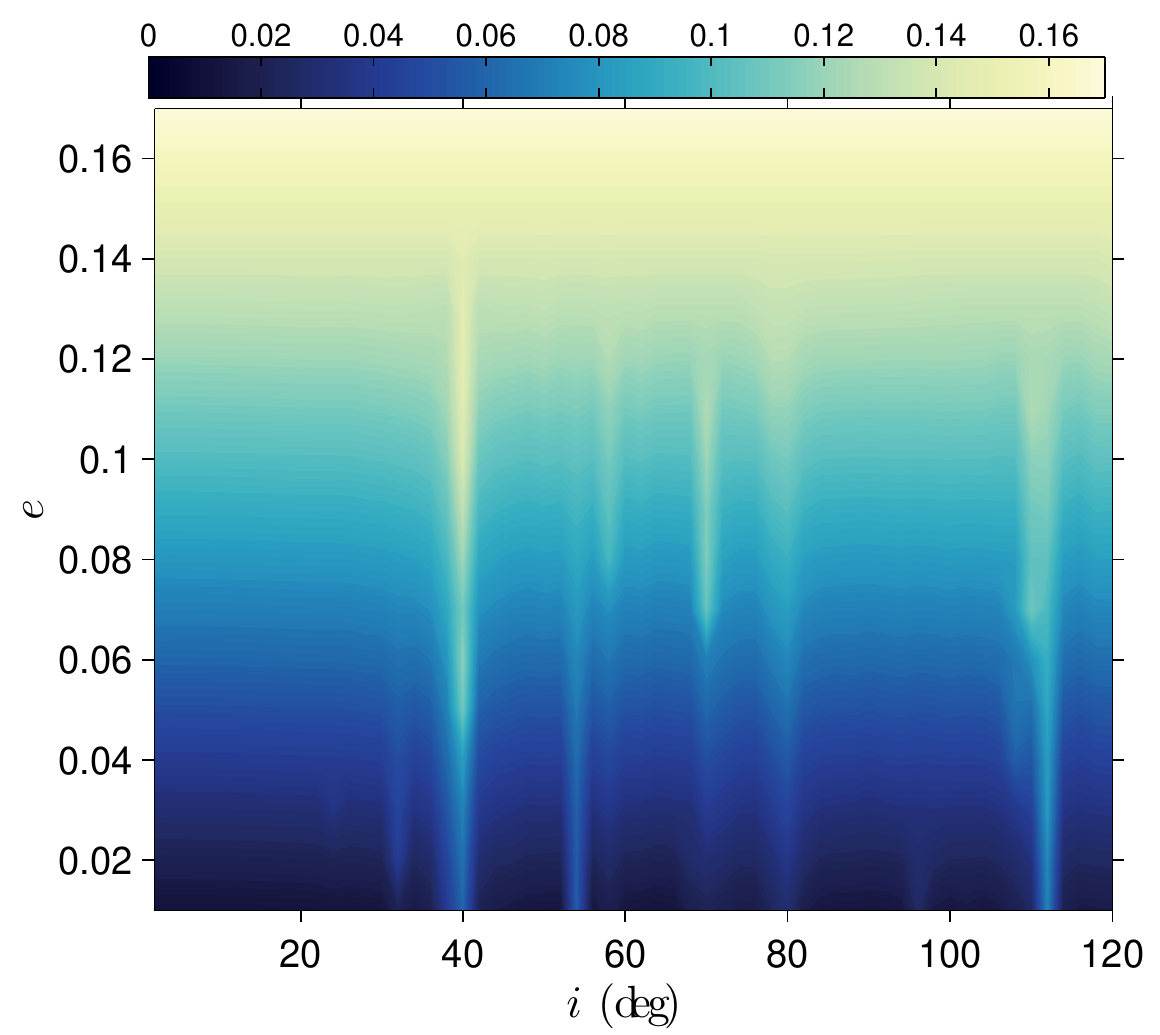}   \hspace{-0.3cm}   \includegraphics[width=0.25\textwidth]{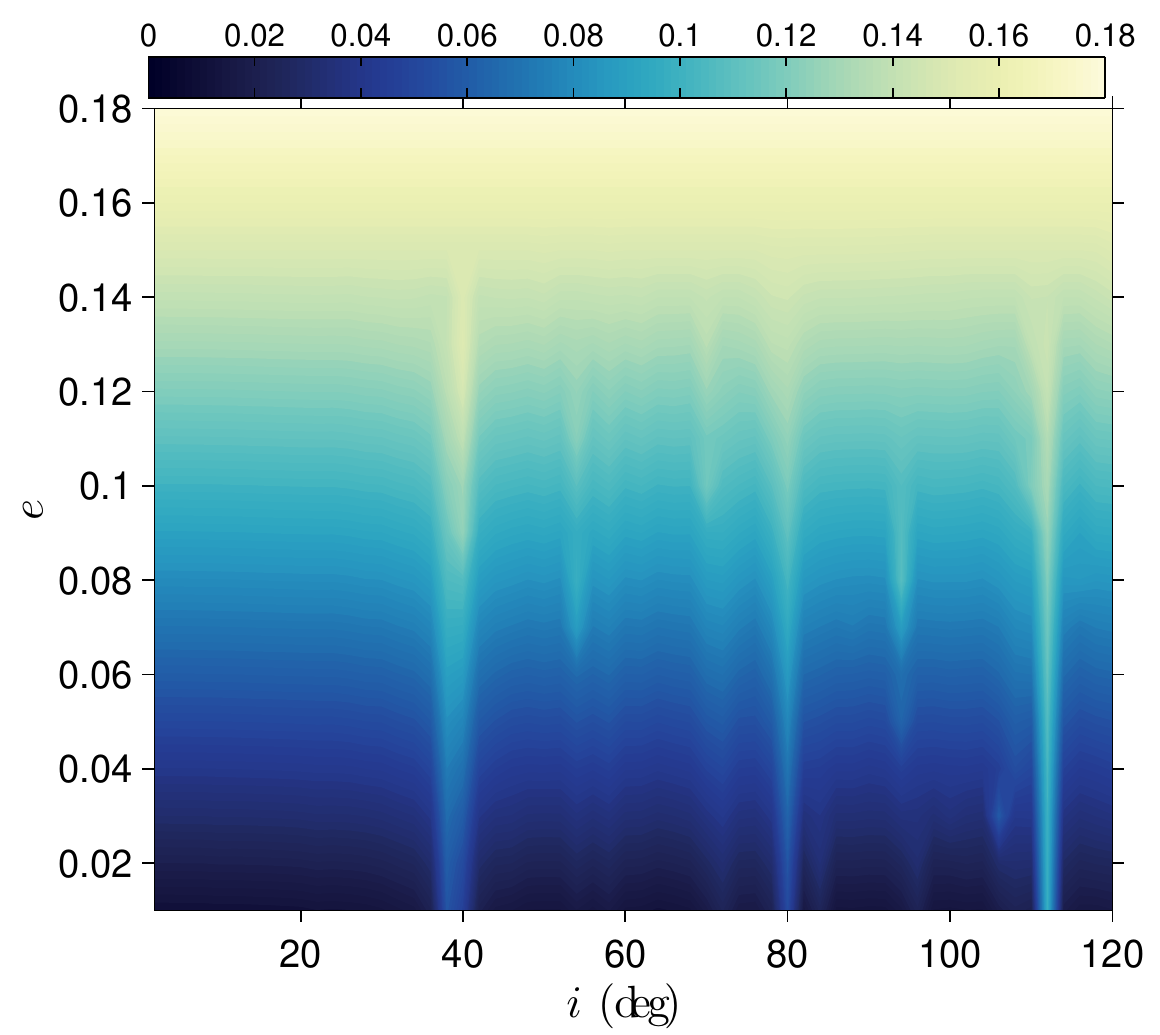} 
        \includegraphics[width=0.25\textwidth]{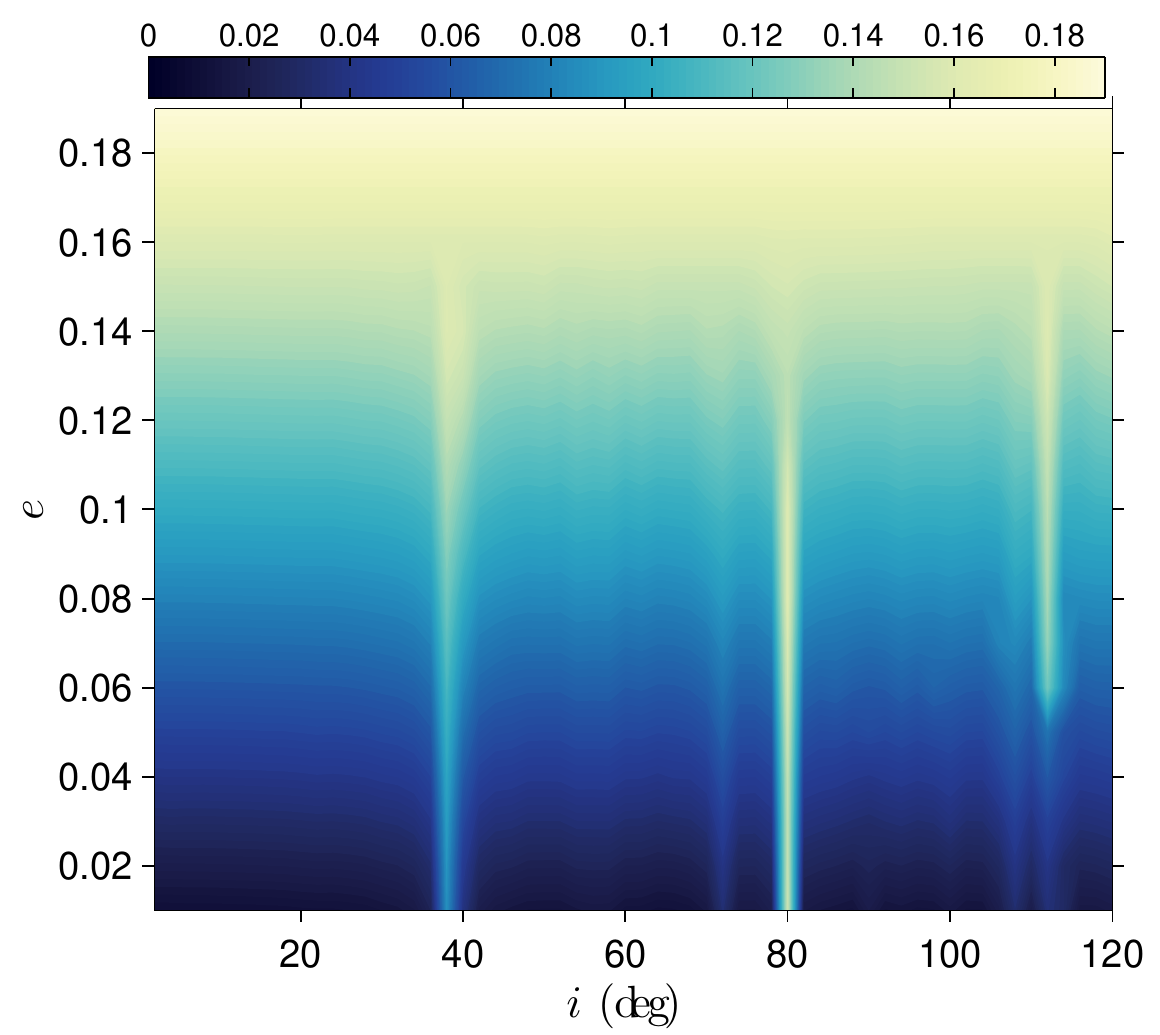}  \hspace{-0.25cm}  \includegraphics[width=0.25\textwidth]{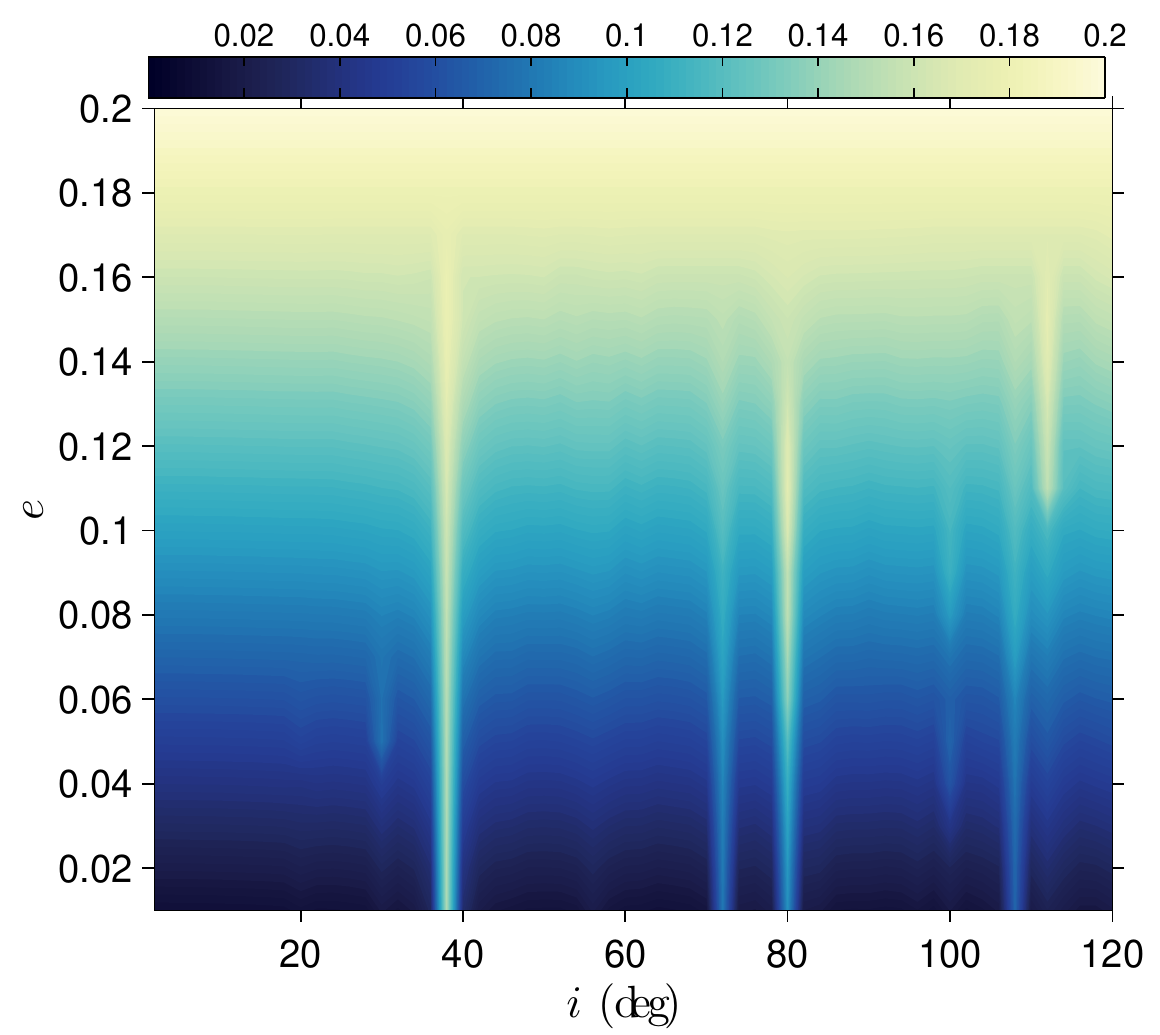}   \hspace{-0.3cm}   \includegraphics[width=0.25\textwidth]{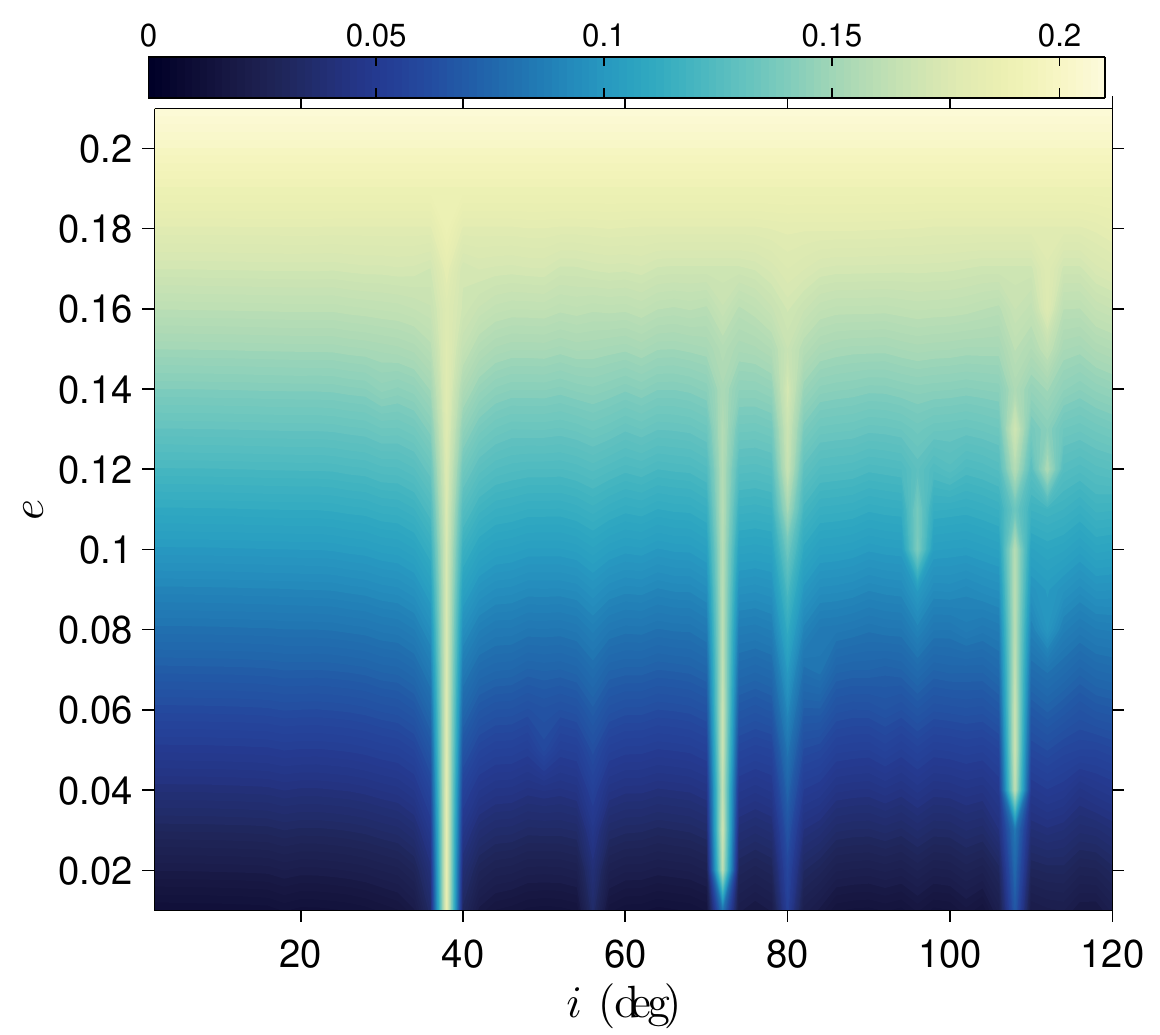}   \hspace{-0.3cm}   \includegraphics[width=0.25\textwidth]{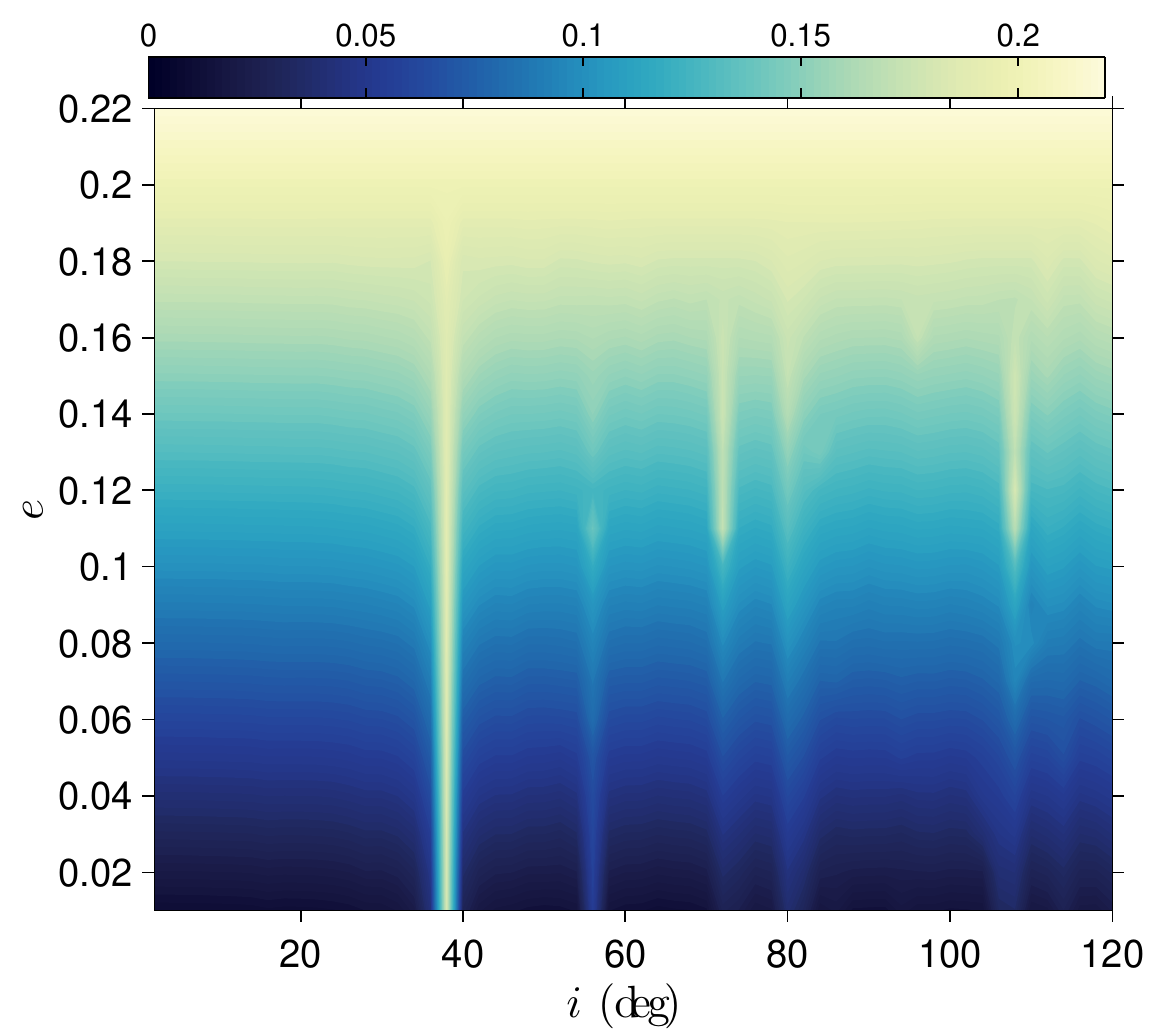} 
          \includegraphics[width=0.25\textwidth]{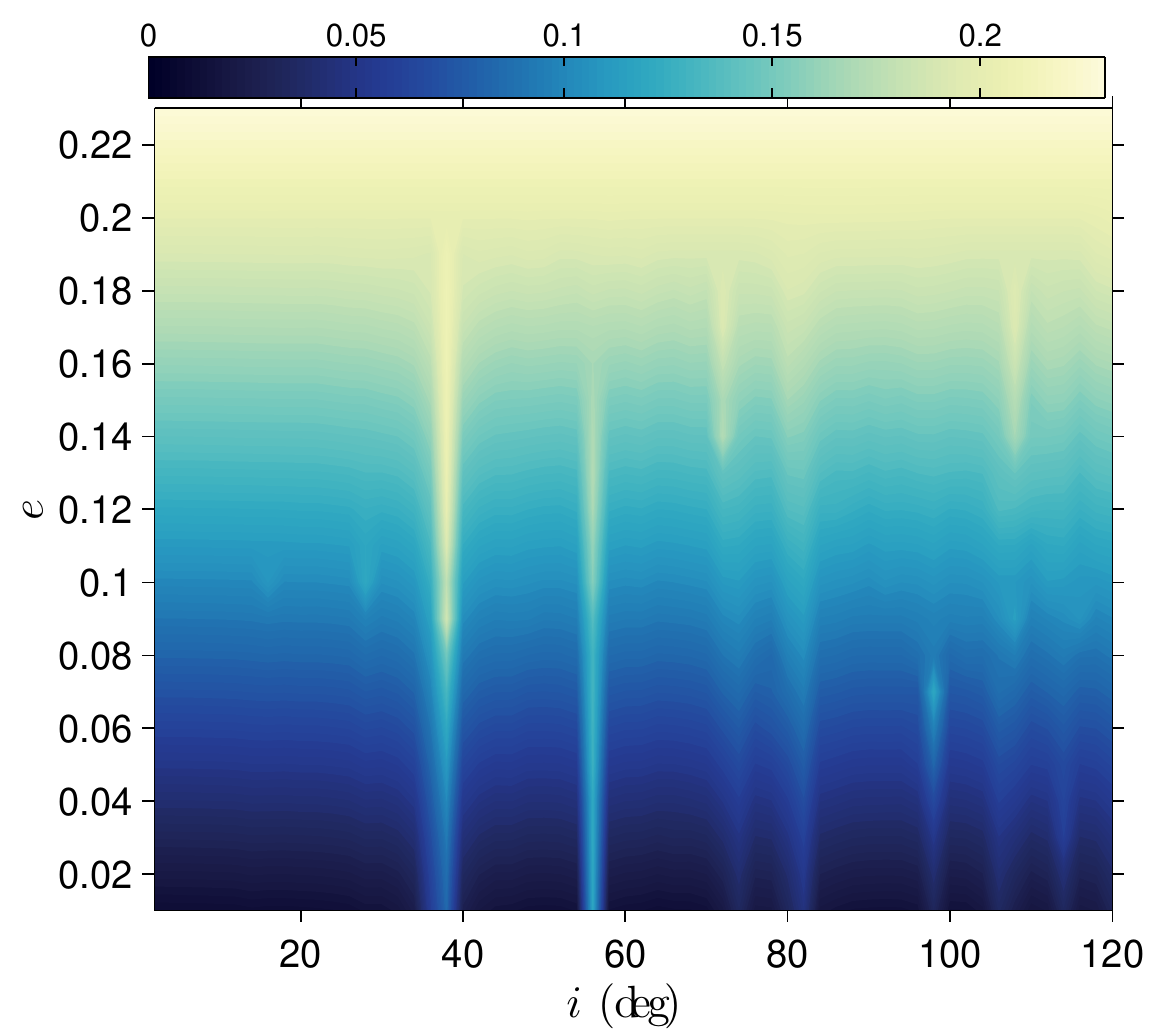}  \hspace{-0.25cm}  \includegraphics[width=0.25\textwidth]{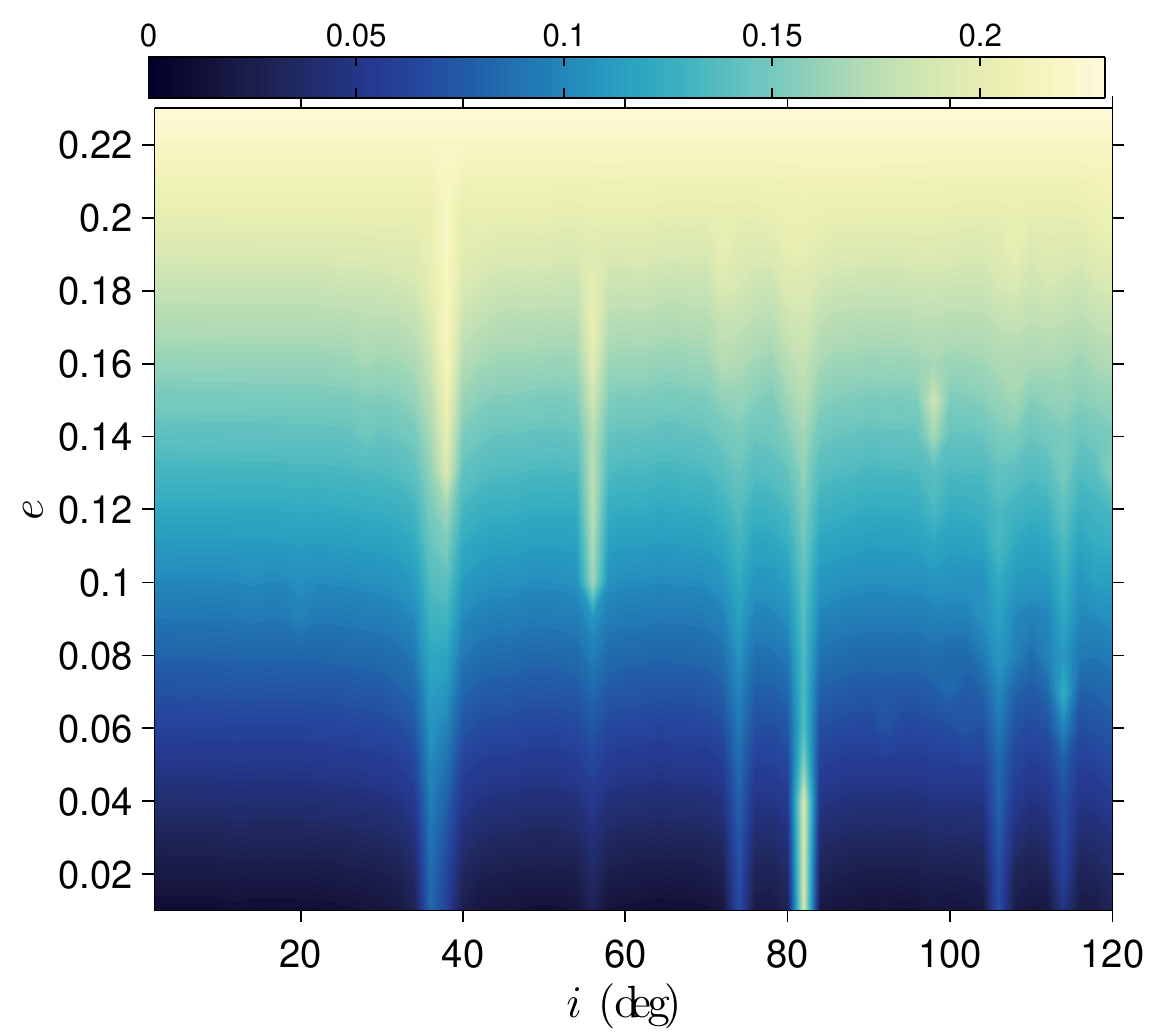}   \hspace{-0.3cm}   \includegraphics[width=0.25\textwidth]{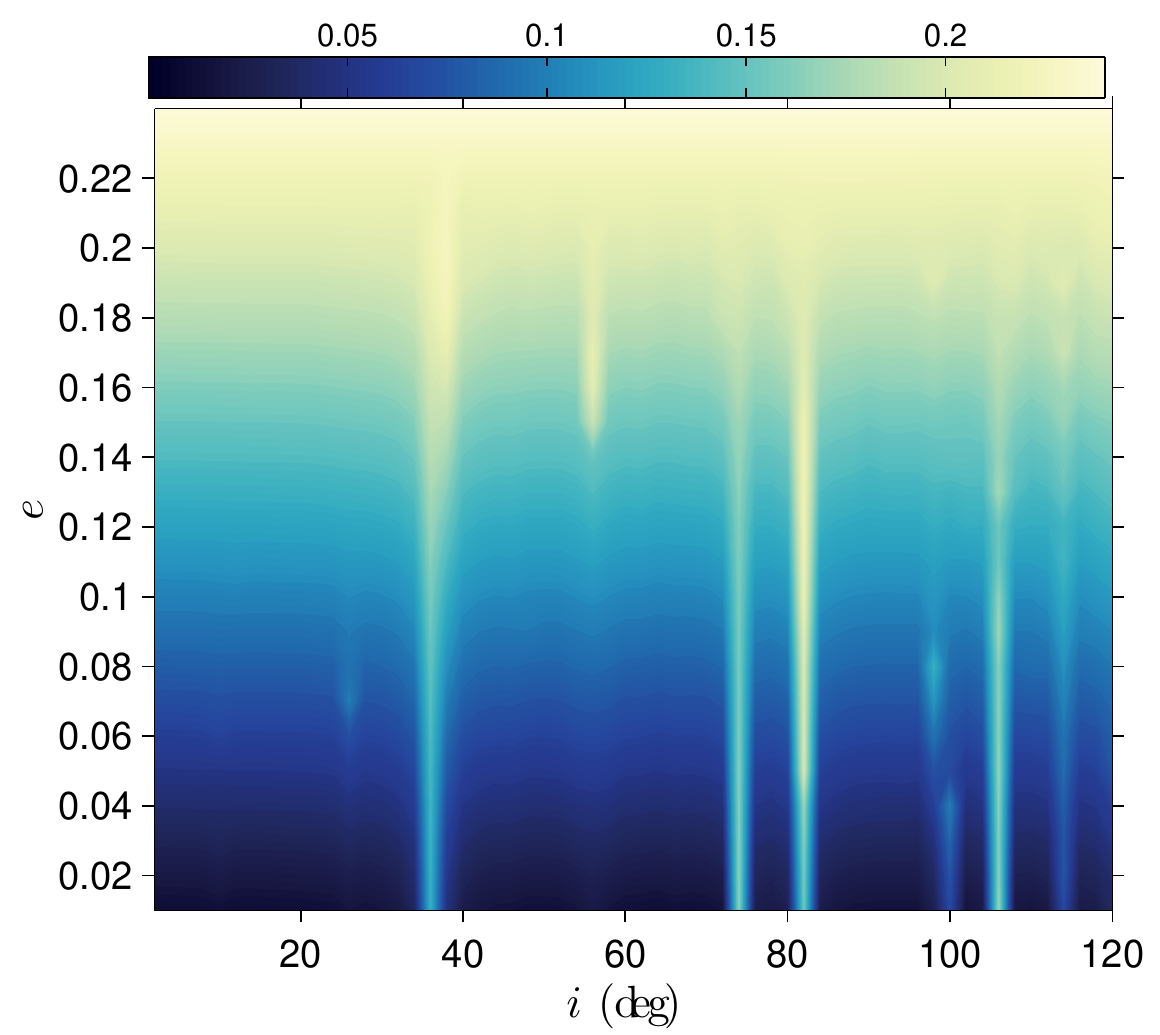}   \hspace{-0.3cm}   \includegraphics[width=0.25\textwidth]{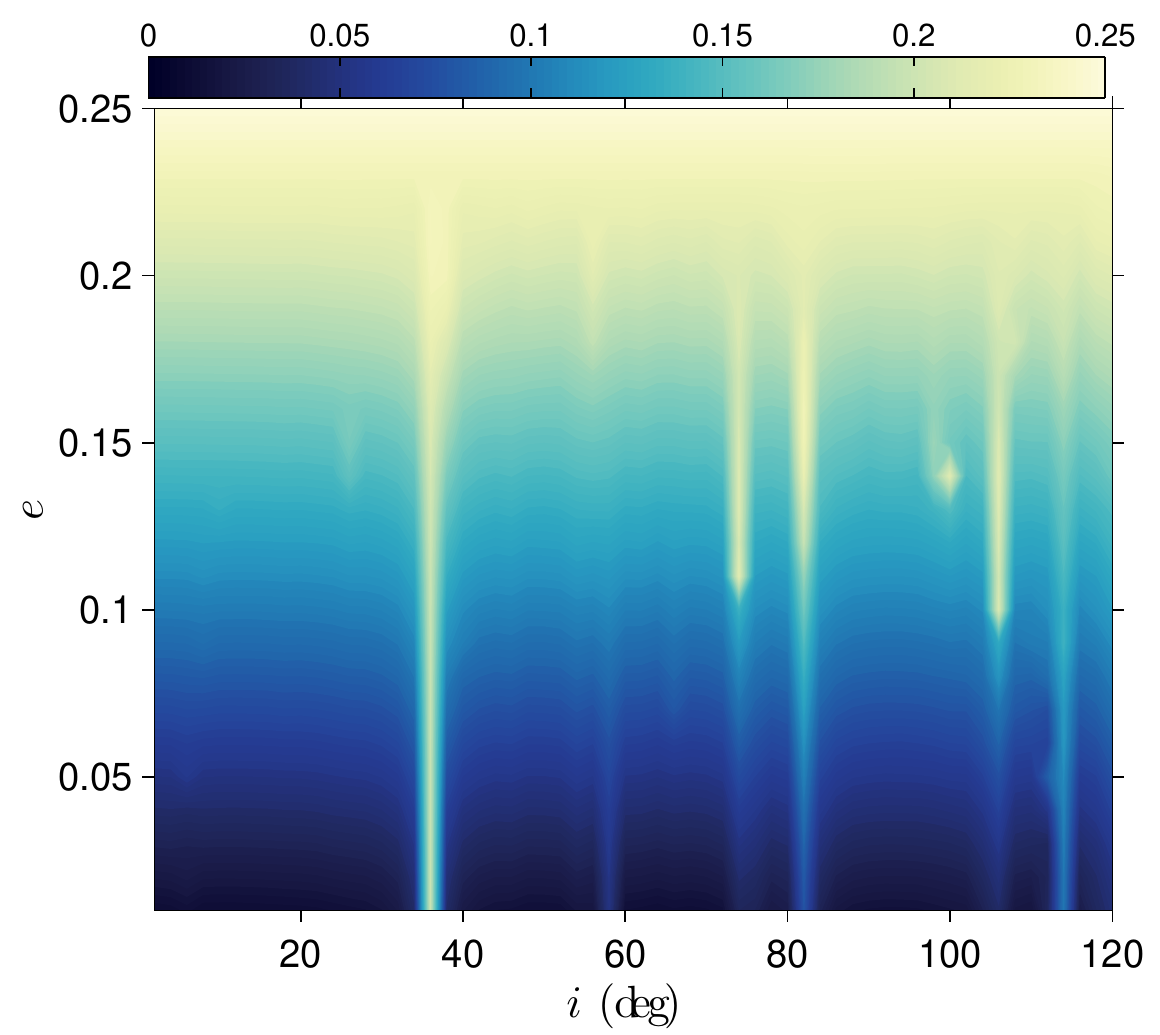} 
            \includegraphics[width=0.25\textwidth]{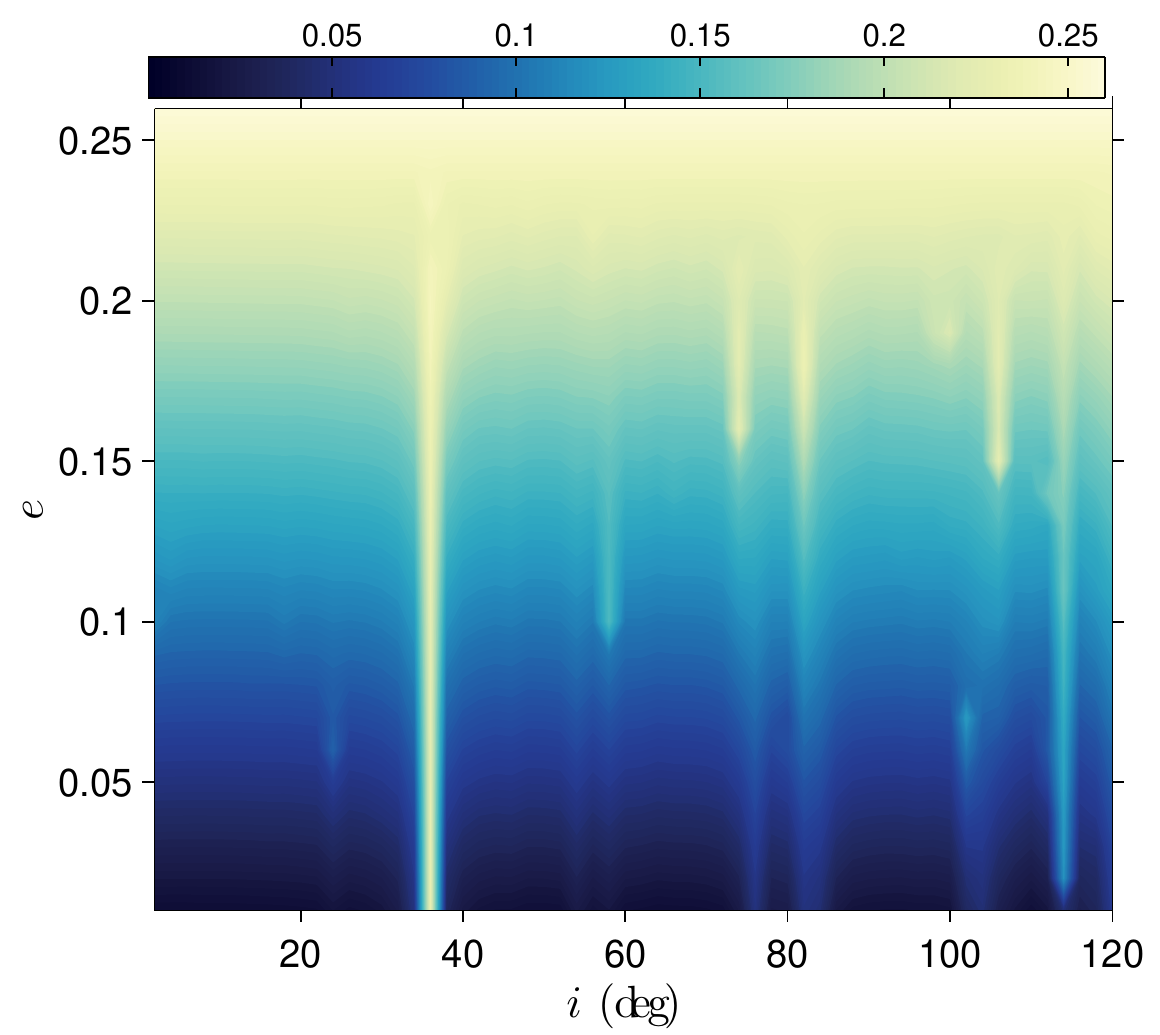}  \hspace{-0.25cm}  \includegraphics[width=0.25\textwidth]{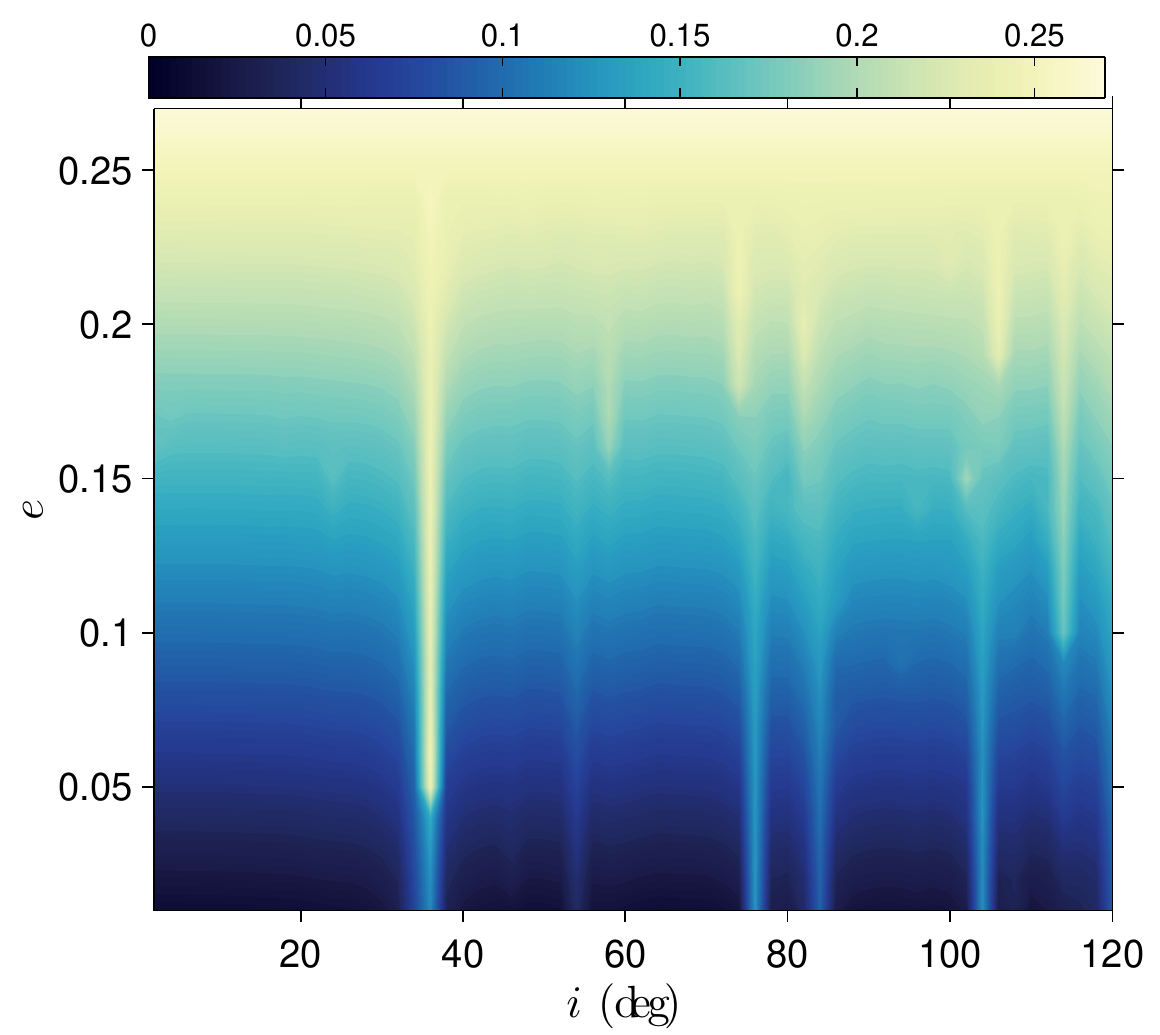}   \hspace{-0.3cm}   \includegraphics[width=0.25\textwidth]{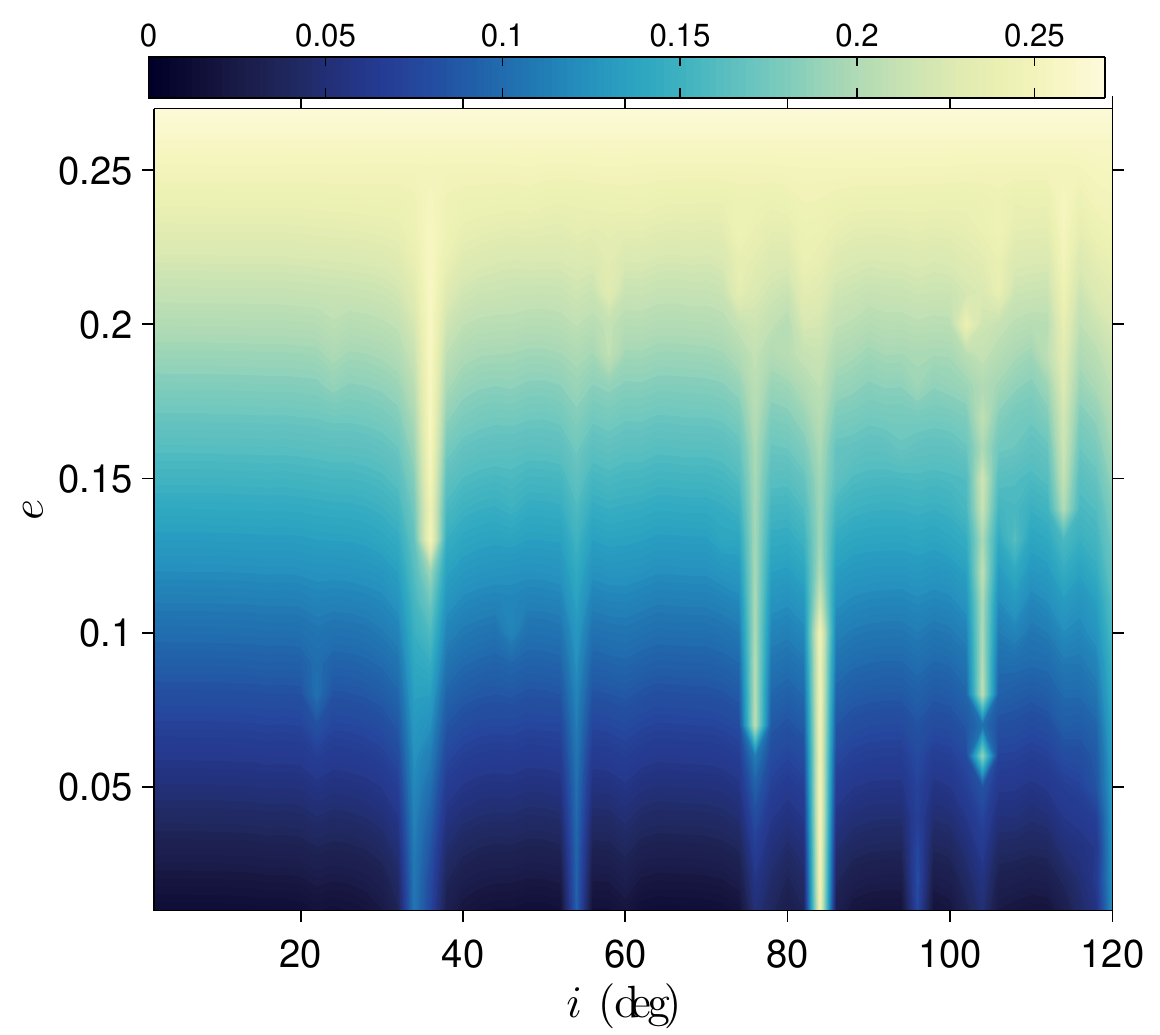}   \hspace{-0.3cm}   \includegraphics[width=0.25\textwidth]{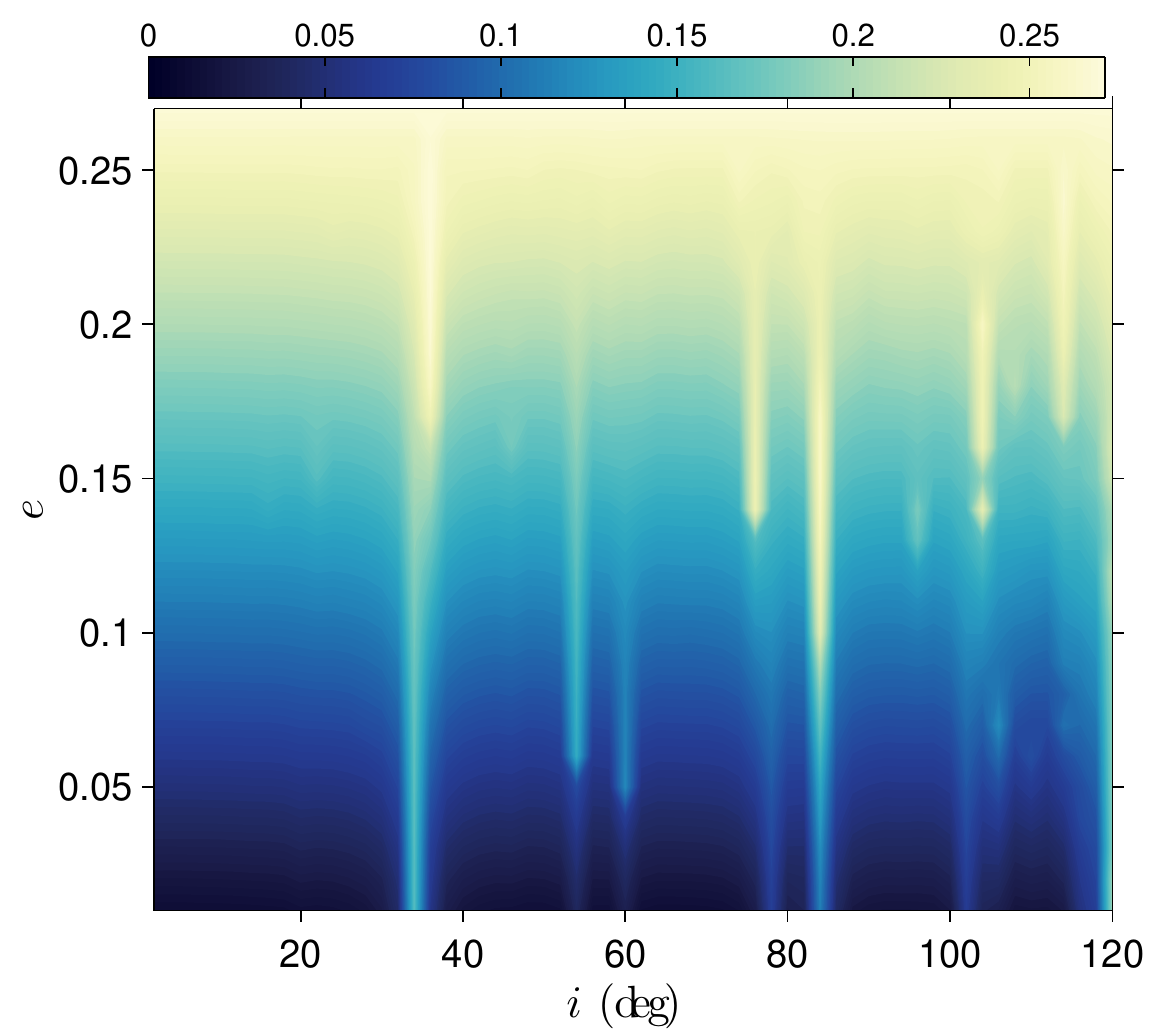} 
            \end{center}
        \caption{Maximum eccentricity computed (color bar) as a function of initial inclination and eccentricity for the initial epoch 2020 and $C_R (A/m)=1$
  m$^2/$kg, assuming $\Omega=0^{\circ}$ and $\omega=0^{\circ}$ at the initial epoch. Each plot depicts the behavior computed starting from a different value of initial semi-major axis. From the top left to the bottom right: $a=R_E+ 700$ km to $a=R_E+3000$ km at a step of 100 km.}\label{fig:ie_emax_2020_high_span_a}
  \end{figure}

\begin{figure}[th!]
  \begin{center}
  \includegraphics[width=0.25\textwidth]{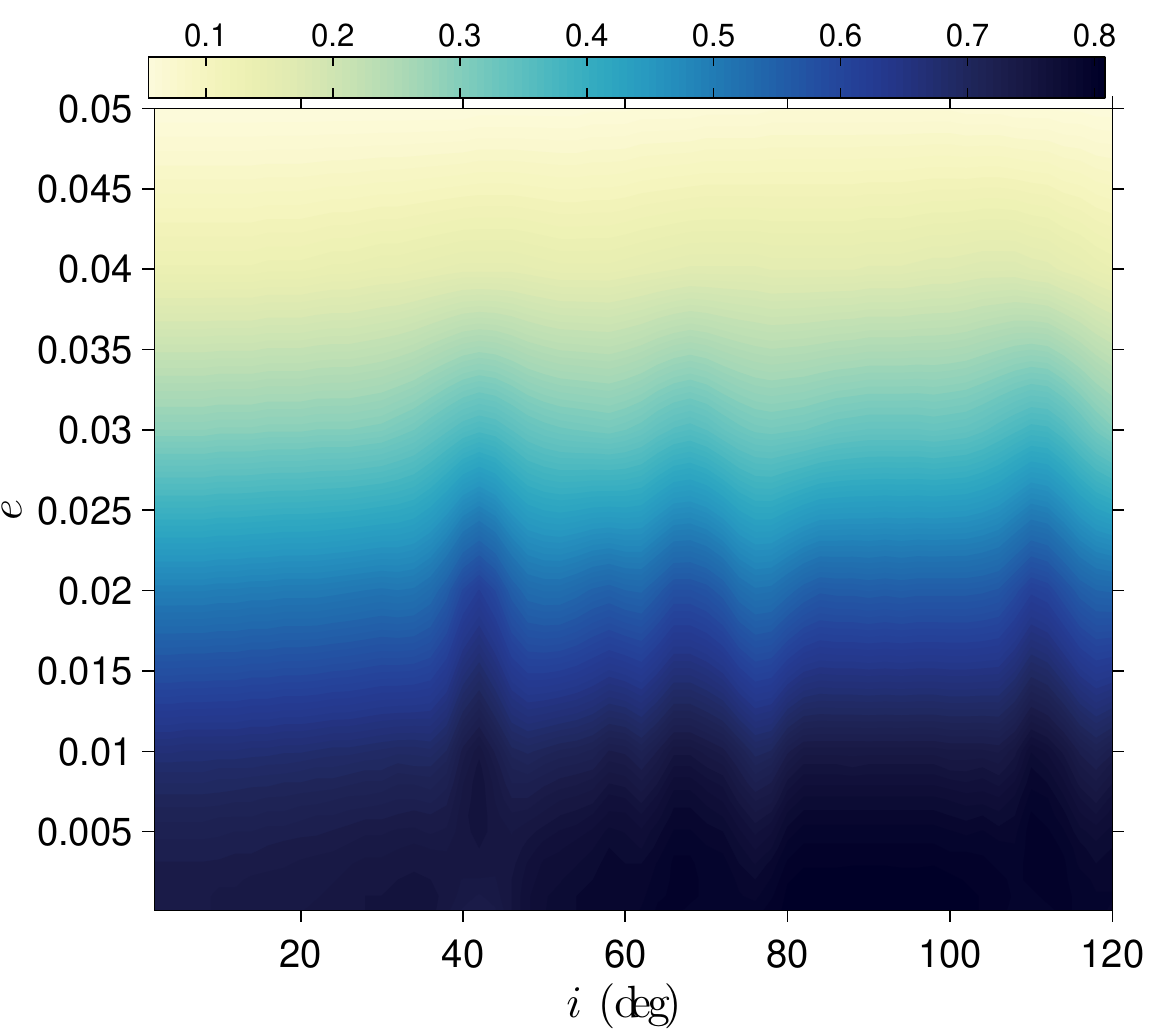}  \hspace{-0.3cm}  \includegraphics[width=0.25\textwidth]{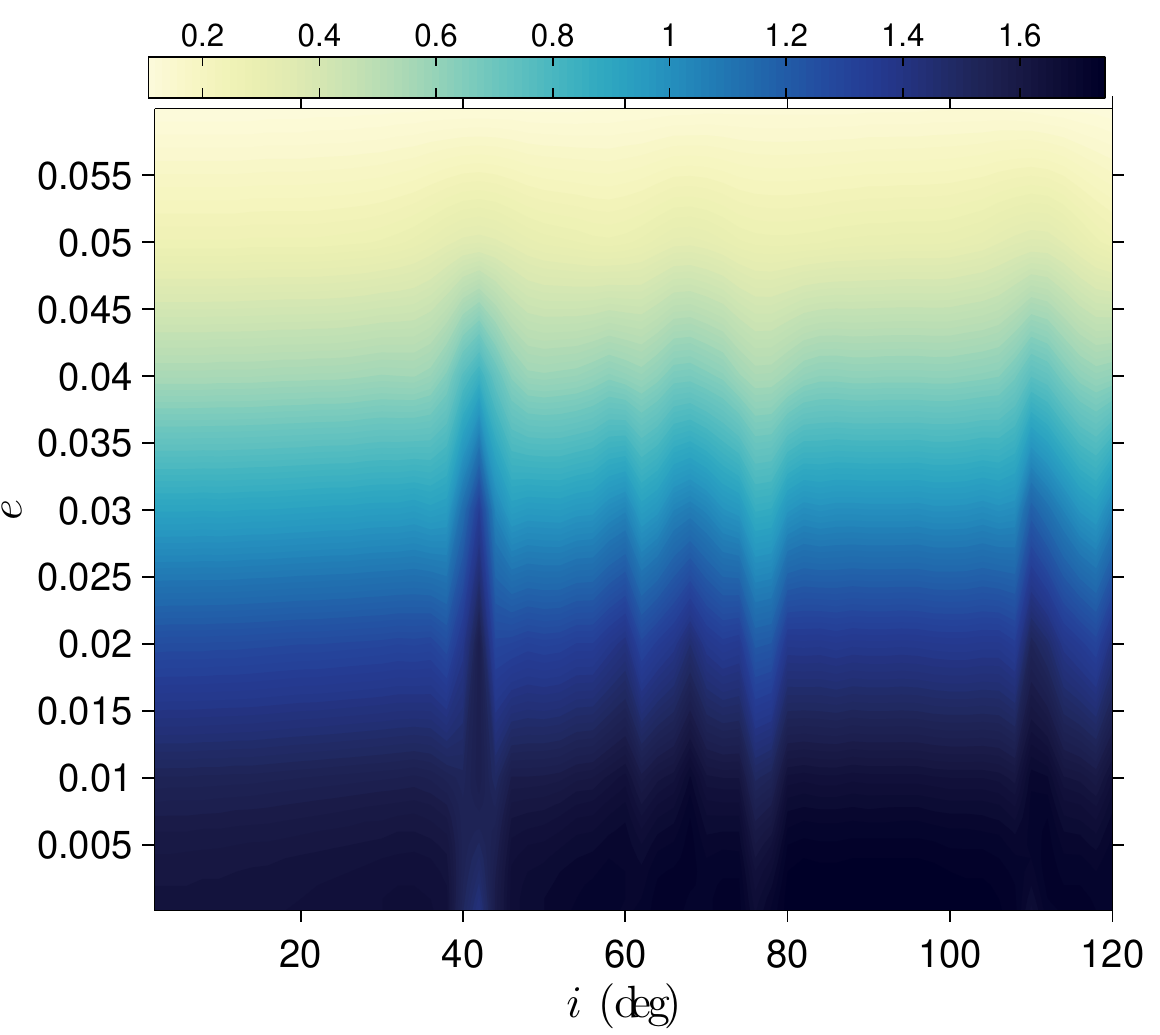}   \hspace{-0.3cm}   \includegraphics[width=0.25\textwidth]{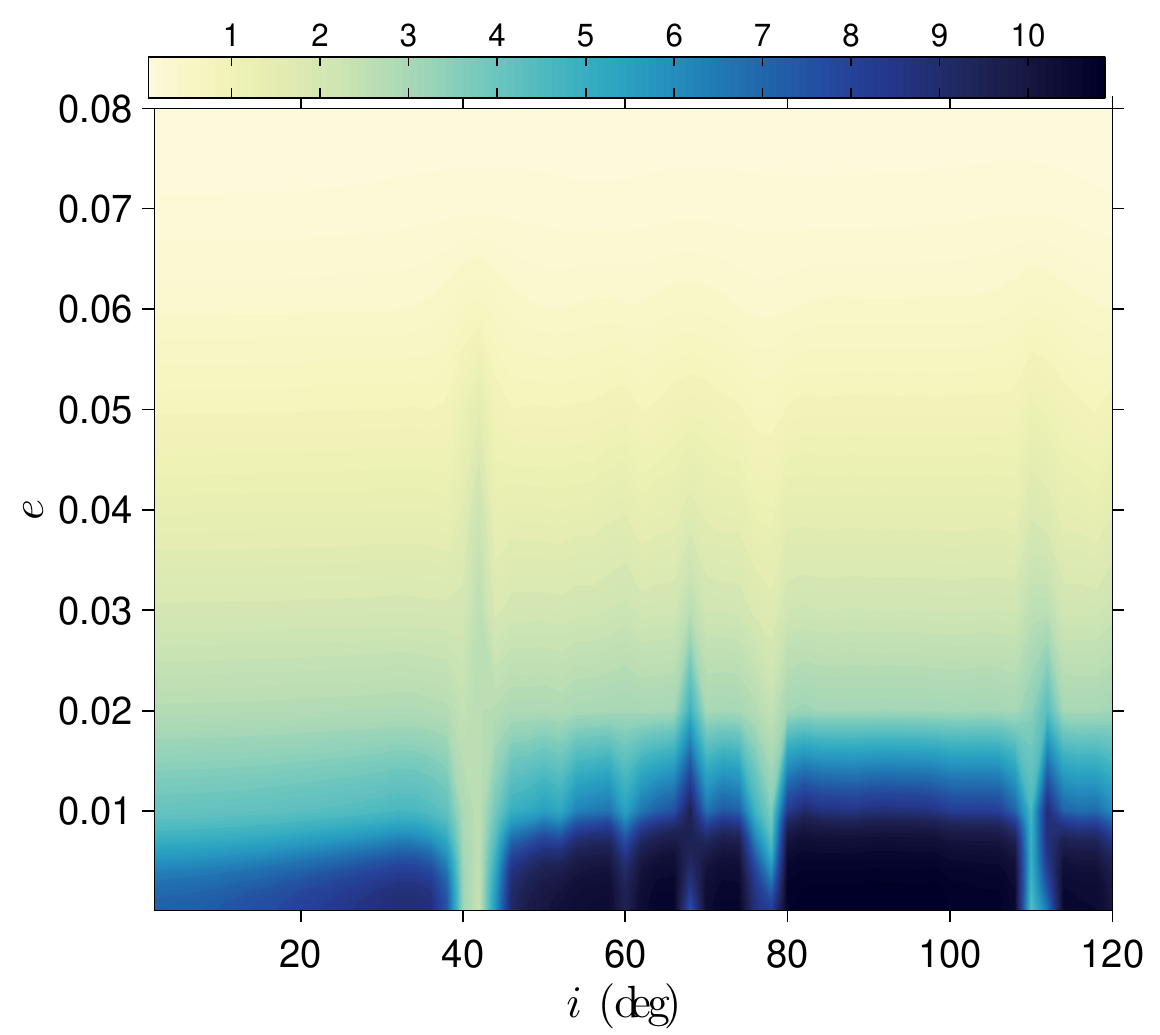}   \hspace{-0.3cm}   \includegraphics[width=0.25\textwidth]{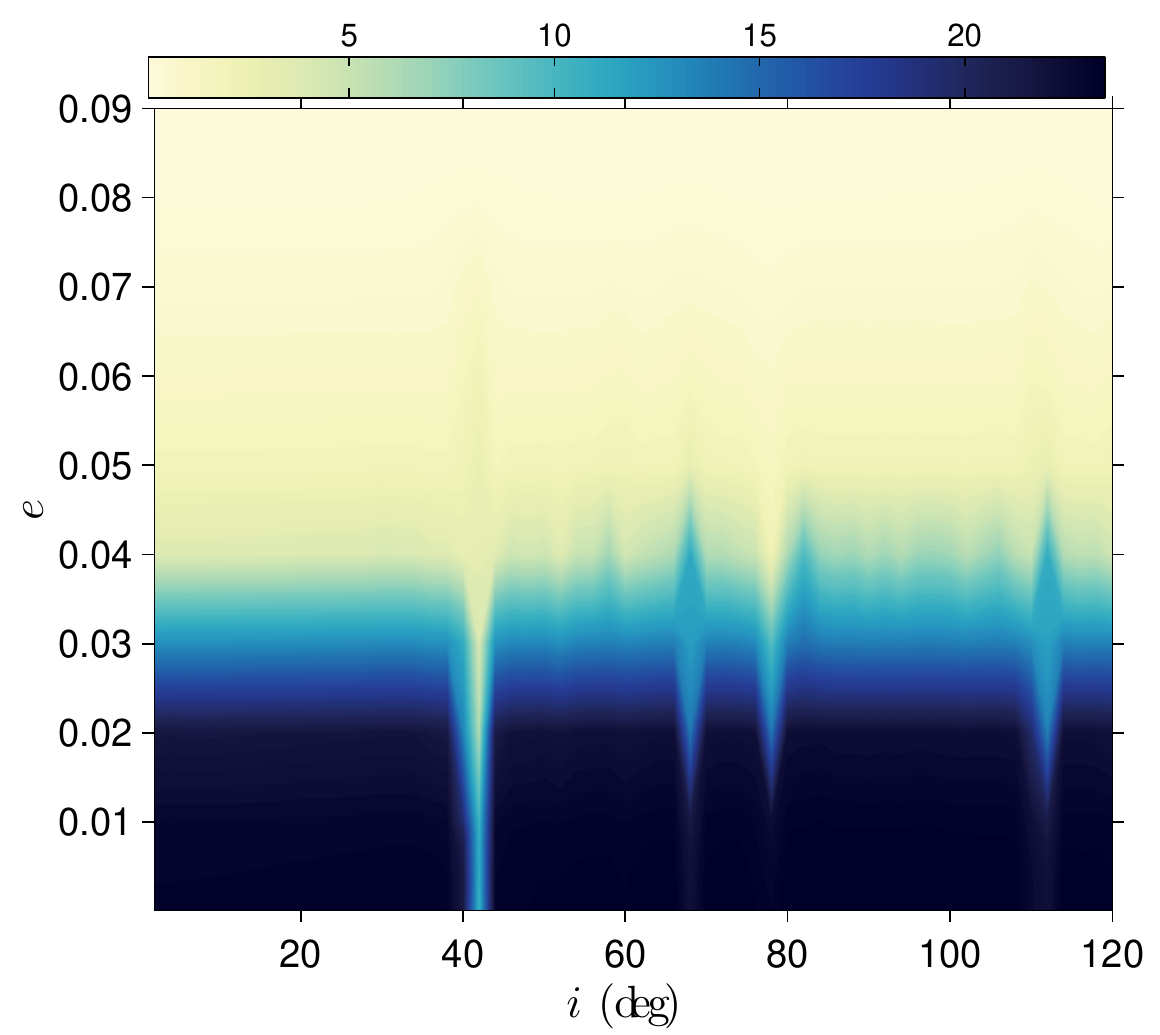} 
    \includegraphics[width=0.25\textwidth]{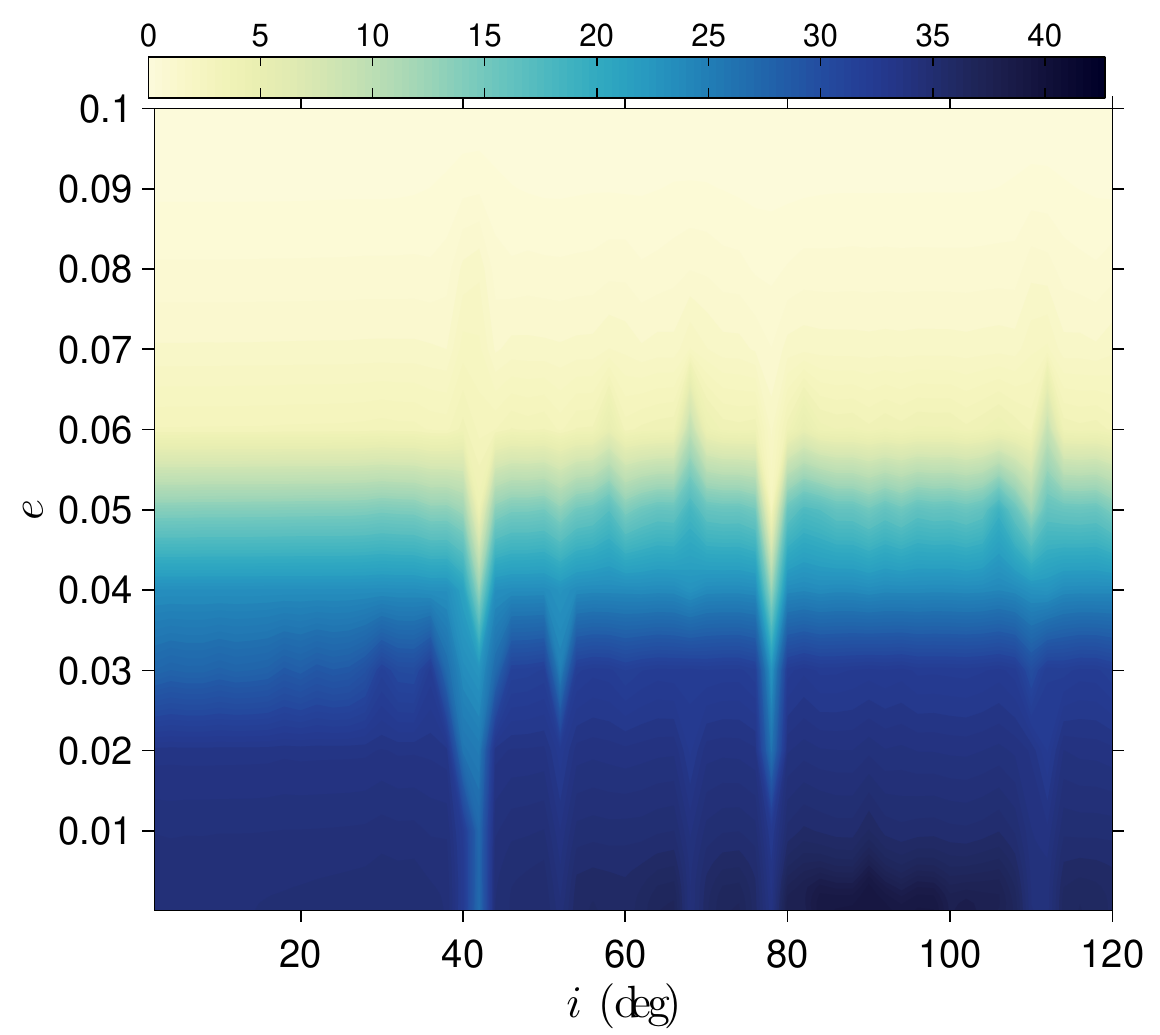}  \hspace{-0.3cm}  \includegraphics[width=0.25\textwidth]{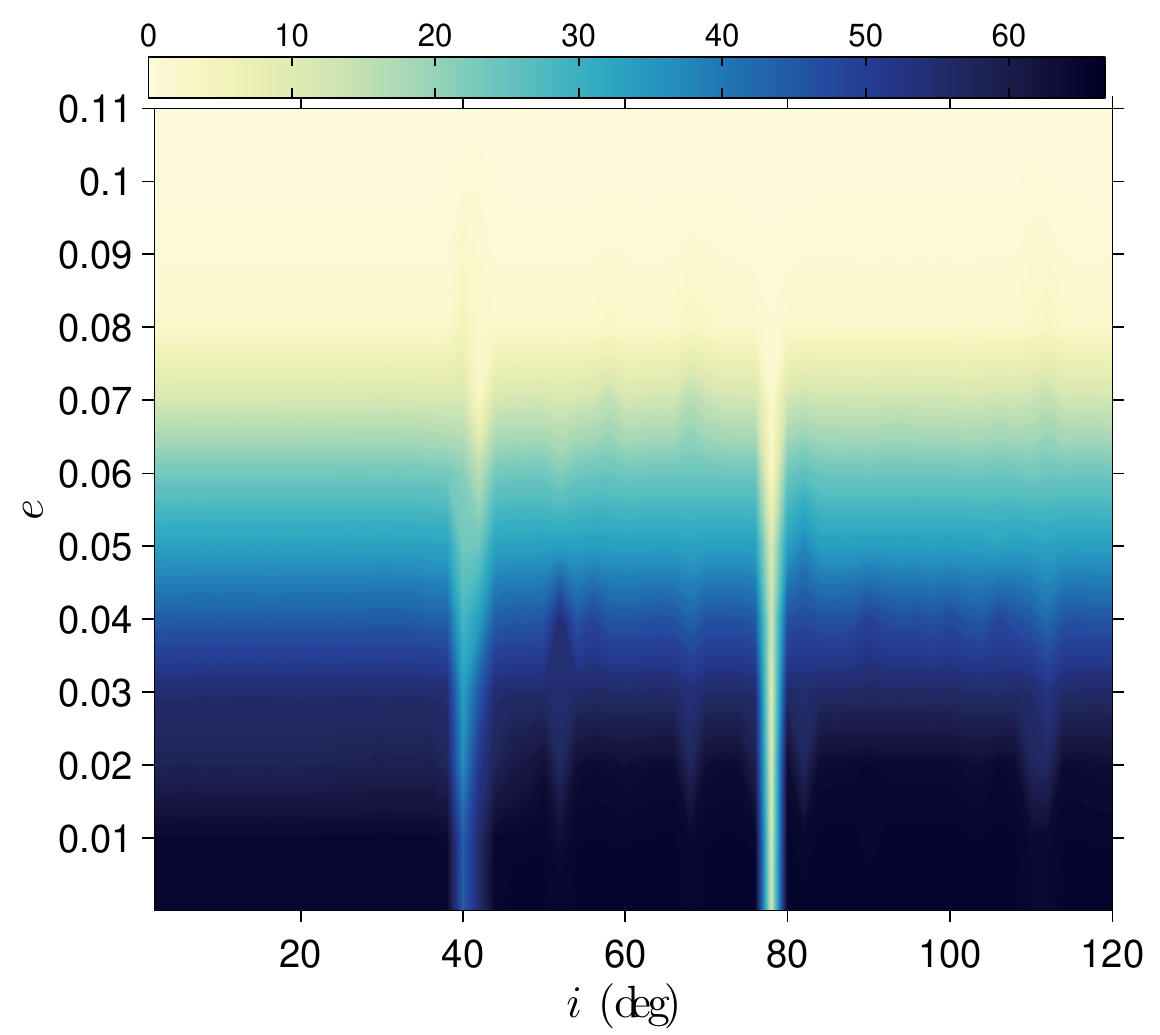}   \hspace{-0.3cm}   \includegraphics[width=0.25\textwidth]{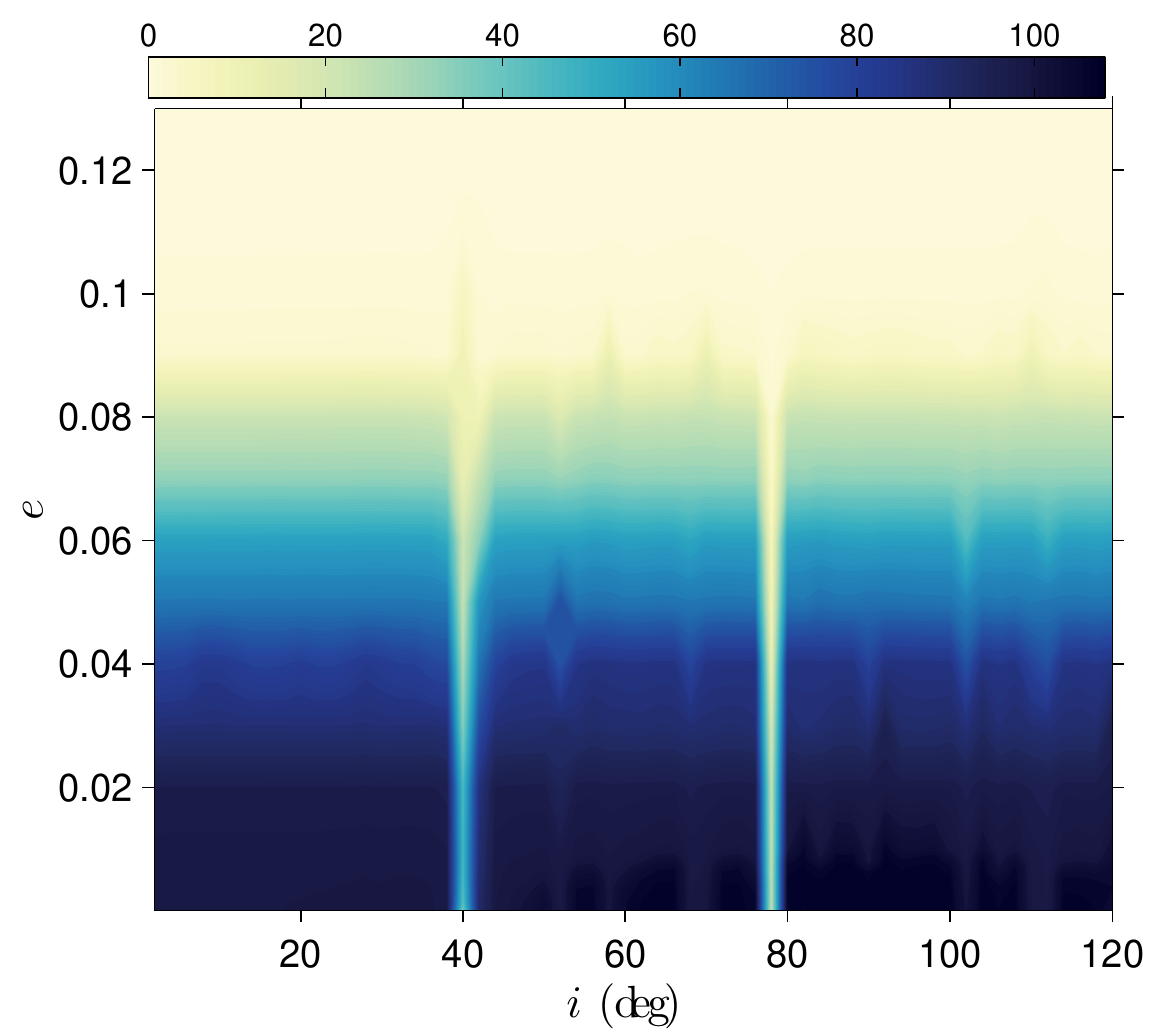}   \hspace{-0.3cm}   \includegraphics[width=0.25\textwidth]{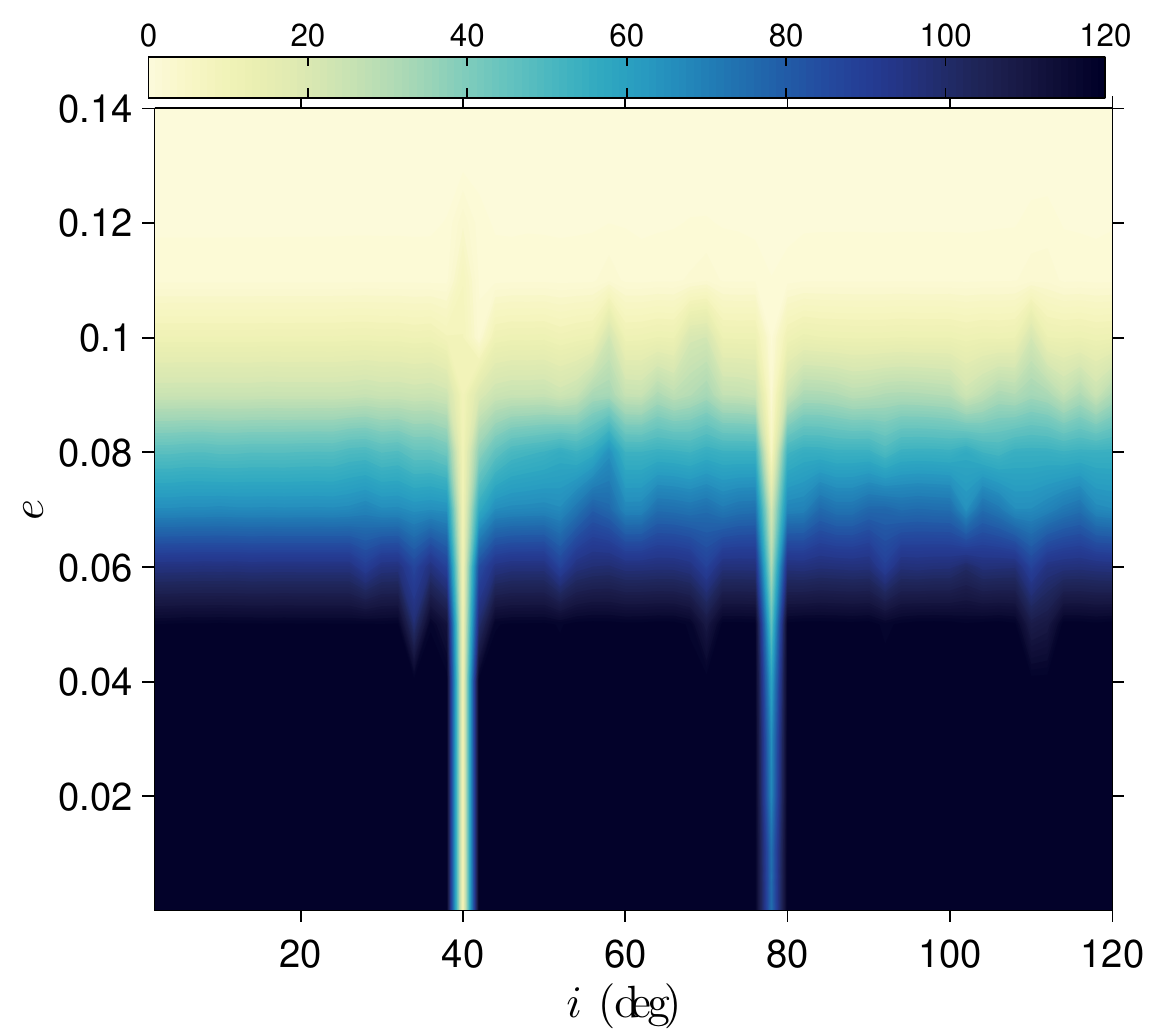} 
      \includegraphics[width=0.25\textwidth]{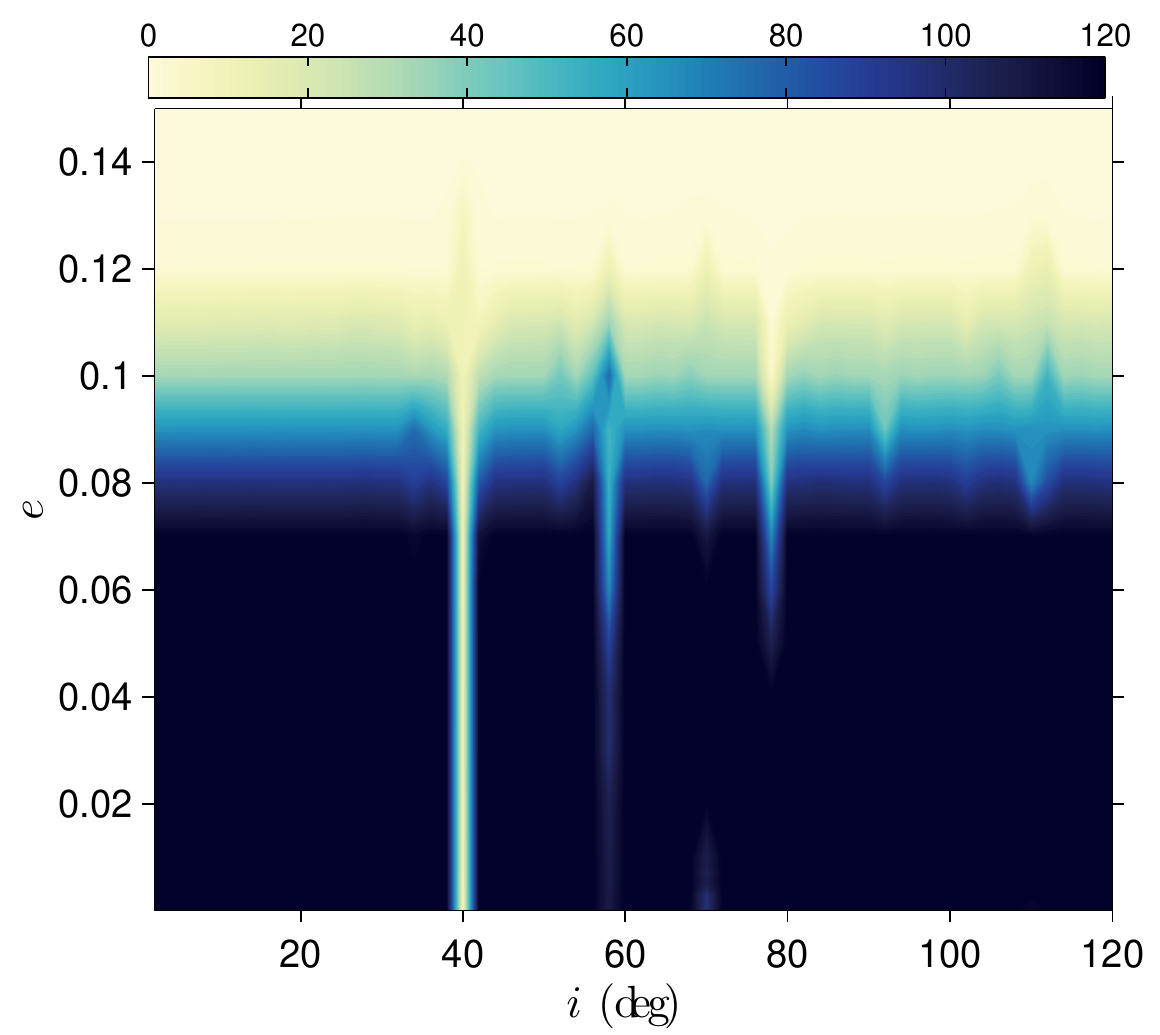}  \hspace{-0.3cm}  \includegraphics[width=0.25\textwidth]{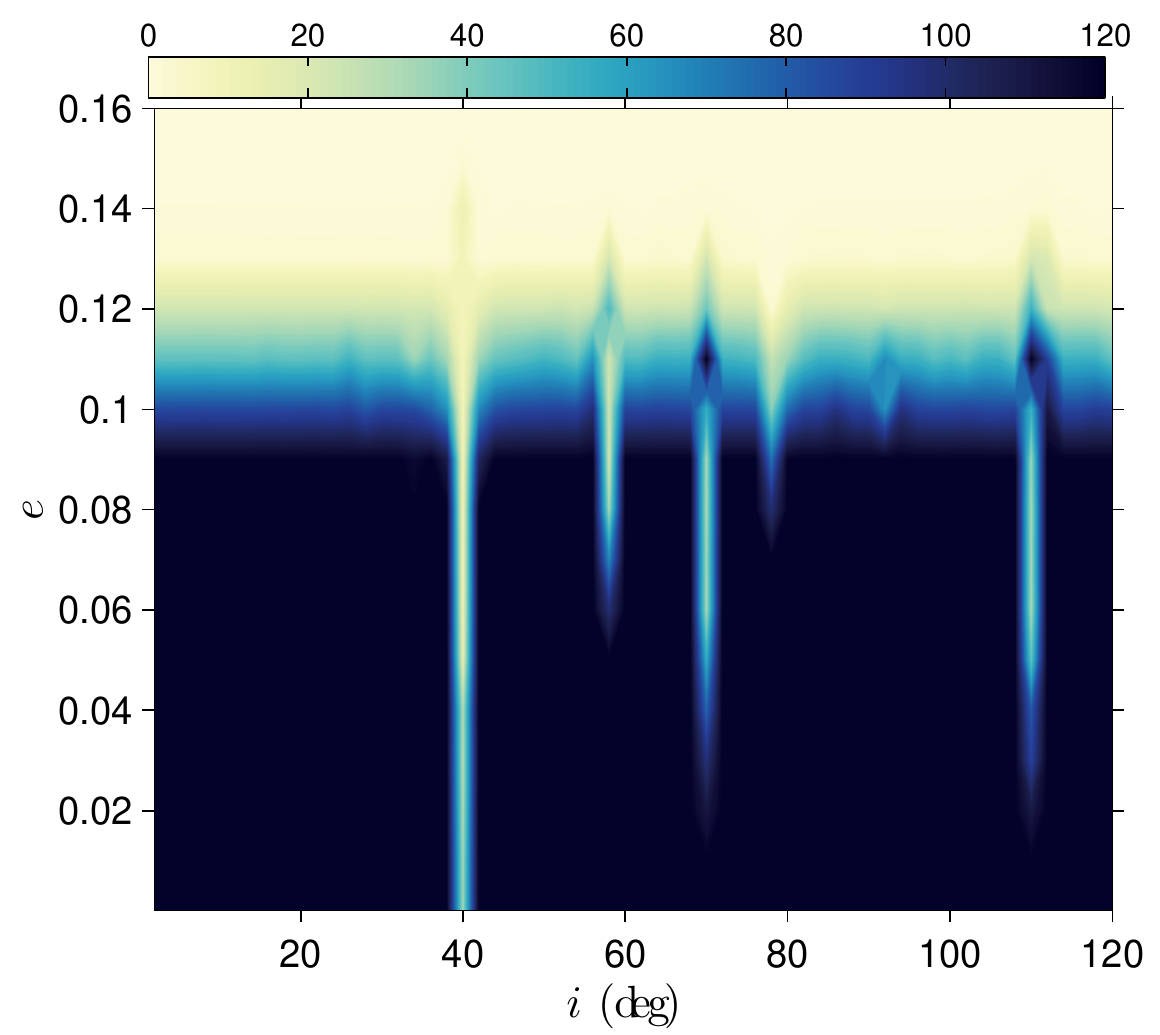}   \hspace{-0.3cm}   \includegraphics[width=0.25\textwidth]{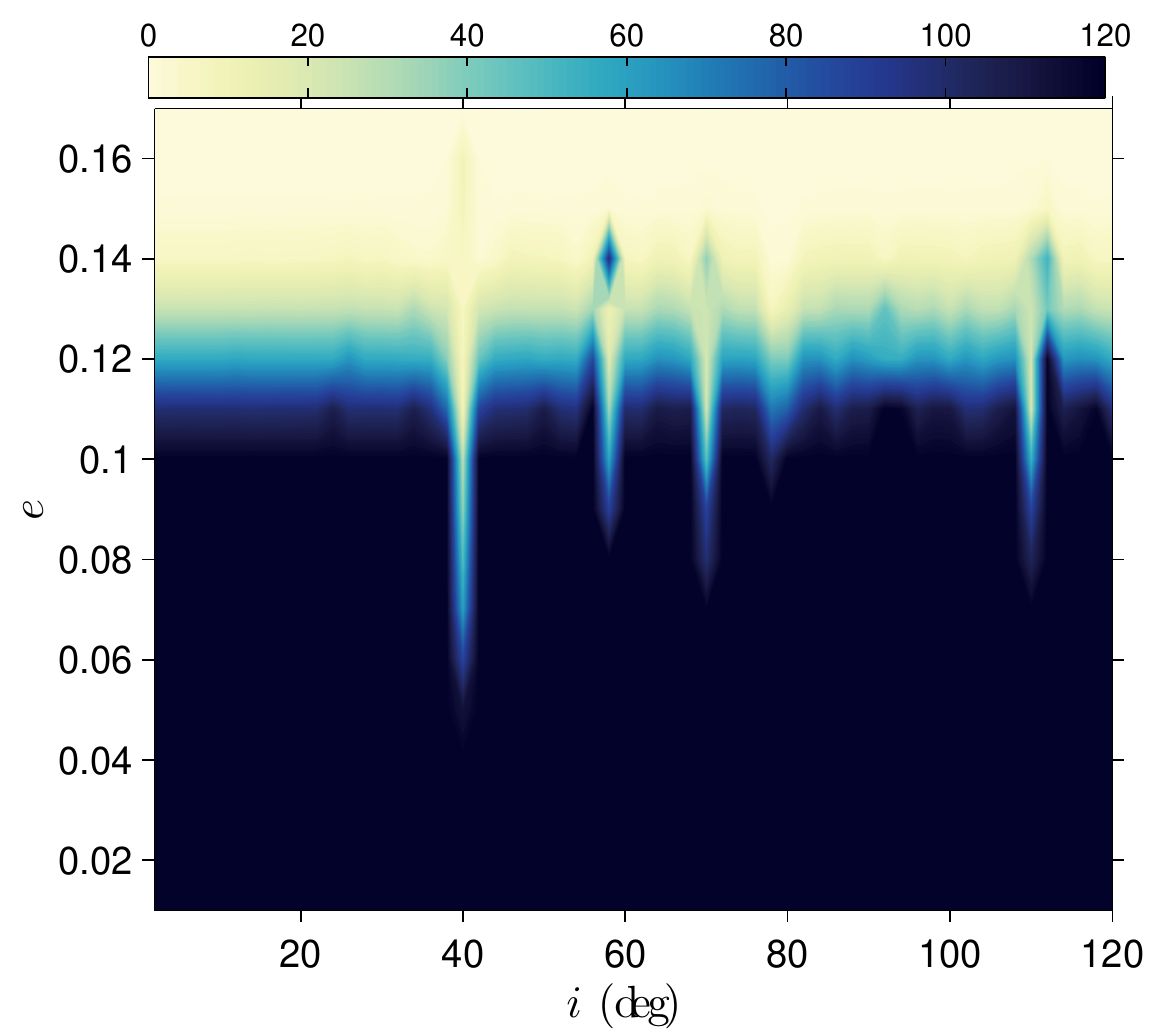}   \hspace{-0.3cm}   \includegraphics[width=0.25\textwidth]{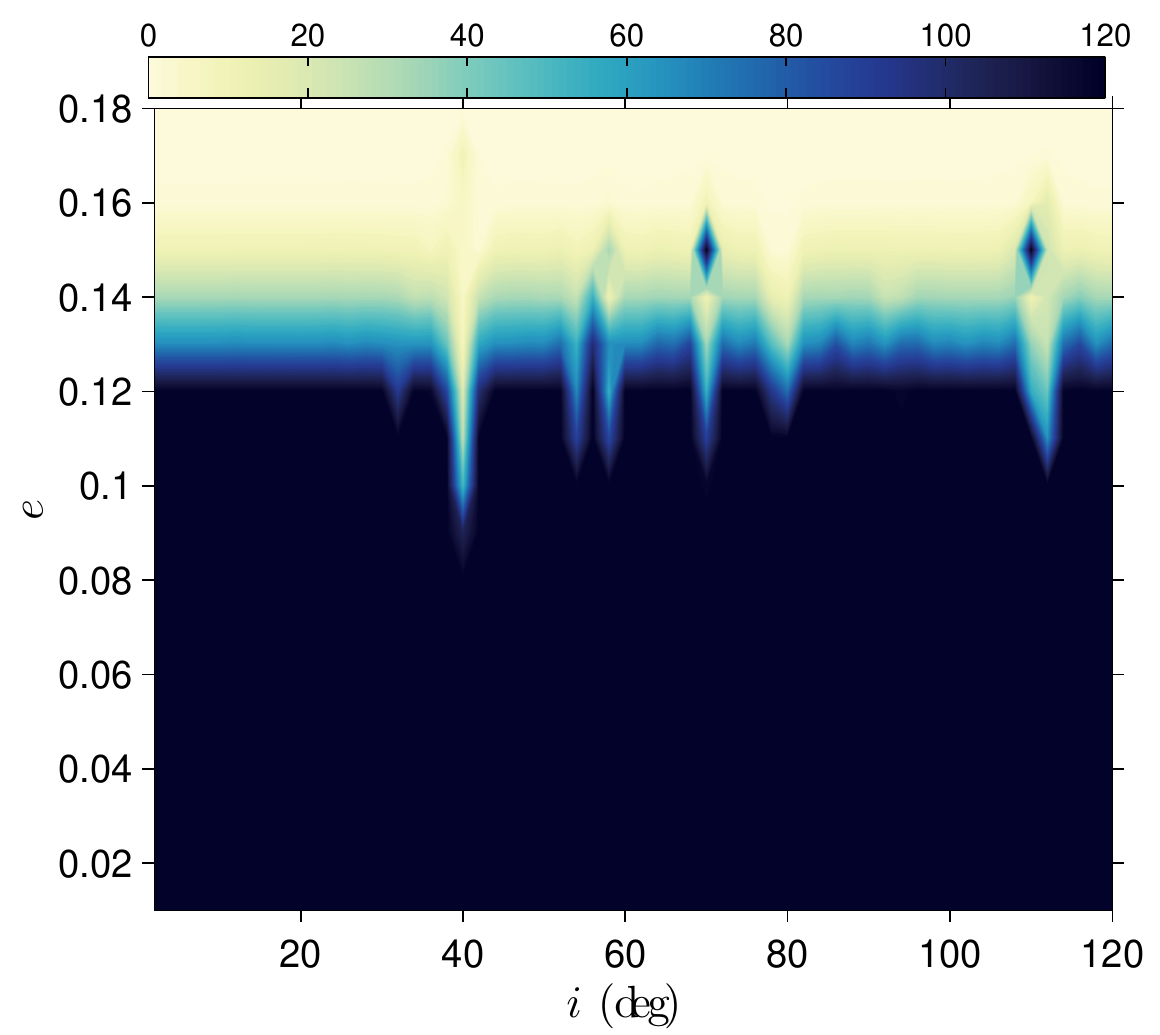} 
        \includegraphics[width=0.25\textwidth]{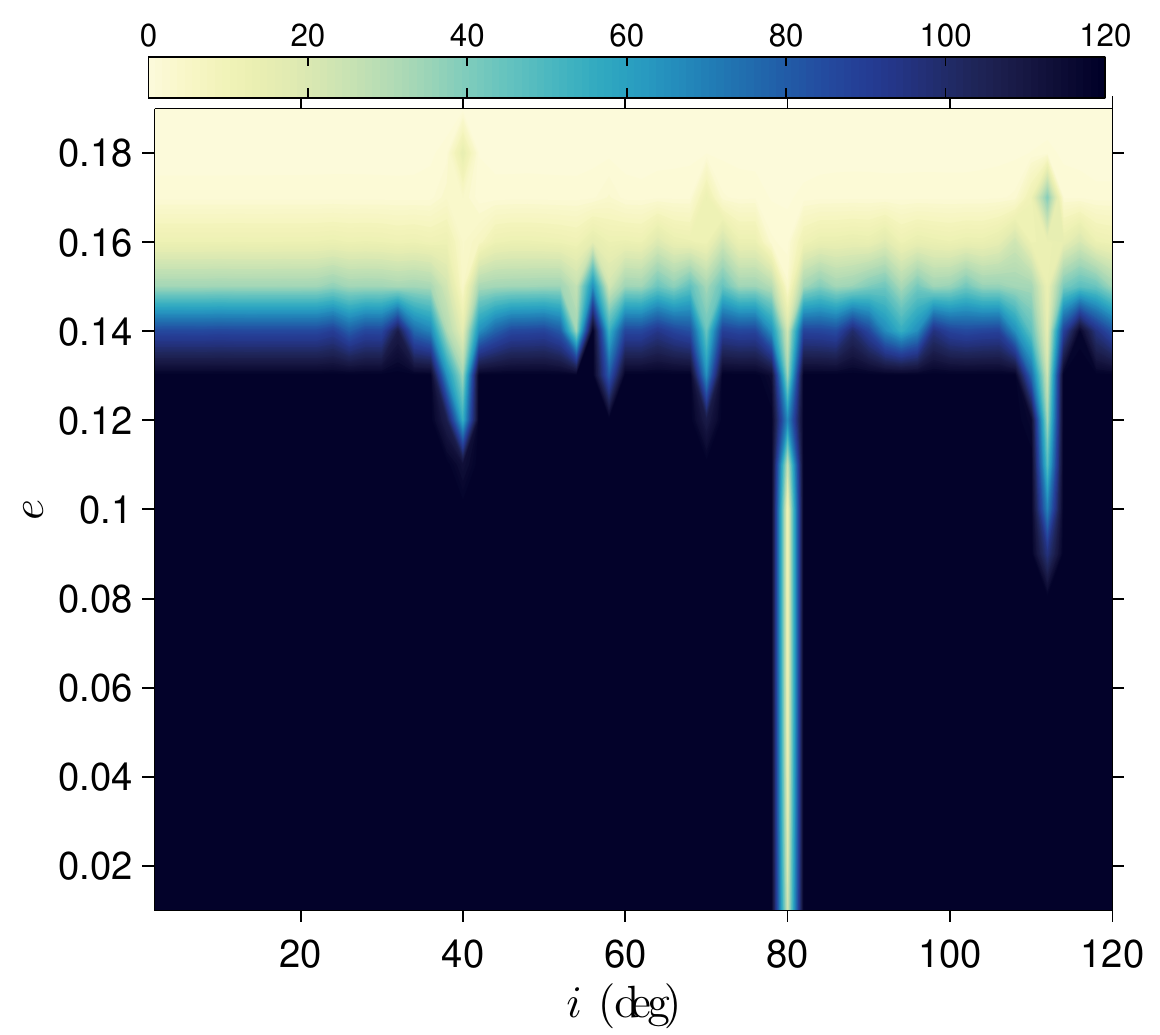}  \hspace{-0.3cm}  \includegraphics[width=0.25\textwidth]{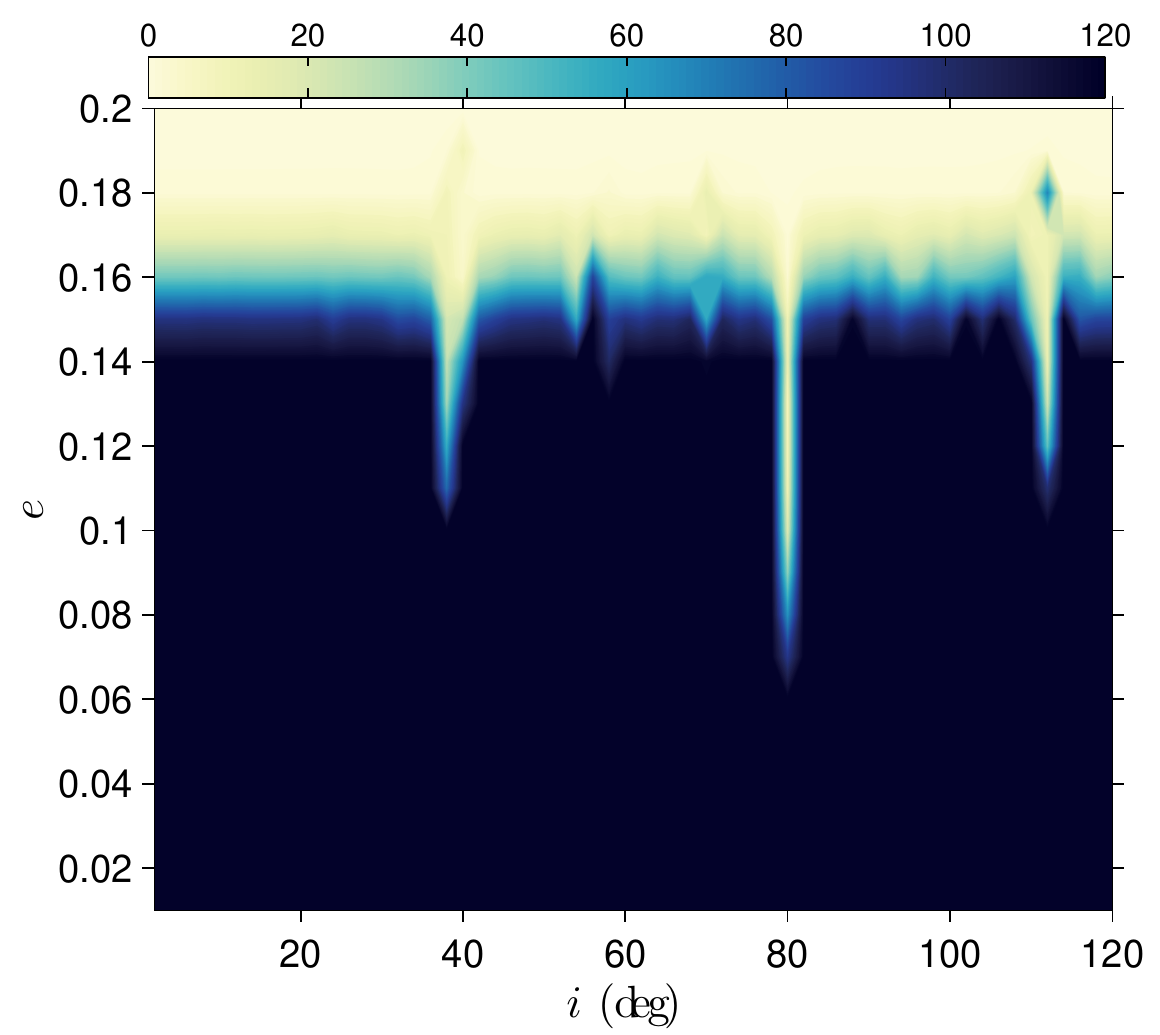}   \hspace{-0.3cm}   \includegraphics[width=0.25\textwidth]{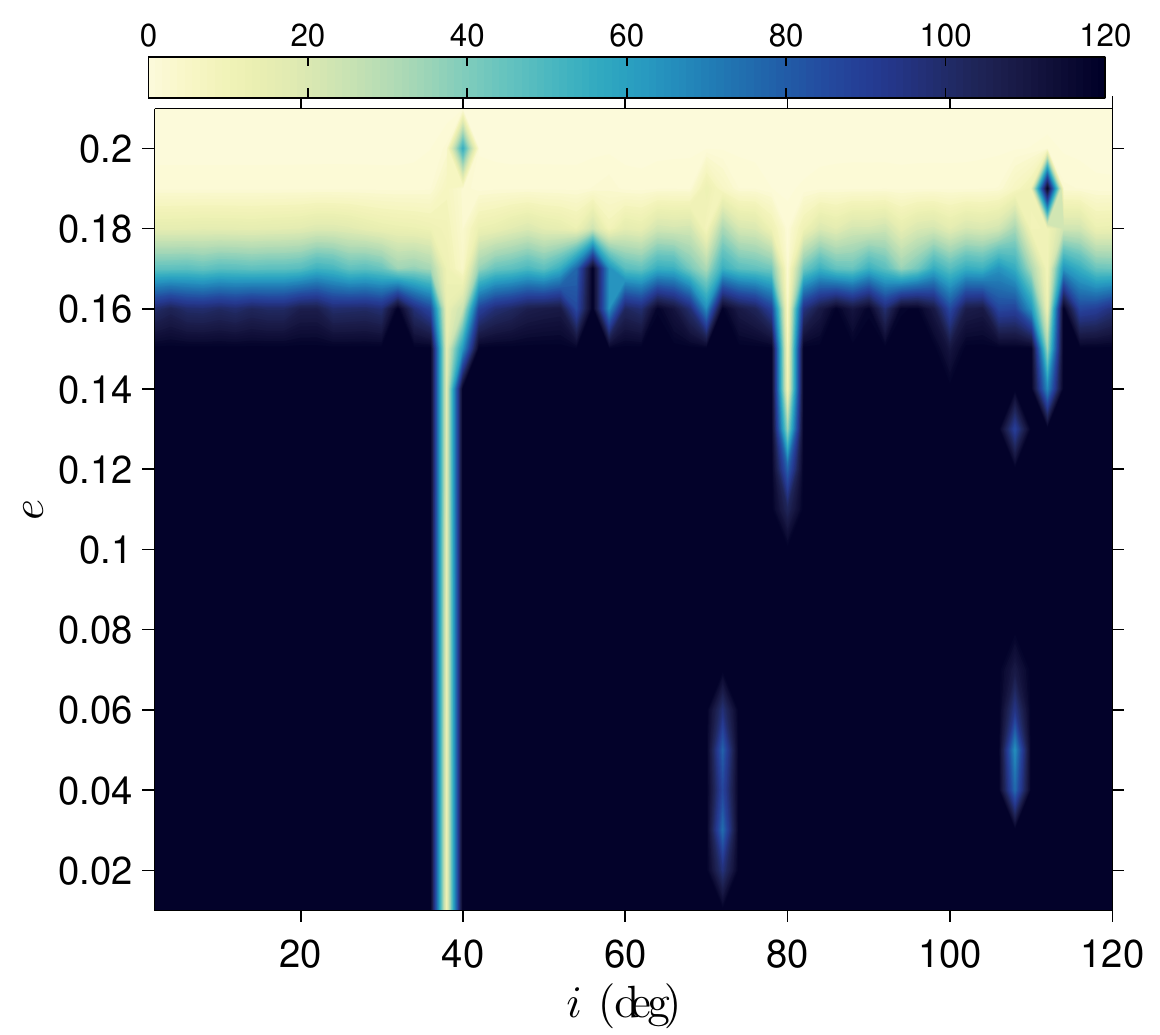}   \hspace{-0.3cm}   \includegraphics[width=0.25\textwidth]{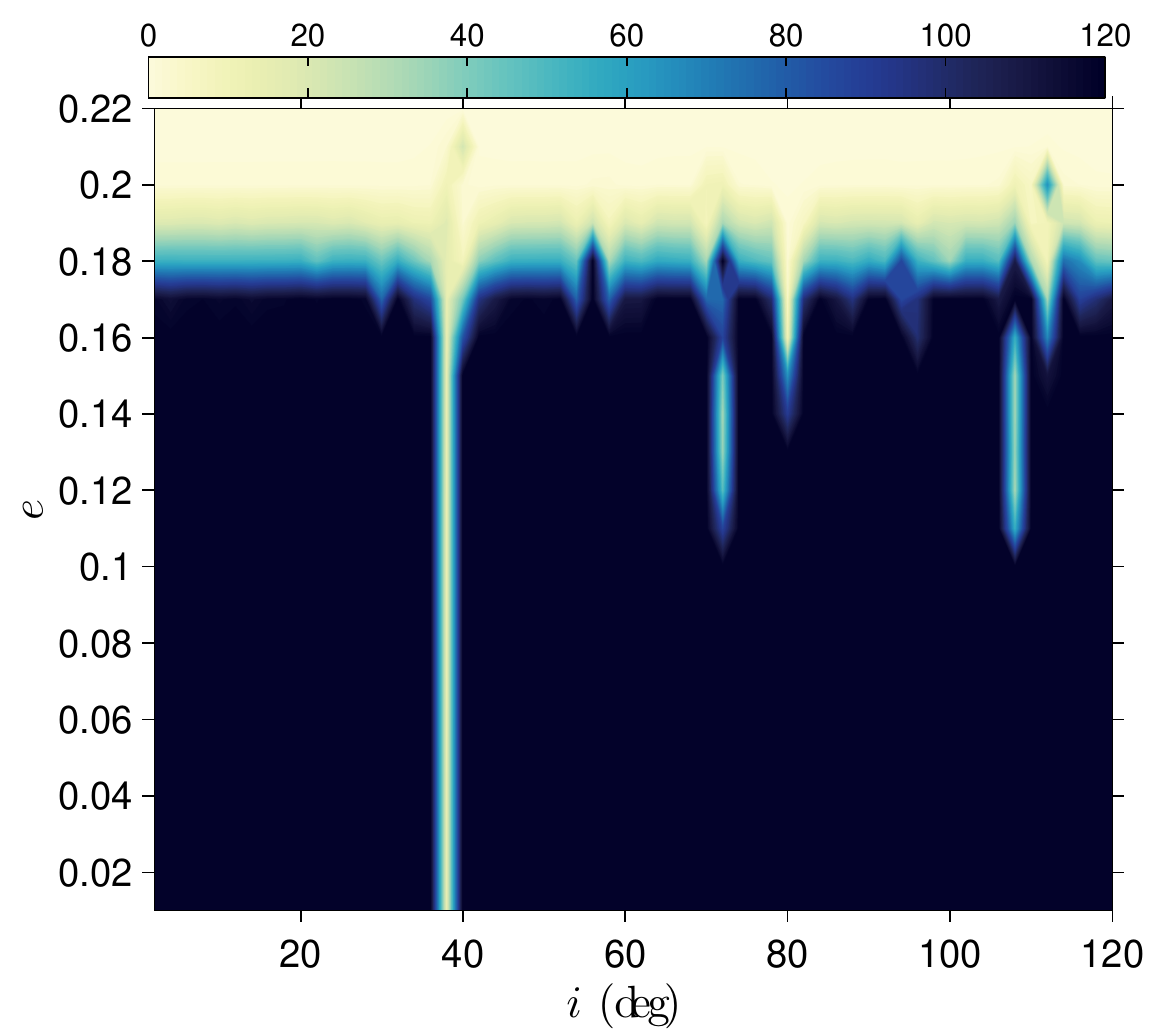} 
          \includegraphics[width=0.25\textwidth]{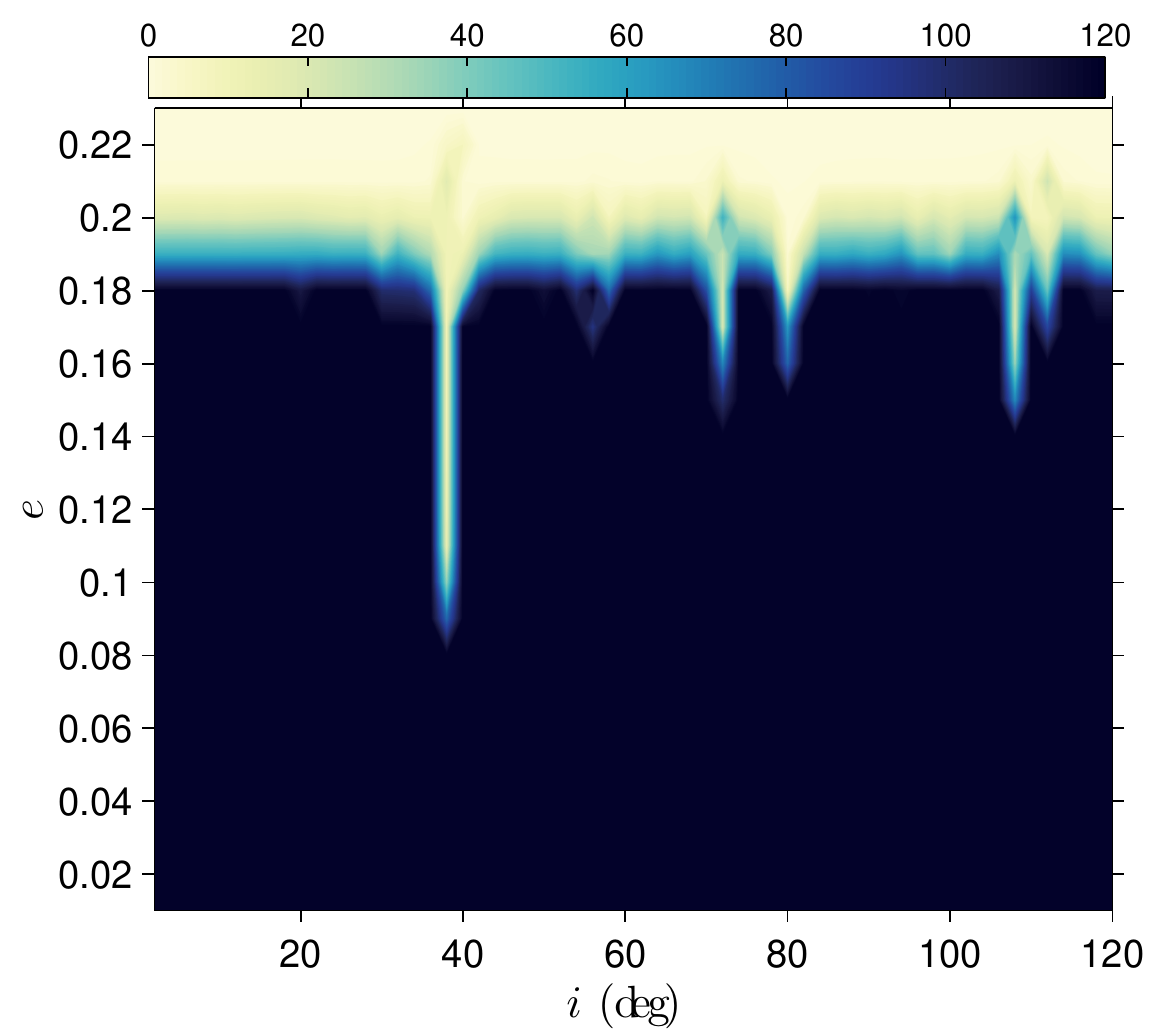}  \hspace{-0.3cm}  \includegraphics[width=0.25\textwidth]{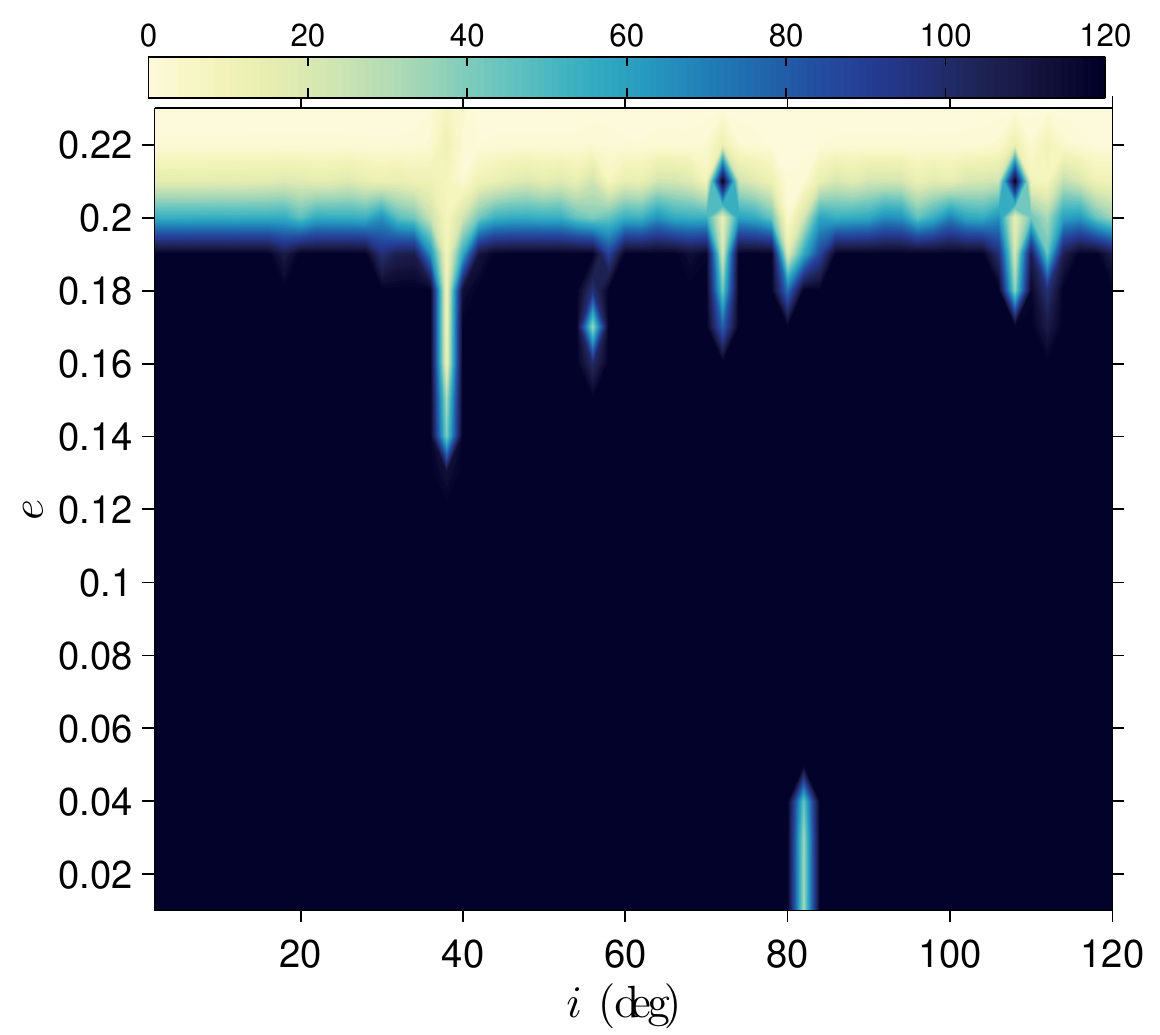}   \hspace{-0.3cm}   \includegraphics[width=0.25\textwidth]{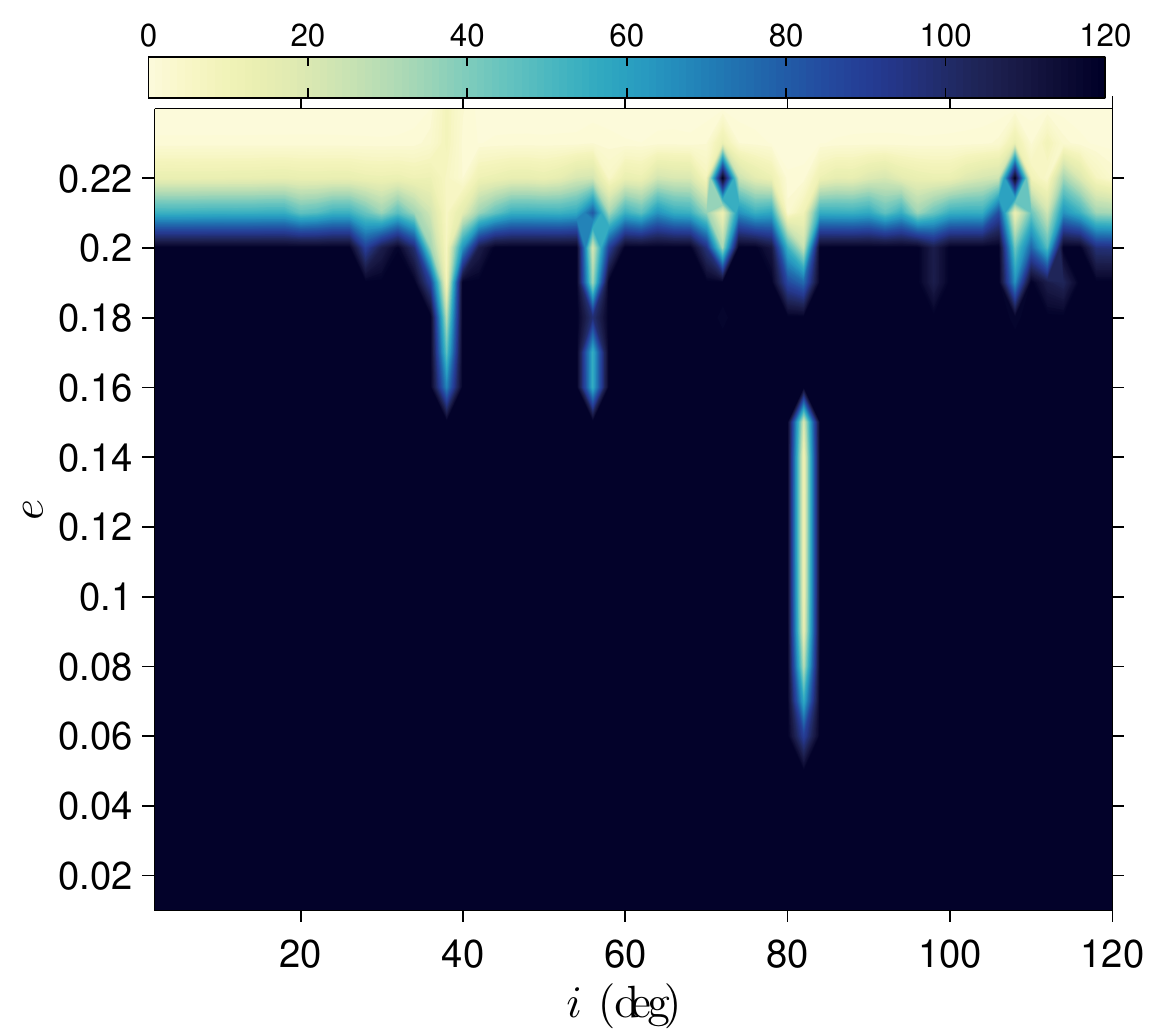}   \hspace{-0.3cm}   \includegraphics[width=0.25\textwidth]{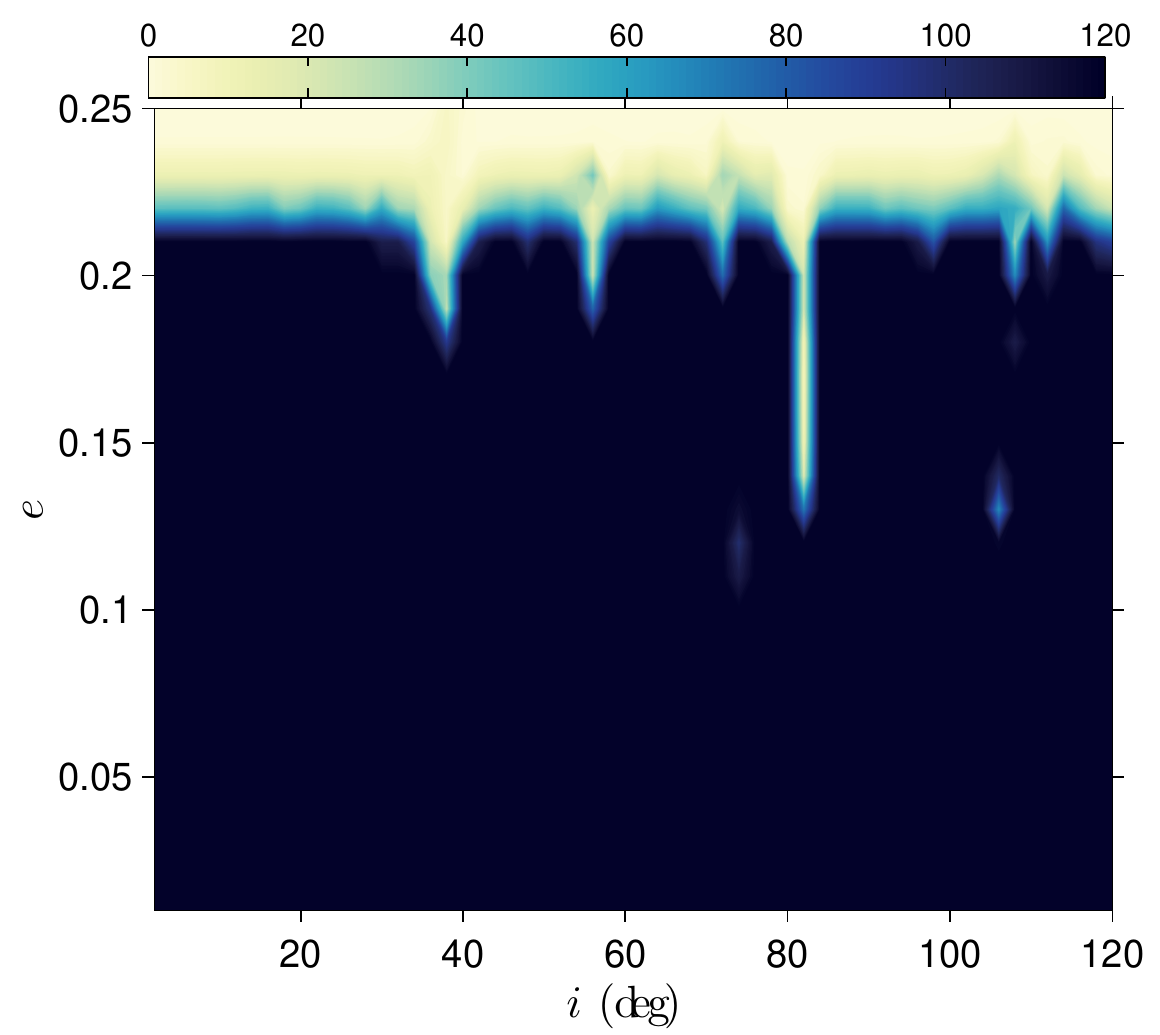} 
            \includegraphics[width=0.25\textwidth]{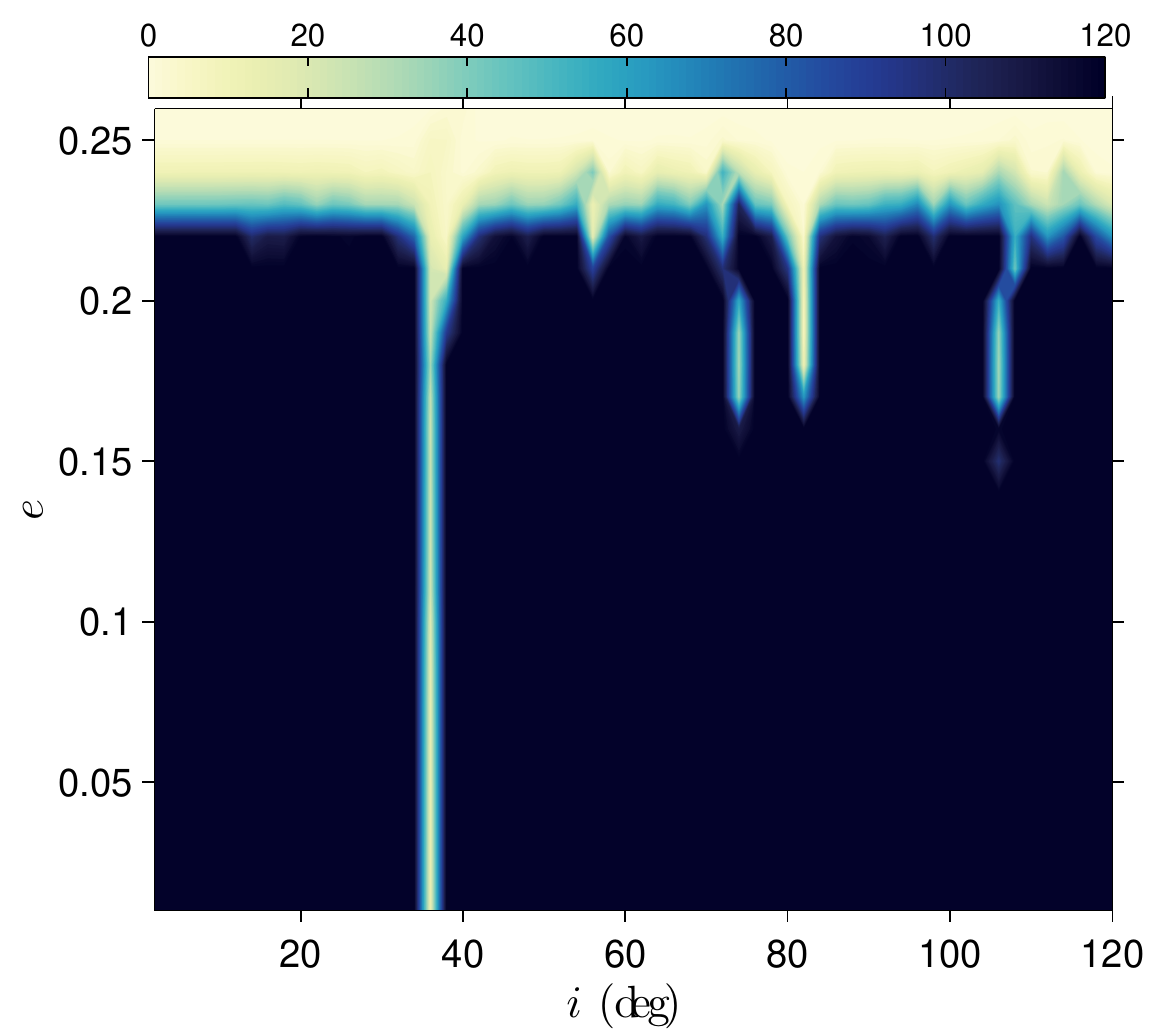}  \hspace{-0.25cm}  \includegraphics[width=0.25\textwidth]{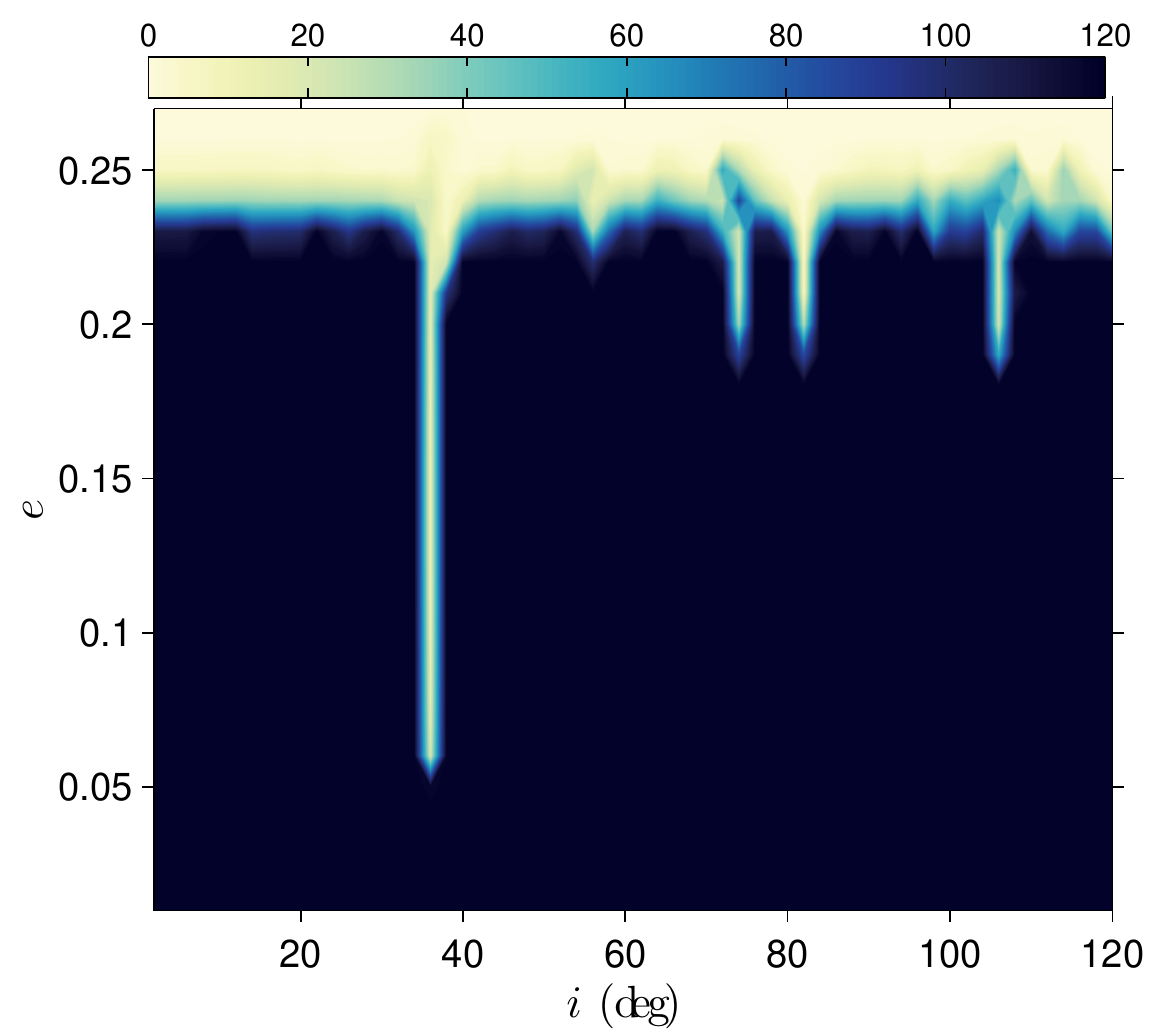}   \hspace{-0.3cm}   \includegraphics[width=0.25\textwidth]{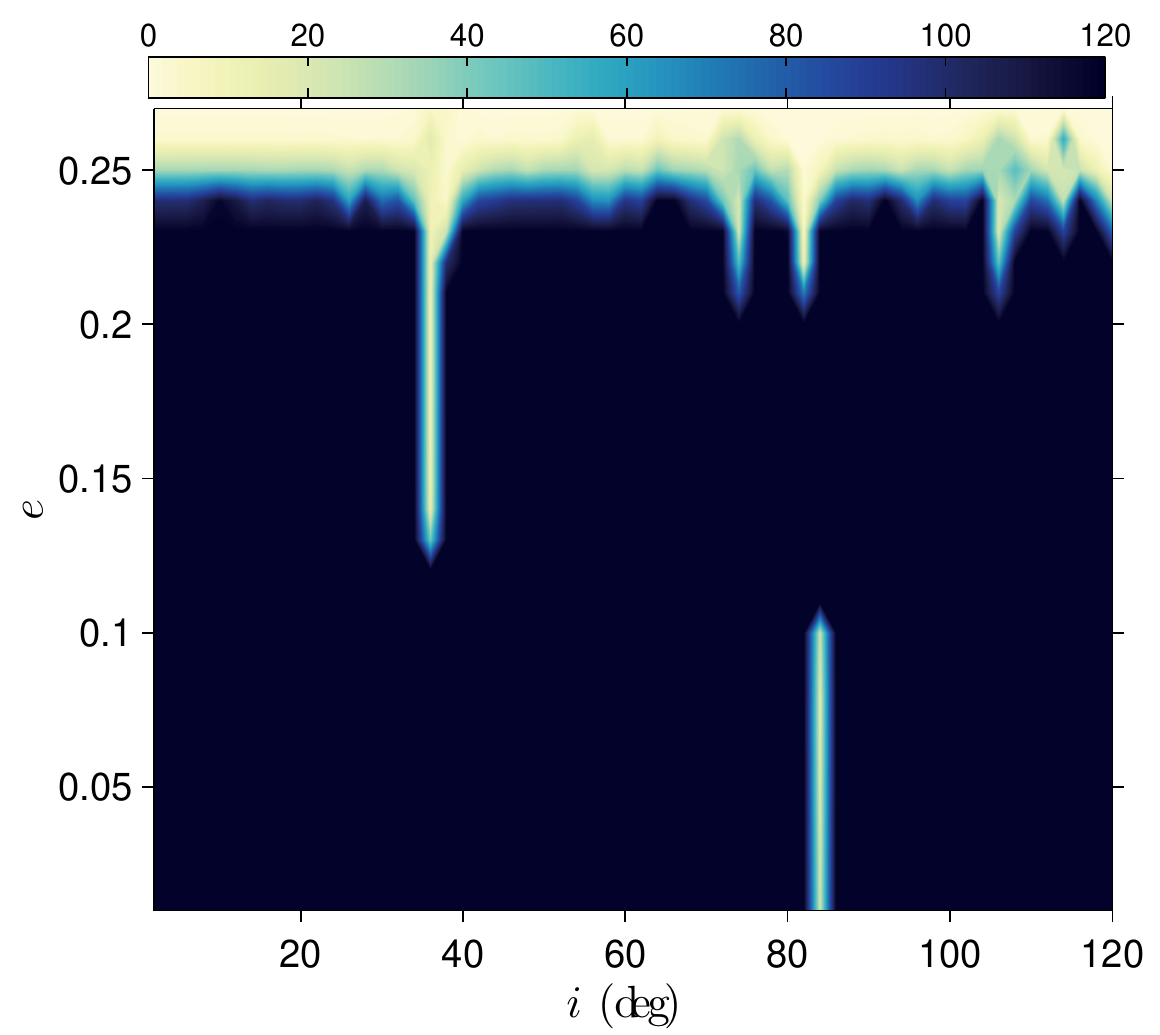}   \hspace{-0.3cm}   \includegraphics[width=0.25\textwidth]{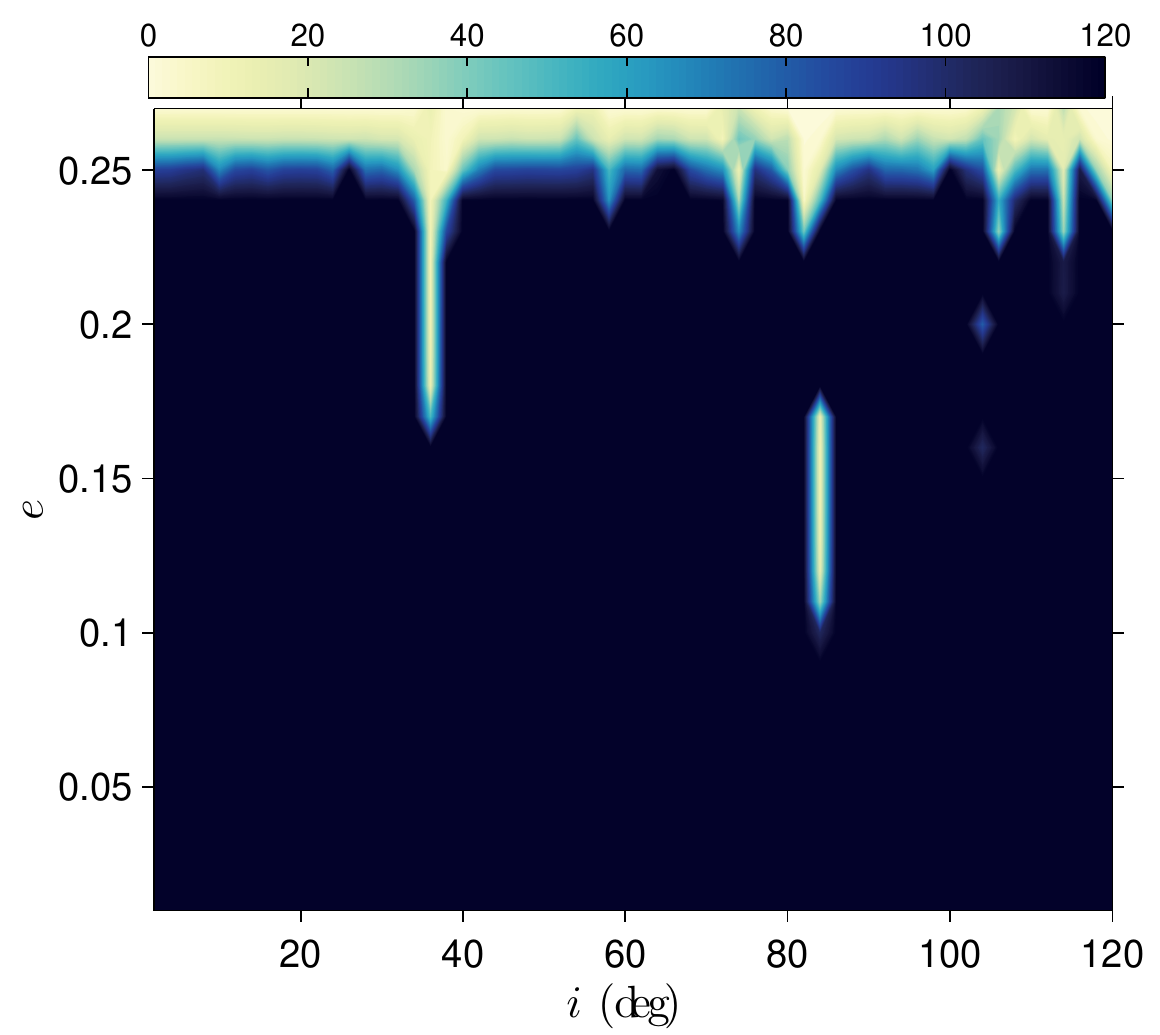} 
            \end{center}
        \caption{Lifetime computed (color bar) as a function of initial inclination and eccentricity for the initial epoch 2020 and $C_R (A/m)=1$
  m$^2/$kg, assuming $\Omega=0^{\circ}$ and $\omega=0^{\circ}$ at the initial epoch. Each plot depicts the behavior computed starting from a different value of initial semi-major axis. From the top left to the bottom right: $a=R_E+ 700$ km to $a=R_E+3000$ km at a step of 100 km.}\label{fig:ie_lifetime_2020_high_span_a}
  \end{figure}

Some illustrative plots for increasing values of initial semi-major
axis, considering the same initial $(\Omega, \omega)$ combination and initial epoch
are shown in Figure \ref{fig:ie_emax_2020_low_span_a} and Figure
\ref{fig:ie_lifetime_2020_low_span_a}, for a standard value of the
area-to-mass ratio, as a function of the initial inclination and eccentricity.
The two figures represent the behavior in terms of
maximum eccentricity and corresponding lifetime, respectively. Analogous plots are displayed in Figure \ref{fig:ie_emax_2020_high_span_a} and Figure
\ref{fig:ie_lifetime_2020_high_span_a} for the same initial epoch and $A/m=1$ m$^2/$kg.

Looking
to the maximum eccentricity behavior, for the majority of the initial
eccentricity values the variation in 120 years is not significant;
i.e., the color associated with a given initial eccentricity corresponds, in
general, to the same initial value. There exist, however, interesting well-defined
regions where the color associated with a
given initial inclination and eccentricity is `brighter' or
`darker'. The effect is much more evident for the high value of area-to-mass ratio (Figure \ref{fig:ie_emax_2020_high_span_a}), but an accurate look can detect the same feature also for the lower area-to-mass ratio case. For instance, there are `yellower' bands or islands in correspondence of $i\approx 40^{\circ}$ or $i\approx 80^{\circ}$ on the fourth and the last line of plots in Figure  \ref{fig:ie_emax_2020_low_span_a}. In all the panels of the same figure, similar behaviors can be noticed. 

Focusing on Figure \ref{fig:ie_lifetime_2020_low_span_a}, where the colorbar reports the lifetime, we can see that for high values of semi-major axis, the variation in eccentricity just noticed for the standard value of area-to-mass ratio is not sufficient to provoke a reduction in lifetime. In other words, perturbation effects different from the atmospheric drag can be exploited only if the eccentricity is high enough, depending on the value of semi-major axis. This is the reason why most of the plots is black, apart from a narrow band corresponding to high values of eccentricity. The width of this eccentricity band is much larger for the higher value of area-to-mass ratio (see Figure  \ref{fig:ie_lifetime_2020_high_span_a}).

\begin{figure}[th!]
  \begin{center}
   \includegraphics[width=0.25\textwidth]{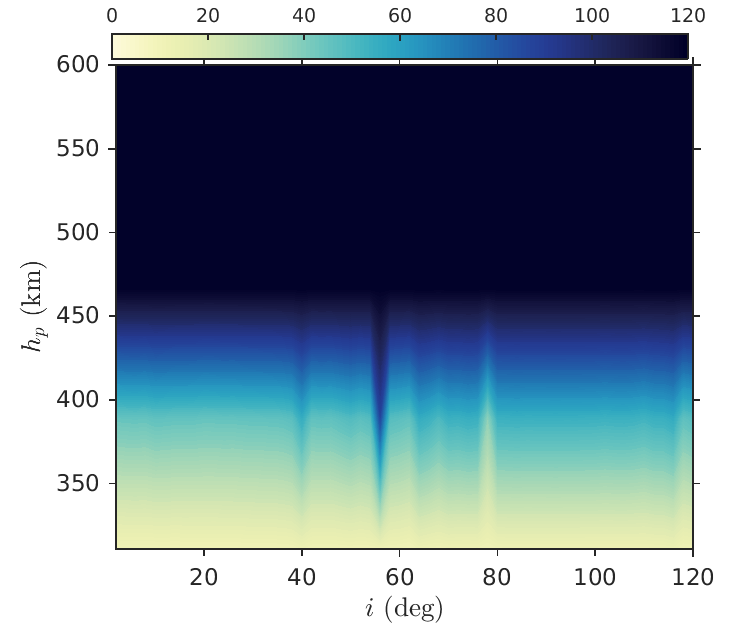} \hspace{-0.25cm}  \includegraphics[width=0.25\textwidth]{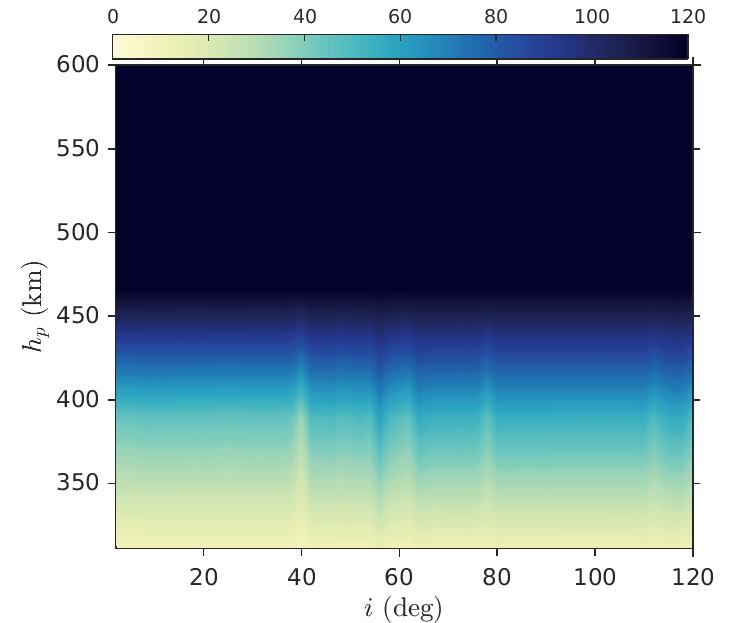}   \hspace{-0.3cm}   \includegraphics[width=0.25\textwidth]{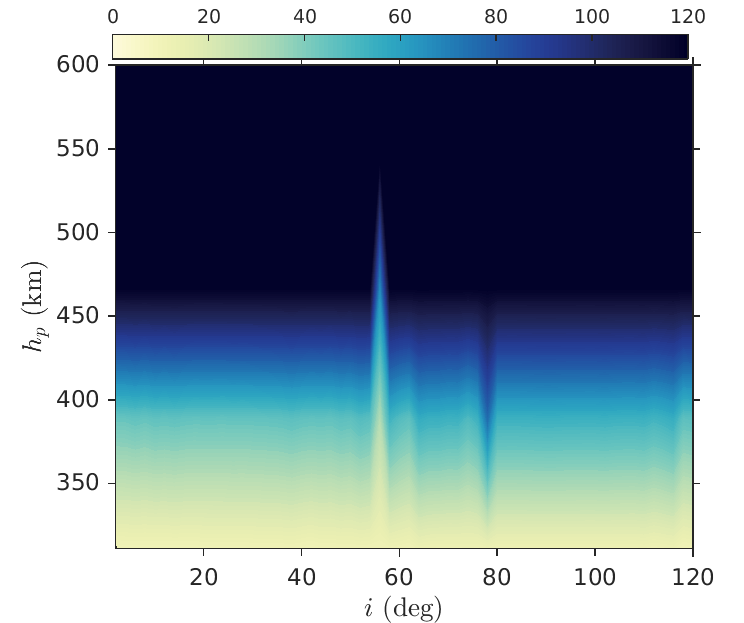}   \hspace{-0.3cm}   \includegraphics[width=0.25\textwidth]{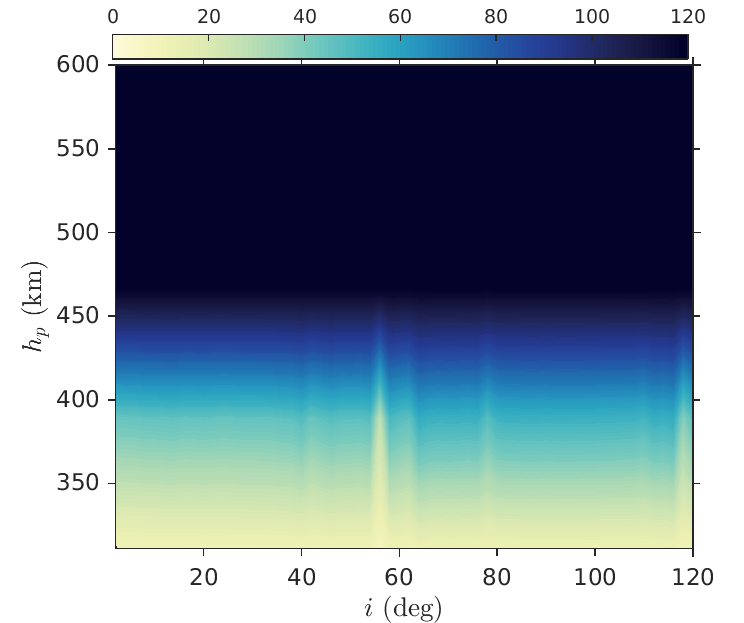} 
   \includegraphics[width=0.25\textwidth]{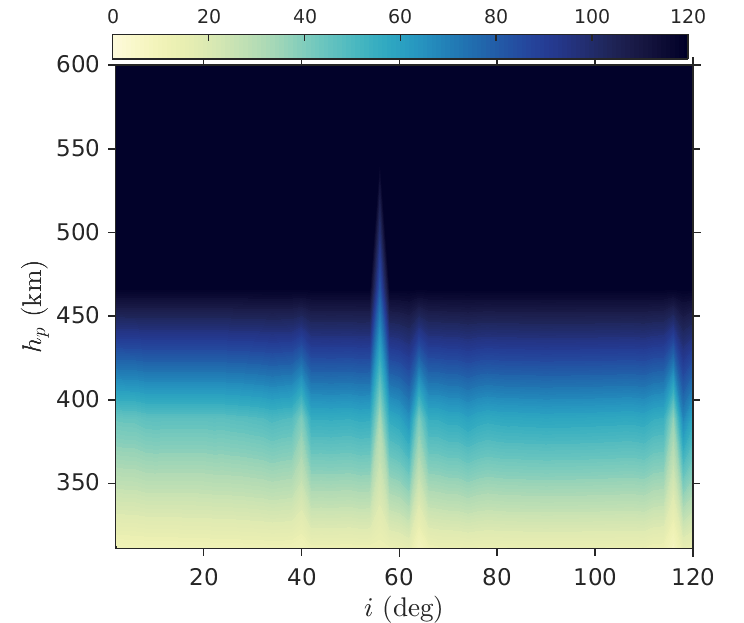} \hspace{-0.25cm}  \includegraphics[width=0.25\textwidth]{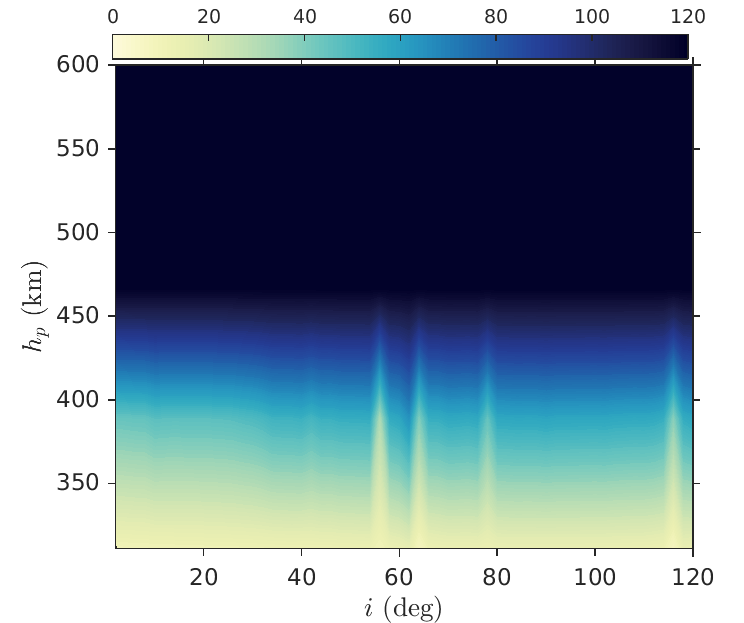}   \hspace{-0.3cm}   \includegraphics[width=0.25\textwidth]{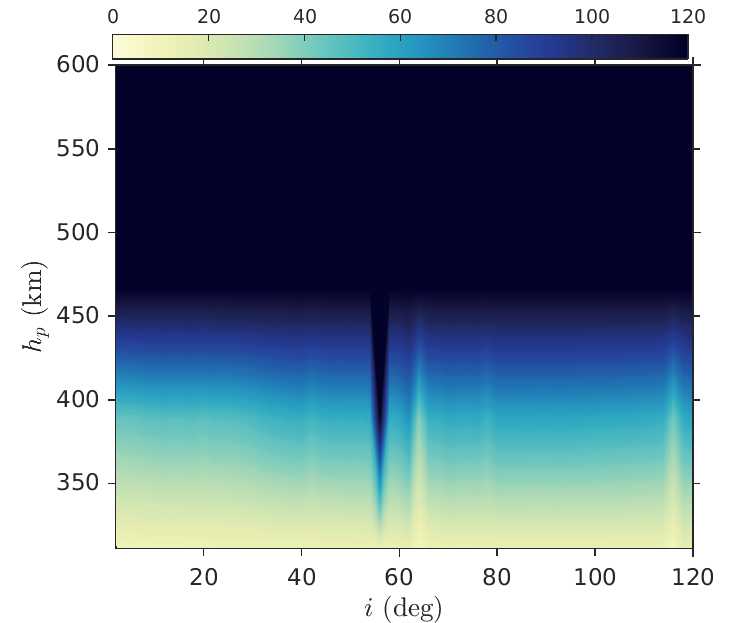}   \hspace{-0.3cm}   \includegraphics[width=0.25\textwidth]{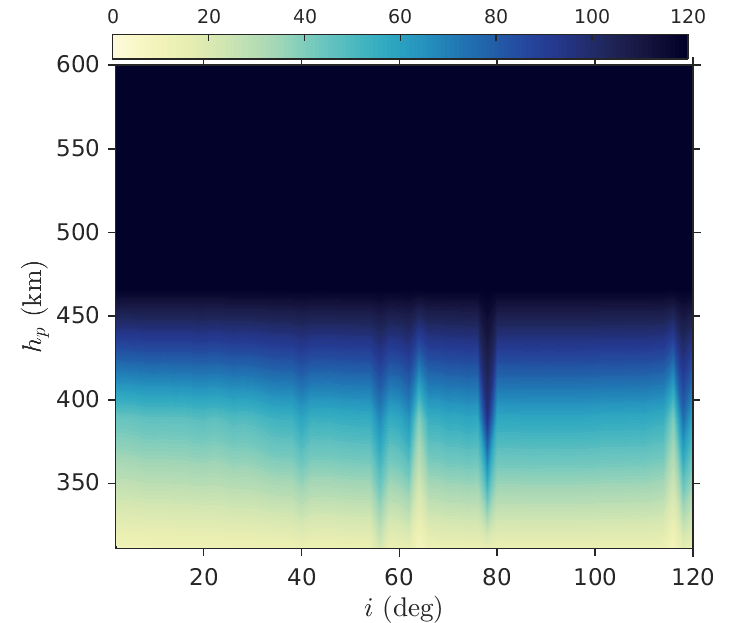}    
      \includegraphics[width=0.25\textwidth]{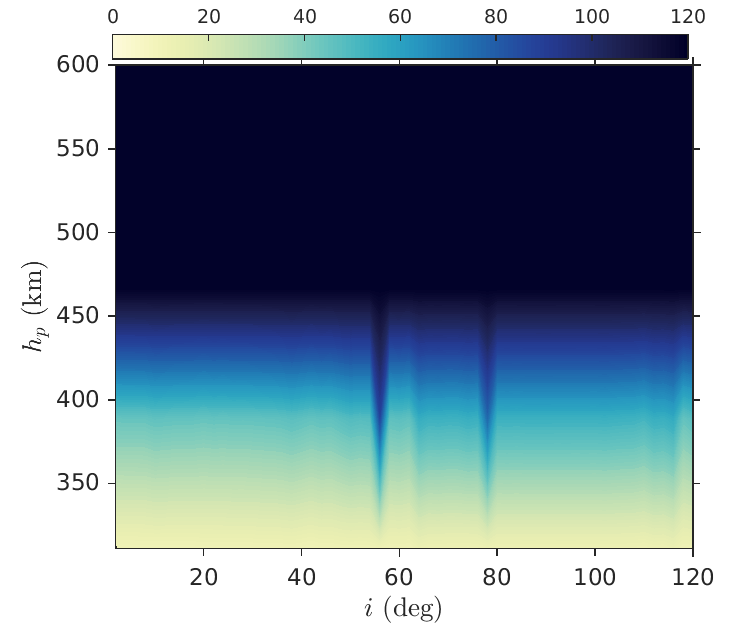} \hspace{-0.25cm}  \includegraphics[width=0.25\textwidth]{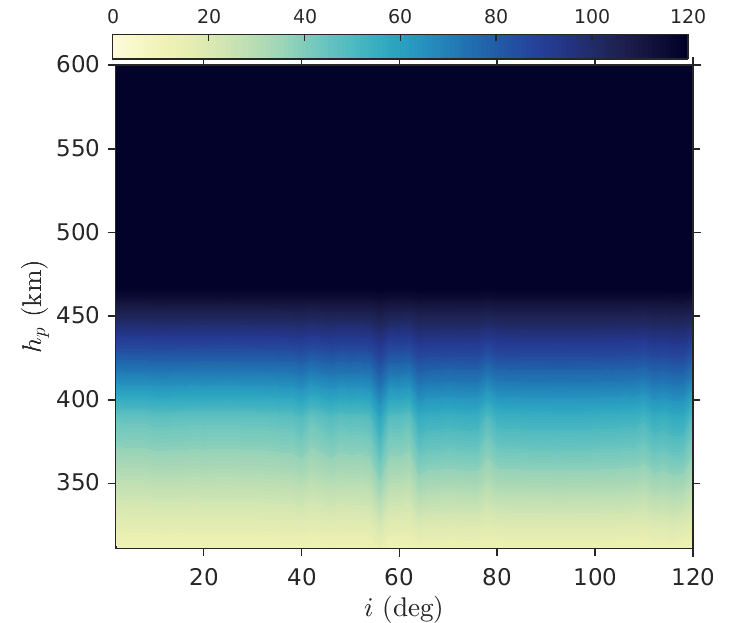}   \hspace{-0.3cm}   \includegraphics[width=0.25\textwidth]{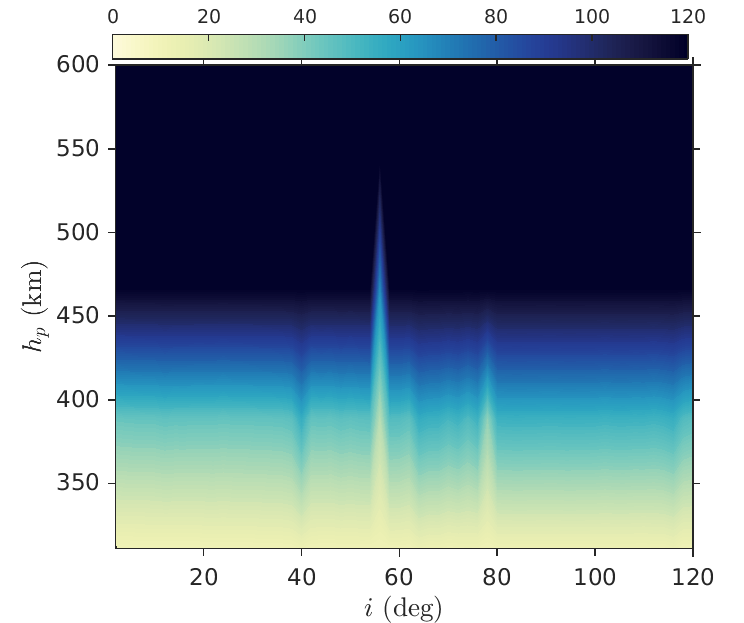}   \hspace{-0.3cm}   \includegraphics[width=0.25\textwidth]{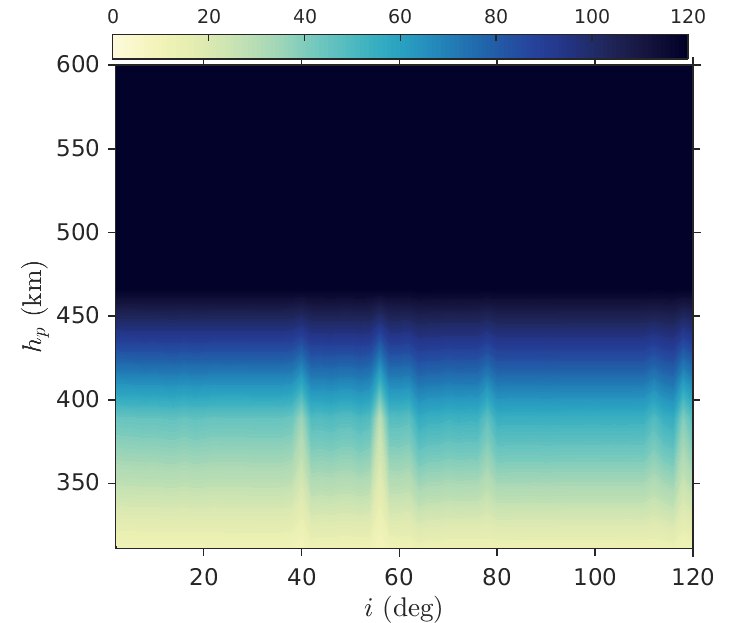}       
        \includegraphics[width=0.25\textwidth]{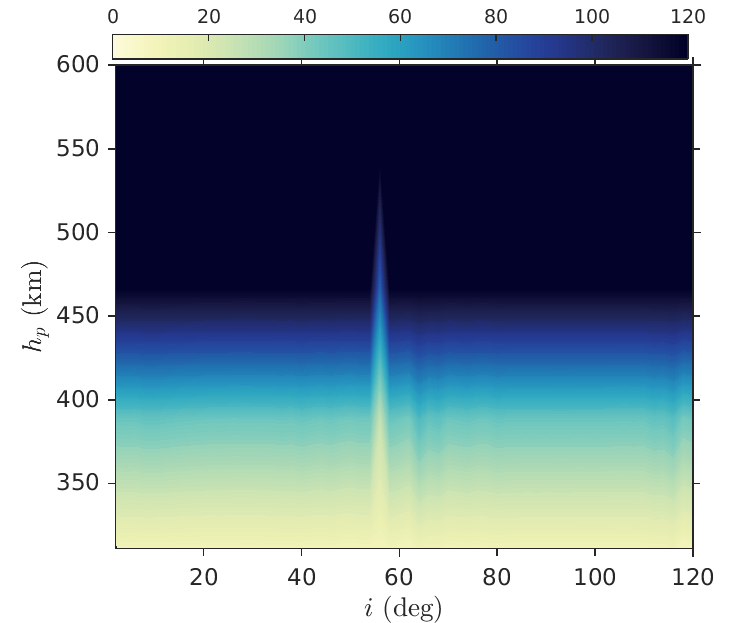} \hspace{-0.25cm}  \includegraphics[width=0.25\textwidth]{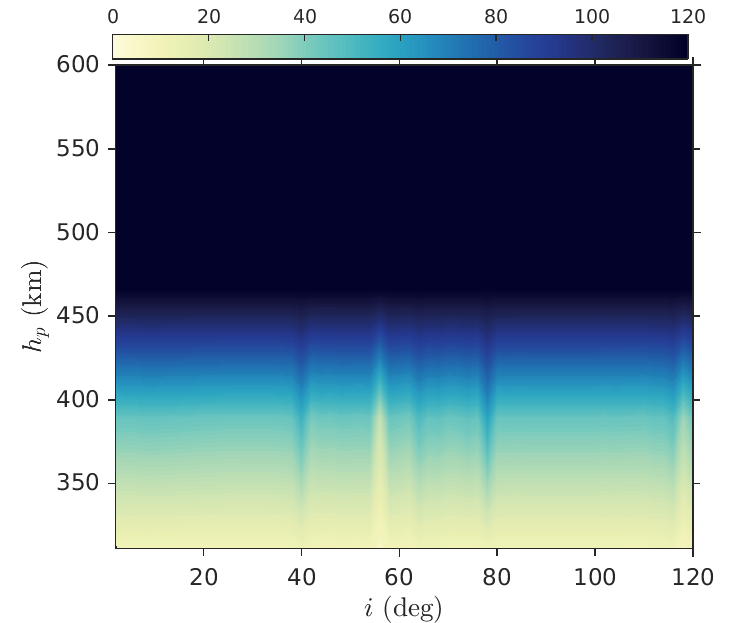}   \hspace{-0.3cm}   \includegraphics[width=0.25\textwidth]{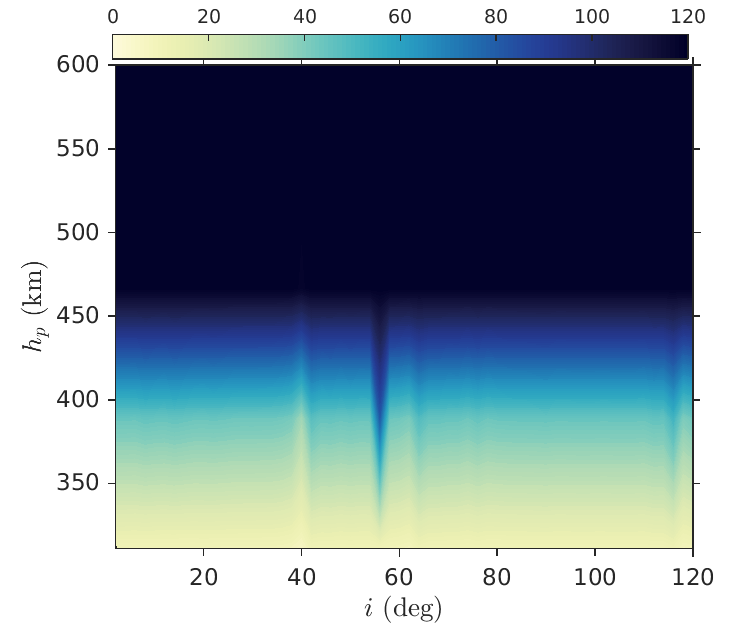}   \hspace{-0.3cm}   \includegraphics[width=0.25\textwidth]{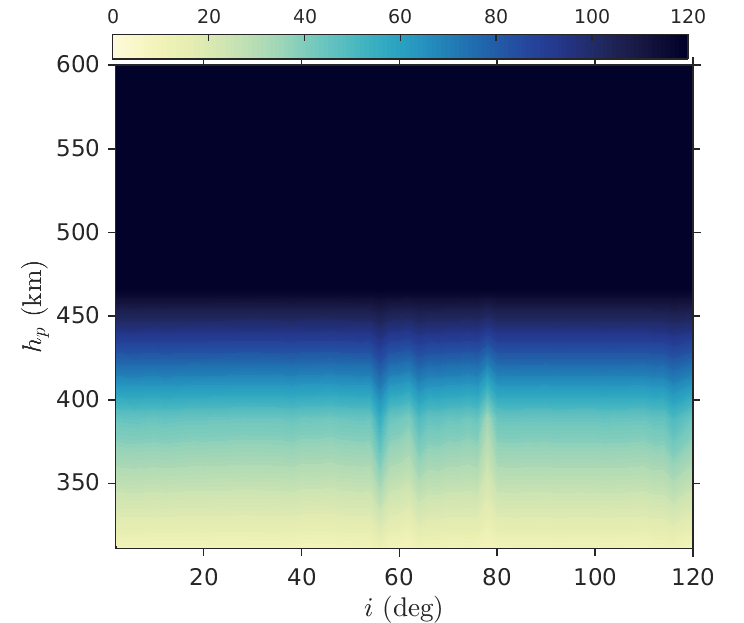}  
              \end{center}
        \caption{Lifetime computed (color bar) as a function of initial inclination and altitude of pericenter for the initial epoch 2020 and $C_R (A/m)=0.024$
  m$^2/$kg, assuming $a=R_E+1400$ km for different initial $(\Omega, \omega)$ configurations. From top to bottom: $\Omega=0^{\circ}$, $\Omega=90^{\circ}$, $\Omega=180^{\circ}$, $\Omega=270^{\circ}$. From left to right: $\omega=0^{\circ}$, $\omega=90^{\circ}$, $\omega=180^{\circ}$, $\omega=270^{\circ}$.} \label{fig:ihp_lifetime_2020_low_spanOo}
  \end{figure}

To better clarify this point, in
Figure \ref{fig:ihp_lifetime_2020_low_spanOo}, for $a=R_E+1400$ km and the standard area-to-mass ratio value,
we show the lifetime behavior as a function of the initial inclination
and value of pericenter altitude for all the 16 $(\Omega, \omega)$
configurations explored.  From this figure, we can see that to adopt the initial pericenter altitude
instead of the initial eccentricity as reference parameter helps in
understanding how deep the spacecraft should enter into the atmosphere
to eventually reenter. Moreover, in each plots of the figure bright
or dark bands stand out in correspondence of well-defined values of
inclinations, as it was already clear for the high area-to-mass ratio case from Figure \ref{fig:ie_emax_2020_high_span_a}. These are the resonant inclinations, associated
with the cyan stars in Figure
\ref{fig:lifetime_elliptic_LEO}. According to whether a resonant
inclination is associated with a lighter or darker shade than that of
the neighboring points, the lifetime is shorter or longer, i.e., the
eccentricity experiences a growth or a reduction. These are the
regions where the perturbation exerted by the geopotential or by the
Moon or by the Sun (both gravitationally or as solar radiation
pressure) are able to alter the general behavior. Finally, the initial
configuration is crucial to obtain a reduction or an increase in
lifetime.

\begin{figure}[th!]
\begin{center}
\includegraphics[width=0.7\textwidth]{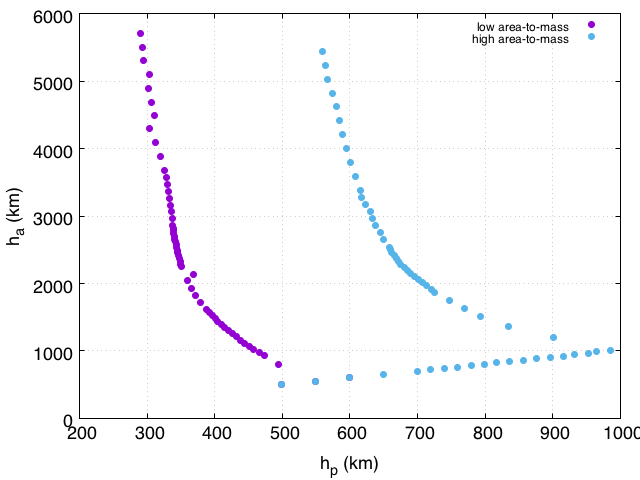} 
\end{center}
\caption{Pericenter $h_p$ and apocenter $h_a$ (km) required to reenter in 25 years for the two values of area-to-mass ratio considered in this study.}\label{fig:hp_ha_25y}
\end{figure}

\begin{figure}[th!]
\begin{center}
\includegraphics[width=0.3\textwidth]{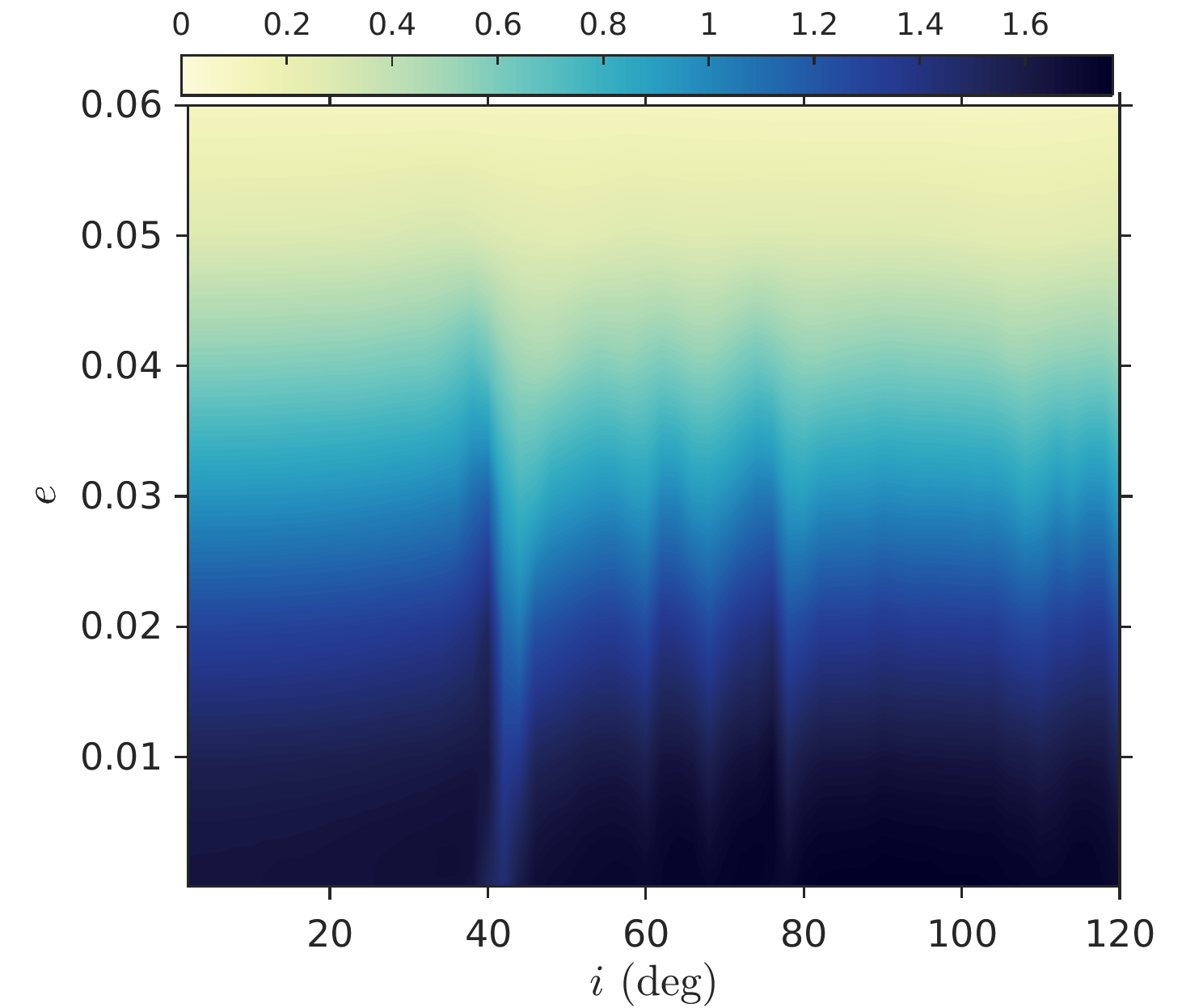} \includegraphics[width=0.3\textwidth]{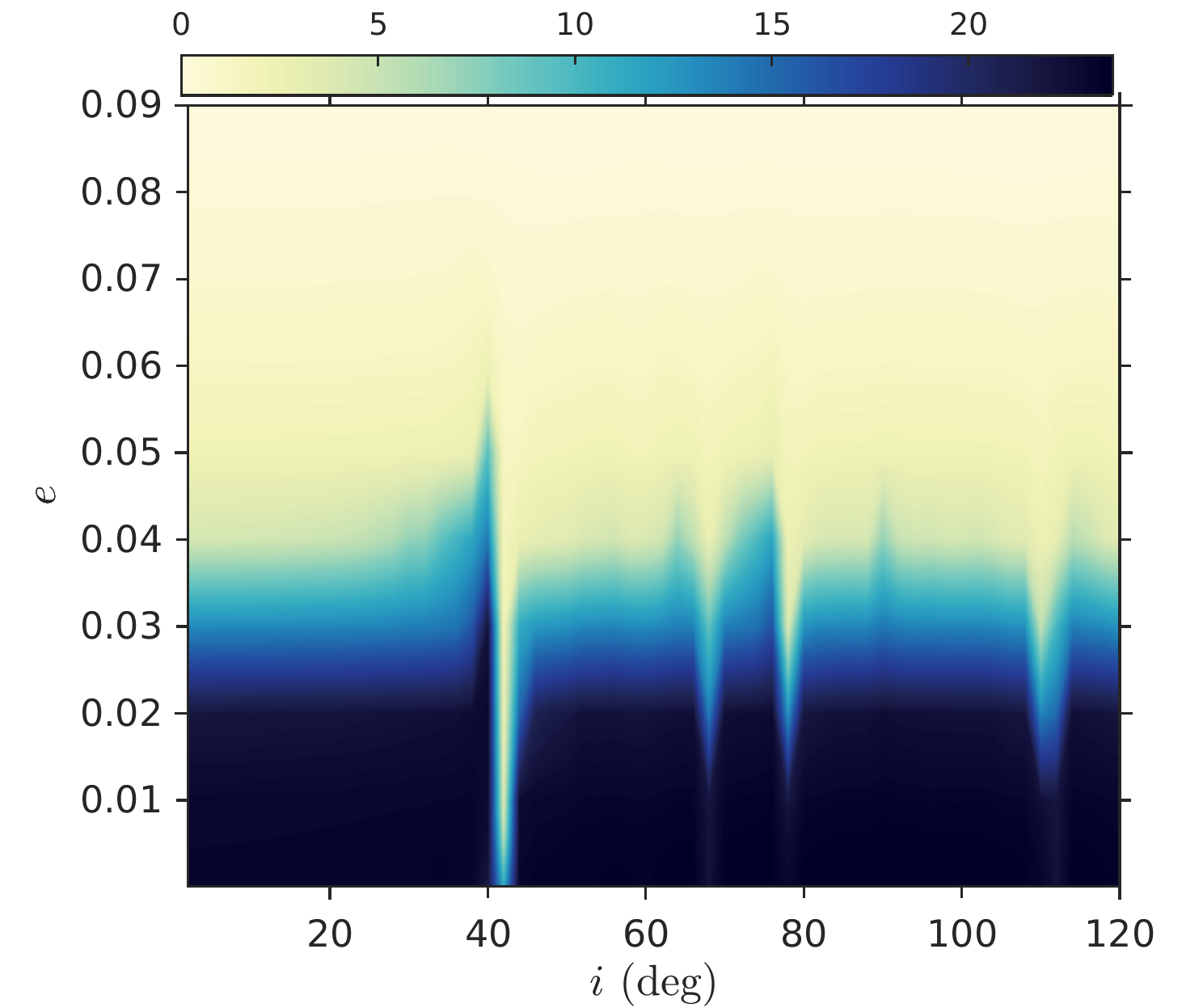} \includegraphics[width=0.3\textwidth]{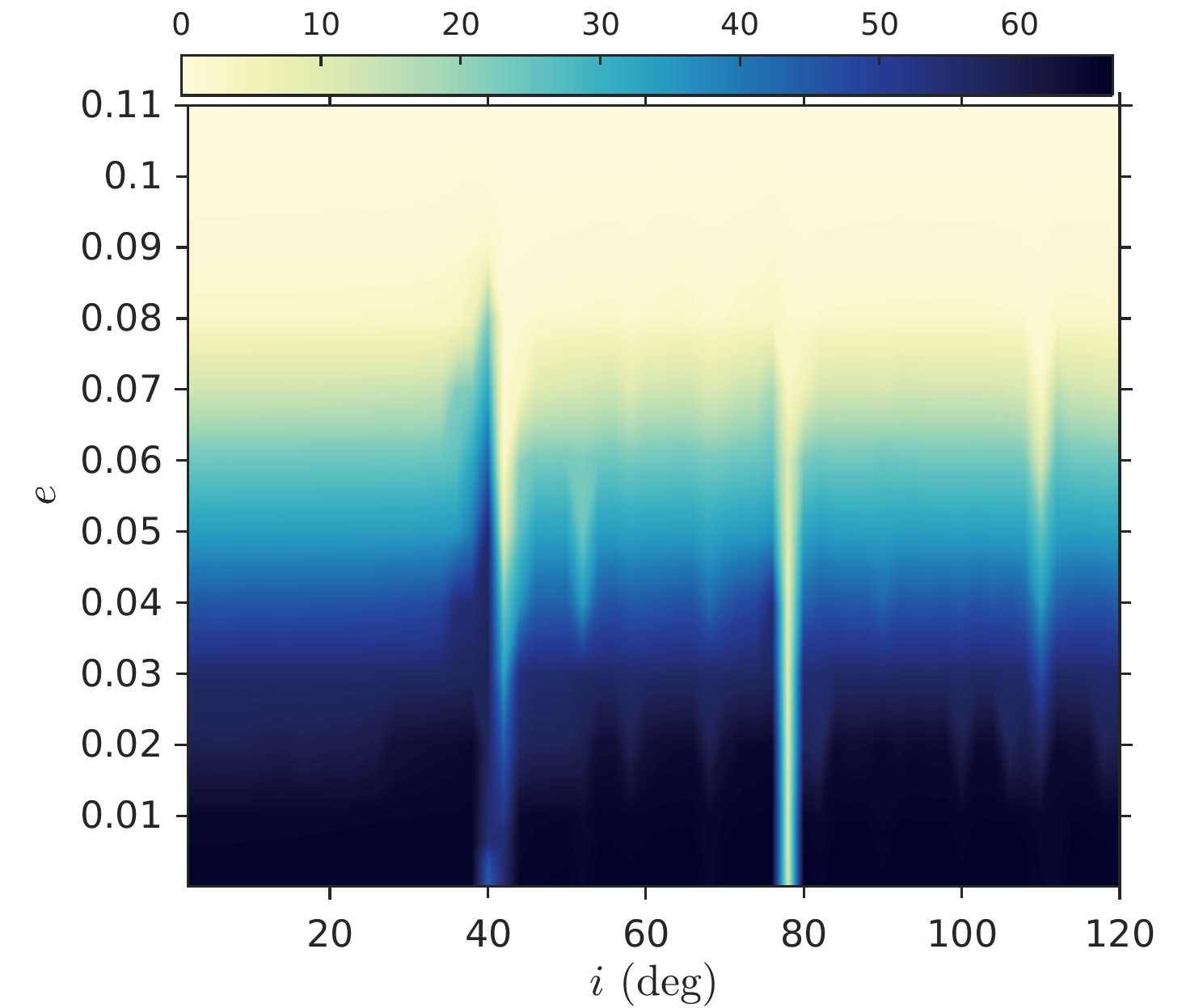} 
\end{center}
\caption{Lifetime computed over 120 years (color bar) as a function of initial inclination and eccentricity for $\Omega=180^{\circ}$ and $\omega=90^{\circ}$ at the initial epoch 2020, with $C_R (A/m)=1$
  m$^2/$kg. Left: $a=R_E + 800$ km; middle: $a=R_E + 1000$ km; right: $a=R_E + 1200$ km.}\label{fig:drag_AM}
\end{figure}

Concerning the results for the high value of area-to-mass ratio considered, a
complete analysis of the results is beyond the scope of this work, but it is worth stressing
that the simulations revealed that the exploitation of a drag
sail might be successful also for circular orbits up to an altitude of about
1200 km, if we allow a reentry time as long as 50 years; up to about 1050 km, if we aim at complying with the 25-year rule. 
Apart from the behavior displayed in Figure \ref{fig:ie_lifetime_2020_high_span_a}, we provide further examples for a different initial $(\Omega, \omega)$ initial configuration in Figure
\ref{fig:drag_AM}, where we show the lifetime computed for $a=R_E + 800$ km,
$a=R_E+1000$ km and $a=R_E + 1200 $ km. In Fig. \ref{fig:hp_ha_25y}, we show the pericenter and apocenter altitudes corresponding to a reentry in less than 25 years for both values of area-to-mass ratio. Notice the higher pericenter values in the case of an enhanced area-to-mass ratio and the behavior of quasi-circular orbit in the same case.

\begin{figure}[th!]
\begin{center}
\includegraphics[width=0.8\textwidth]{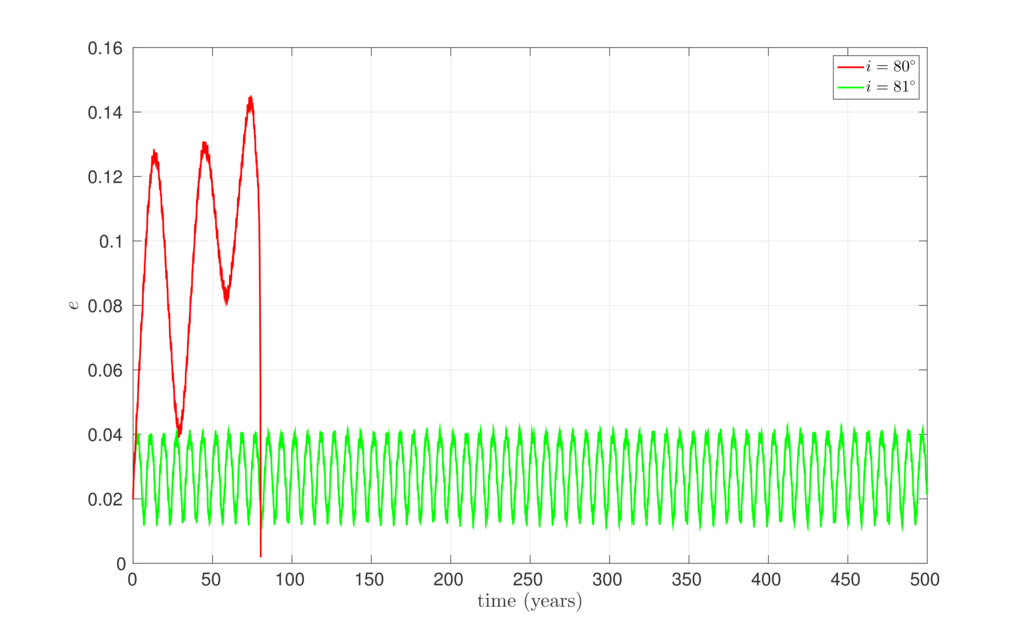} 
\end{center}
\caption{Eccentricity evolution considering $C_R (A/m)=1$  m$^2/$kg, $a=R_E+1950$ km, $e=0.02$, $\Omega=270^{\circ}$, $\omega=270^{\circ}$ at the initial epoch 2020. The behavior computed starting from two neighboring values of inclination are shown: $i=80^{\circ}$ (red curve) which corresponds to a SRP resonance, and $i=81^{\circ}$ (green curve) which does not. }\label{fig:example_Am}
\end{figure}

\begin{figure}[th!]
\begin{center}
\includegraphics[width=0.4\textwidth]{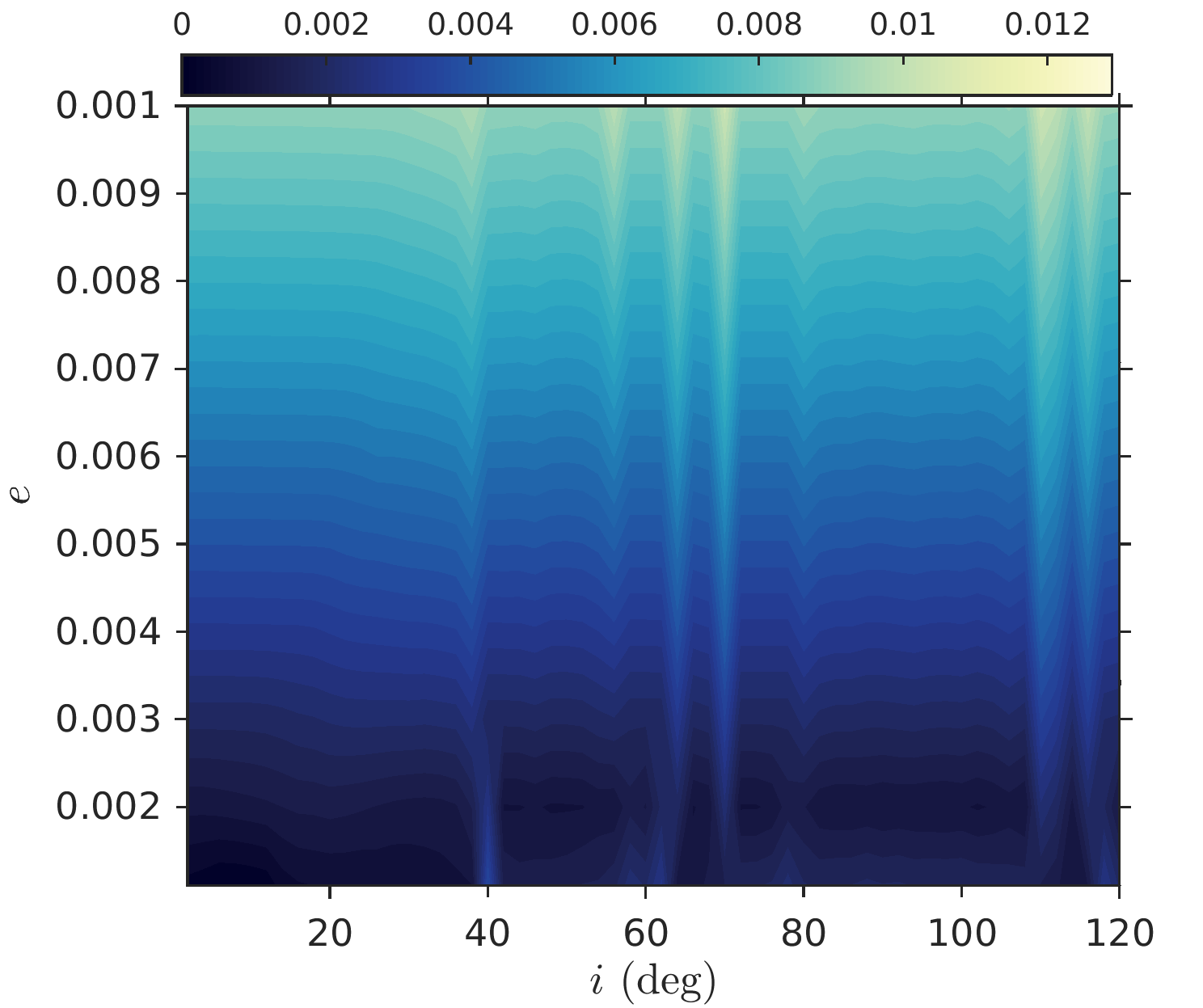} \includegraphics[width=0.4\textwidth]{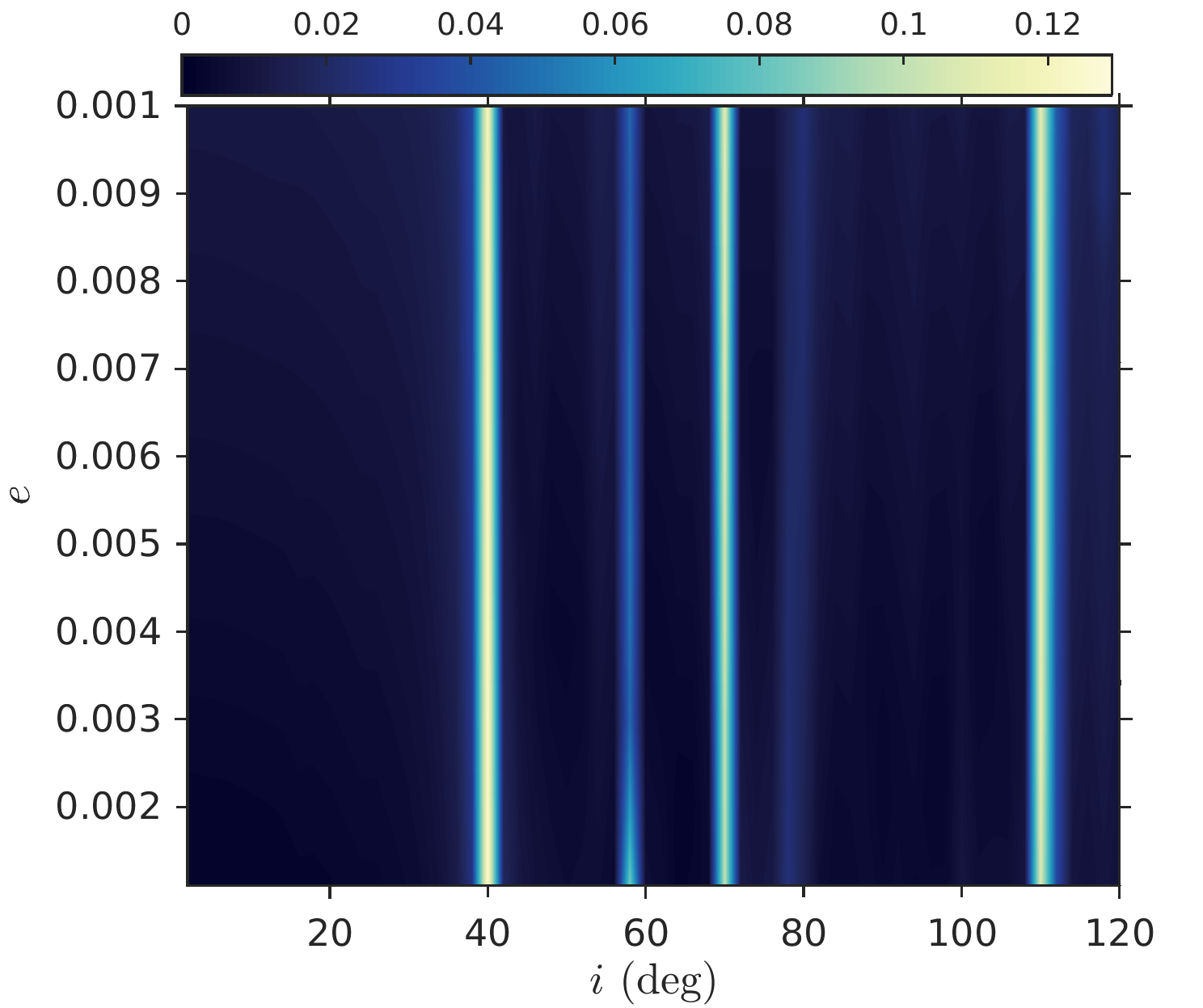} 
\end{center}
\caption{Maximum eccentricity computed over 120 years (color bar) for quasi-circular orbits as a function of initial inclination and eccentricity for $a=R_E + 1560$ km, $\Omega=90^{\circ}$, $\omega=0^{\circ}$ at the initial epoch 2020. Left: $C_R (A/m)=0.024$
  m$^2/$kg; right:  $C_R (A/m)=1$
 m$^2/$kg. Note the change of scale in the color bars between the left and the right panels.}\label{fig:two_Am}
\end{figure}

From the point of view of the orbital dynamics, the two main differences between the usage of a drag or a solar sail are the altitude where they can be exploited, and the fact that the drag sail can be considered for any value of inclination. Instead,
in order to properly exploit the SRP perturbation (i.e., the solar sail)
the corresponding resonant inclination bands shall be targeted. In this
case, for circular orbits, it is possible to achieve reentry for values of
semi-major axis up to about 2900 km (see third panel in the last line of Figure \ref{fig:ie_lifetime_2020_high_span_a}). An example of the SRP effect is provided in Figure
\ref{fig:example_Am}, where we show the long-term evolution of the eccentricity
for $C_R (A/m)=1$  m$^2/$kg, $a=R_E+1950$ km, $e=0.02$, $\Omega=270^{\circ}$,
$\omega=270^{\circ}$, and two different values of initial inclination at the
initial epoch 2020. In the case of the resonant inclination,  $i=80^{\circ}$,
the reentry can be achieved. In Figure \ref{fig:two_Am}, we compare the maximum
eccentricity behavior for a same initial condition for a satellite which is not
equipped with a sail, and a satellite which is.  The opening of strong reentry 
corridors at the resonant inclinations are clearly evident in the area-enhanced 
case of the right panel.

The number of reentry solutions computed for the same initial epoch 2020 for the two values of area-to-mass ratio is 83402 and 490920, respectively.

The reader interested in different maps can find other cases in \cite{A2017a} and \cite{A2017b}.

\subsection{On the Geopotential Expansion}\label{sec:BomPom}

\begin{figure}[ht!]
  \begin{center}
       \includegraphics[width=0.3\textwidth]{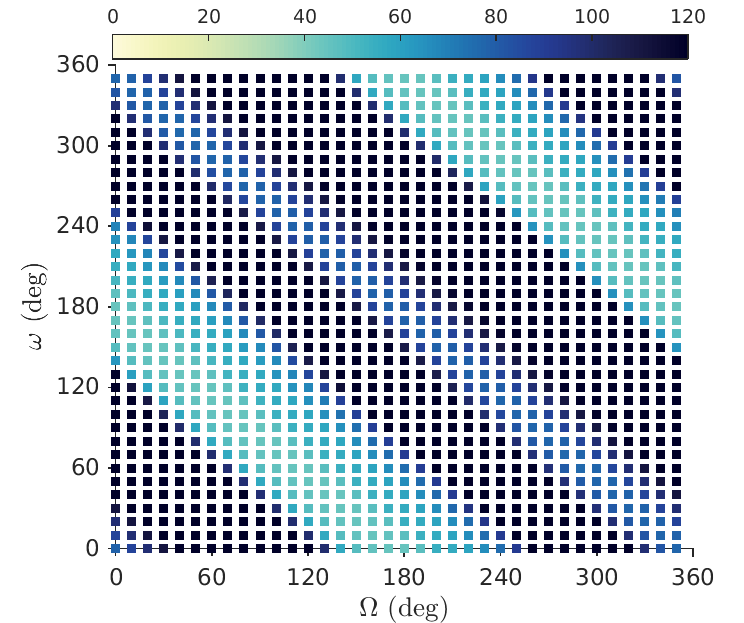}   \includegraphics[width=0.3\textwidth]{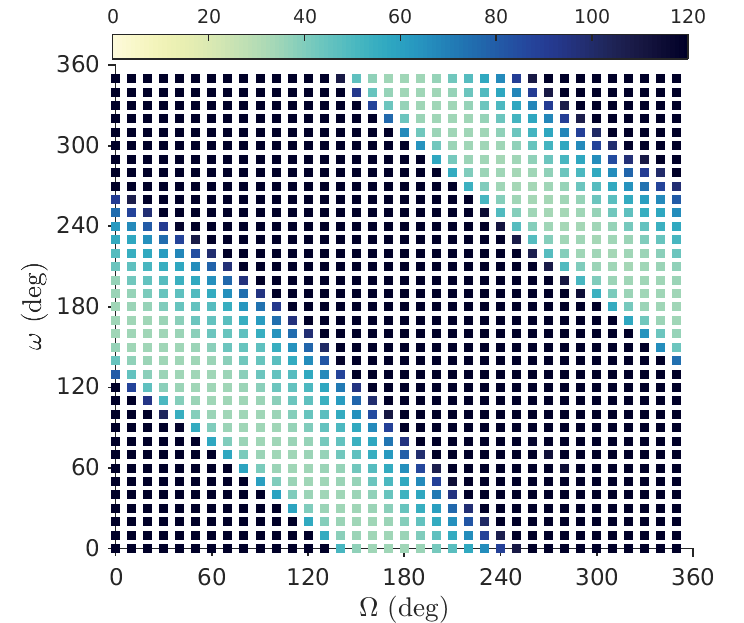}     \includegraphics[width=0.3\textwidth]{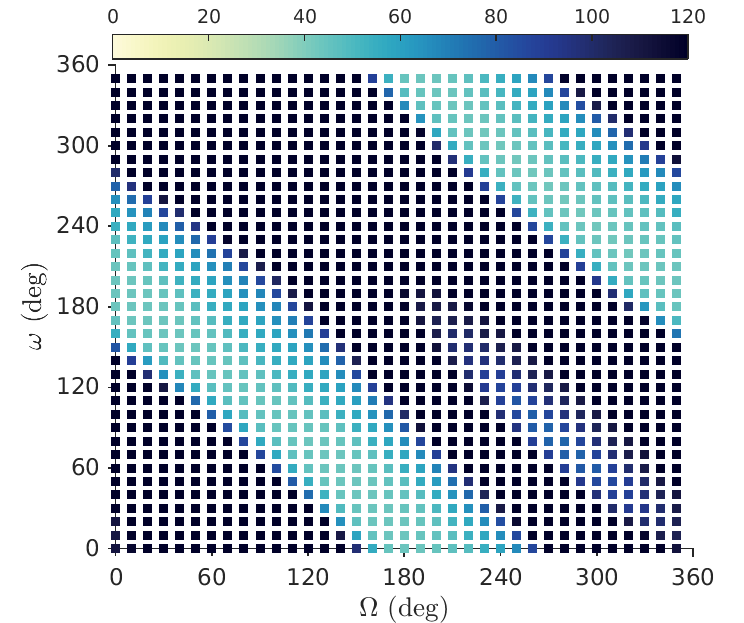}
    \includegraphics[width=0.3\textwidth]{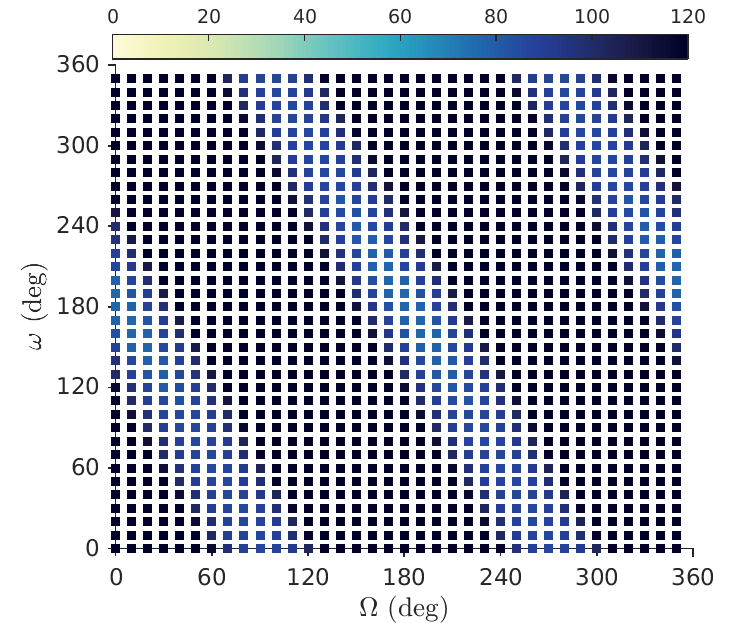}    \includegraphics[width=0.3\textwidth]{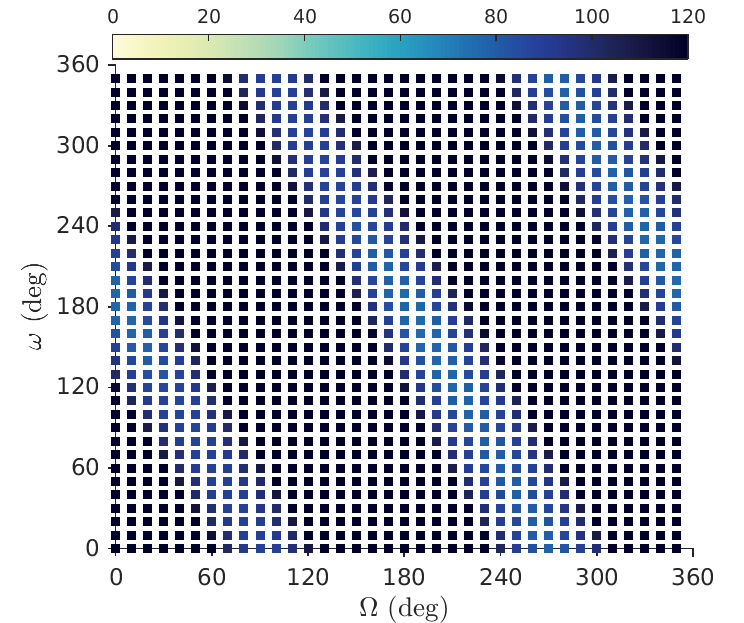}  \includegraphics[width=0.3\textwidth]{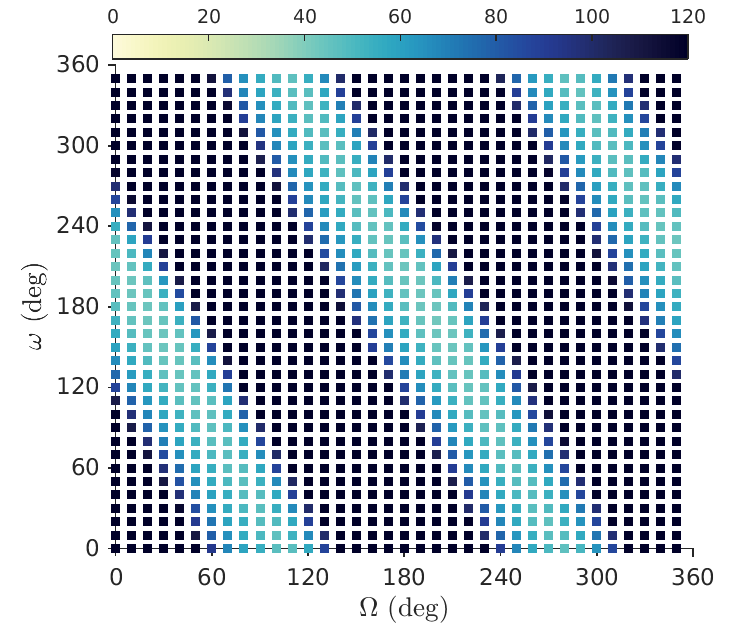}
\caption{Lifetime computed as a function of the initial configuration $(\Omega, \omega)$,
  for $a=R_E+1600 $km, initial $e=0.145$, epoch 2020,
  $C_R (A/m)=0.024\,$m$^2/$kg. Top: initial inclination
  $i=40^{\circ}$. Bottom: initial inclination $i=56^{\circ}$. The different evolution due to the $2\times 2$ (left), $3\times 3$ (middle)
  or $5\times5$ (right) geopotential in the dynamical model is compared.}\label{fig:lifetime1600_ell}
\end{center}
\end{figure}

Finally, we performed a sensitivity study in order to evaluate
  the role, in the dynamical evolution, of the terms of degree and order 
higher than 2 in the geopotential expansion. In
Figure \ref{fig:lifetime1600_ell}, we show the lifetime computed as a
function of the initial $(\Omega, \omega)$ configuration considering
at the initial epoch 2020 $a=7978.14$ km, $e=0.145$, $i=40^{\circ}$
and $i=56^{\circ}$, and three different dynamical models. In all cases, the
dynamical model includes the atmospheric drag, SRP and lunisolar
perturbations,  but the geopotential is expanded up to degree and
order 2, 3, and 5, respectively. Both $\Omega$
and $\omega$ are sampled at a step of 10$^{\circ}$ in the range $[0^{\circ}, 360^{\circ})$.
We show the outcome
corresponding to a high value of initial
eccentricity to emphasize the combined effect of one of the three perturbations with
the atmospheric drag. The figure shows that a dynamical model which considers a
geopotential expansion up to only degree and order 2, but even 3, does
not depict accurately the dynamics and the long-term behavior in eccentricity, when this is induced to vary in a relative significant way by a given resonant perturbation.

\begin{figure}[thbp!]
\begin{center}
\includegraphics[width=0.8\textwidth]{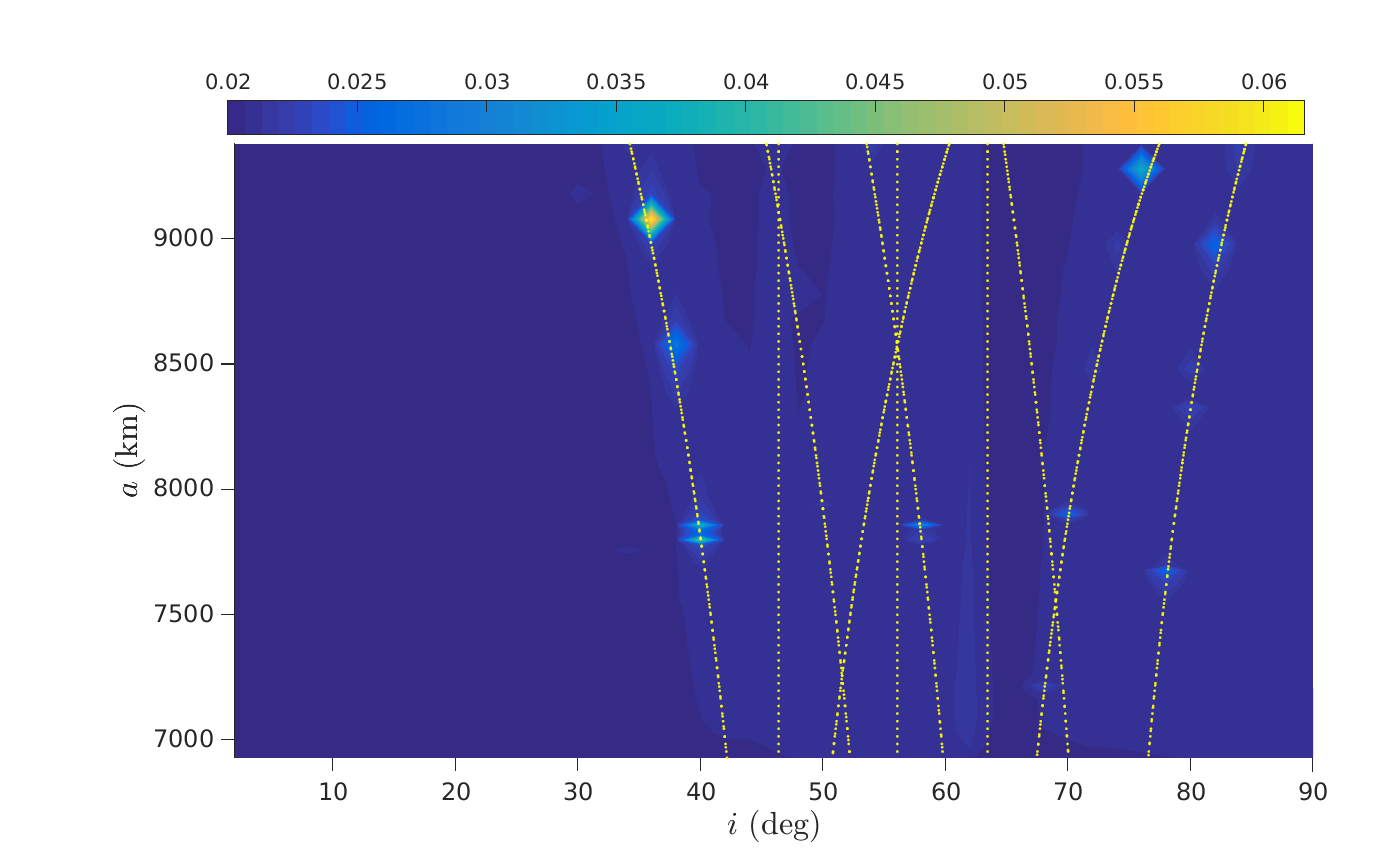} 
\caption{As a function of the initial inclination and semi-major axis, the maximum eccentricity value (color bar) computed in 120 years starting from $e=0.02$, $\Omega=0^{\circ}$, $\omega=0^{\circ}$. for the initial epoch 2018, considering $C_R(A/m)=0.012$ m$^2$/kg and prograde orbits. The yellow lines denote the dominant resonances determining the long-term behavior. The yellow color of such lines does not mean that the corridors are always highly unstable.}\label{fig:ai_res_1}
\end{center}
\end{figure}

\begin{figure}[thbp!]
\begin{center}
\includegraphics[width=0.8\textwidth]{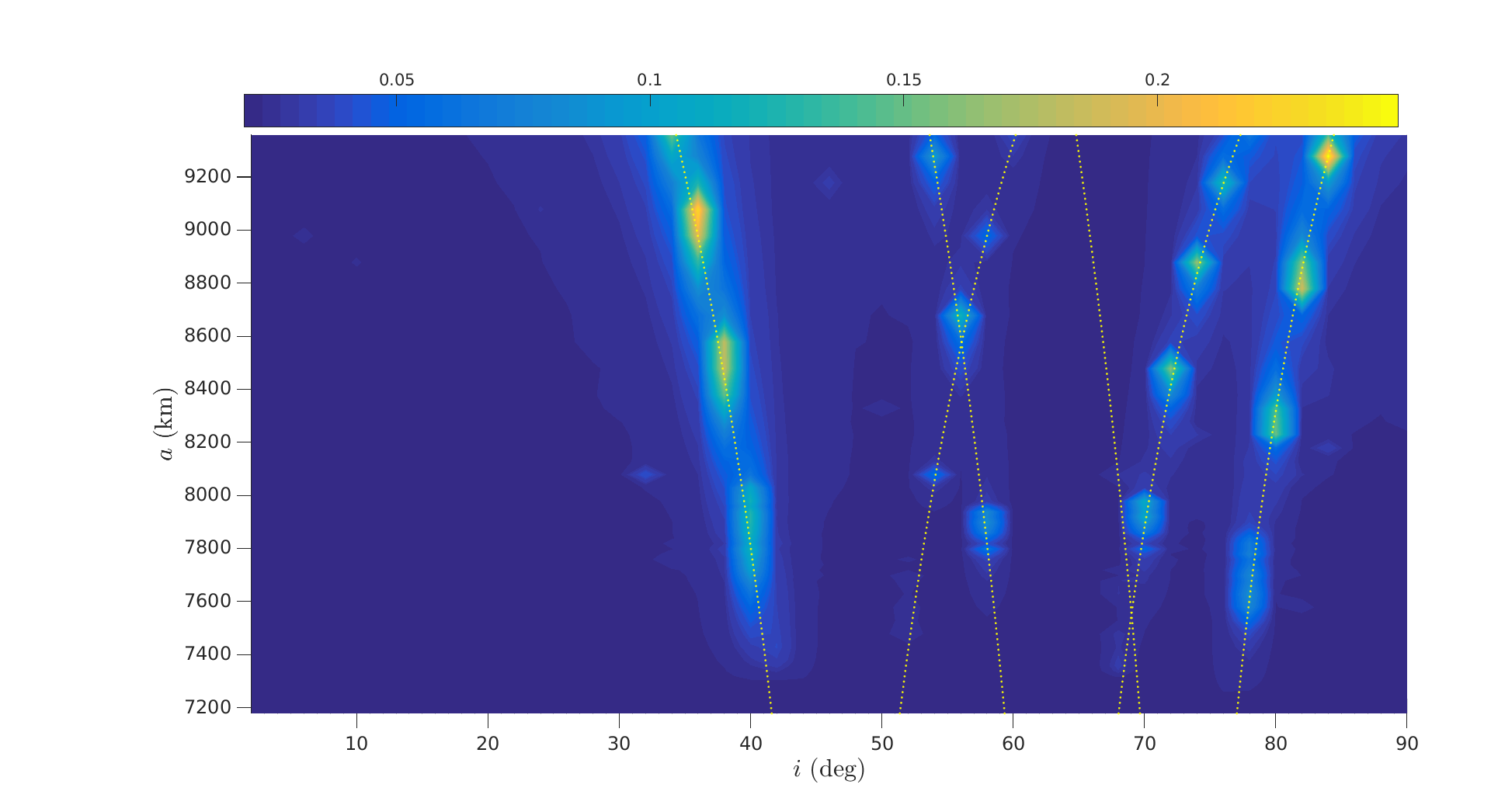} \includegraphics[width=0.8\textwidth]{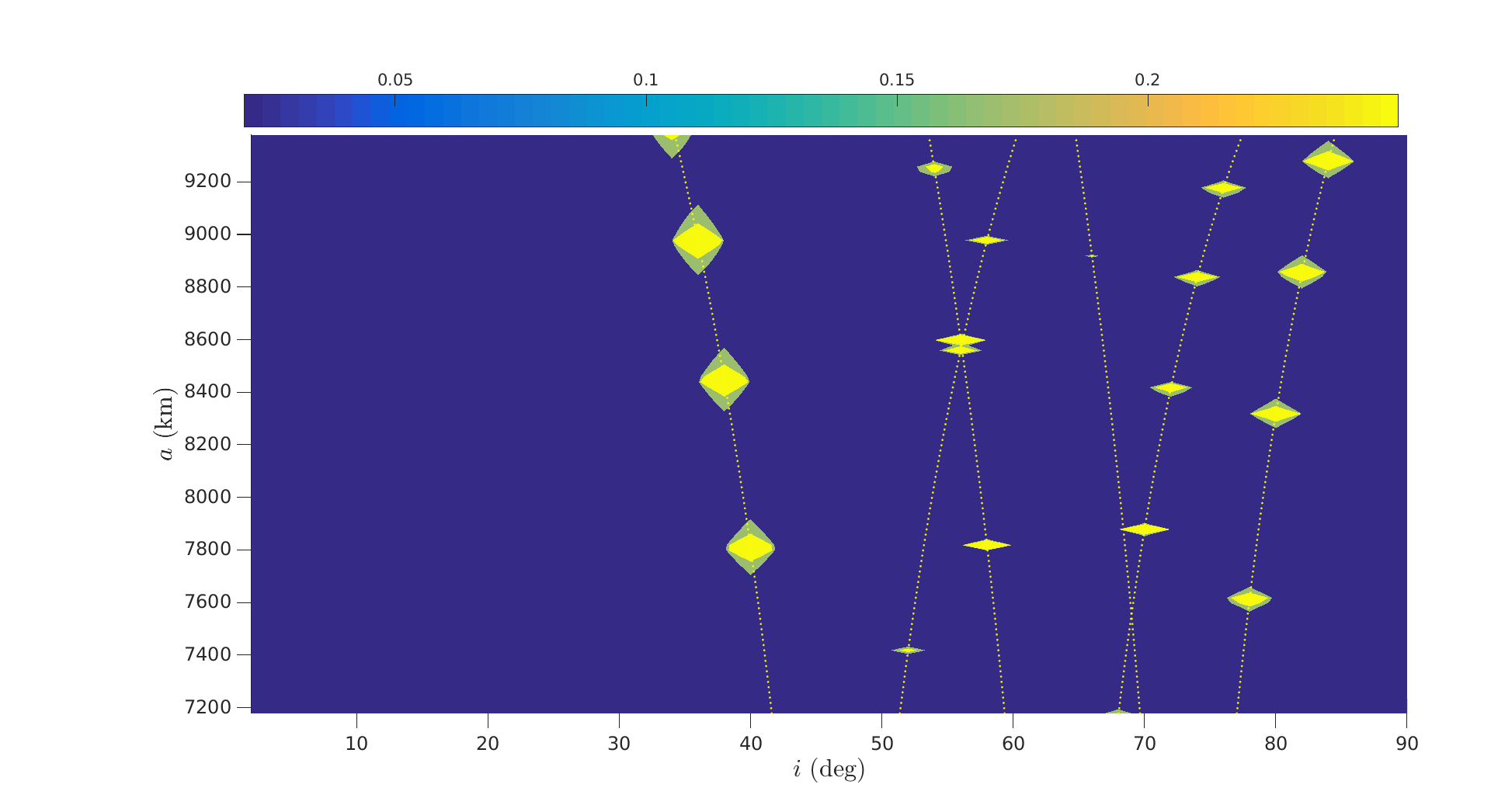} 

\caption{On the top, as a function of the initial inclination and semi-major
axis, the maximum eccentricity value (color bar) computed in 120 years starting
from $e=0.02$, $\Omega=0^{\circ}$, $\omega=0^{\circ}$. for the initial epoch
2020, considering $C_RA/m=1$ m$^2$/kg and prograde orbits. On the bottom, the
analytical estimate of the maximum variation in eccentricity is shown. The yellow lines
denote the dominant resonances determining the long-term
behavior. The yellow color of such lines does not mean that the corridors are always highly unstable.}\label{fig:ai_res_2}

\end{center}
\end{figure}

\section{Resonances in LEO}\label{sec:res}

From a theoretical point of view, the simulations performed have allowed to
recognize that the LEO region is permeated by a strong network of dynamical
resonances, as recently highlighted also for the Medium Earth Orbit region
(see, e.g., \cite{aR15}-\cite{A16}). The main difference with respect to the
previous studies on GNSS constellations is that in LEO the strongest advantage
can be obtained not only from lunisolar resonances, but also from high-degree
zonal harmonics and solar radiation pressure. 

We recall that high-degree zonal harmonics, lunisolar perturbations, and solar radiation pressure
cause long-term periodic variations in
eccentricity, which become quasi-secular when a resonance involving
the rate of $\Omega$ and $\omega$ occurs. In LEO,
for the two values of area-to-mass ratio adopted, we can assume that $\omega$ and
$\Omega$ vary only because of the oblateness of the Earth, and thus the
corresponding rates depend only on $(a,e,i)$, that is, 
$$
\dot\Omega =
-\frac{3}{2}\frac{J_2R^2_En}{a^2(1-e^2)^2}\cos i,\quad \quad \quad
\dot\omega
= \frac{3}{4}\frac{J_2R_En}{a^2(1-e^2)^2}\left(5\cos^2i-1\right),
$$
being $J_2$ the second zonal term and $n$ the mean motion of the
satellite (see, e.g., \cite{B}). This 
explains the inclination bands described in Section \ref{Cartography}.

In Figure \ref{fig:ai_res_1} and Figure \ref{fig:ai_res_2} (top panel), we
show, the maximum value, numerically computed, for the eccentricity in 120 years 
as a function of
the initial inclination and semi-major axis, for prograde orbits starting from $e=0.02$, $\Omega=0^{\circ}$,
$\omega=0^{\circ}$, $A/m=0.012$ m$^2$/kg and $A/m=1$ m$^2$/kg, respectively. In Figure \ref{fig:ai_res_1}, we show the locations of the following resonances:
\begin{itemize}
\item singly-averaged solar gravitational resonances  (e.g., \cite{H80})
\begin{itemize}
\item $\dot\psi=2\dot\omega+\dot\Omega-2 n_{S} \approx 0$,          
  \item $\dot\psi=2\dot\omega+2\dot\Omega-2n_S    \approx 0$,
\end{itemize}
where $n_S$ is the apparent  mean motion of the Sun;
\item doubly-averaged lunisolar gravitational perturbations (e.g., \cite{H80})
\begin{itemize}
\item $\dot\psi=\dot\omega \approx 0$,
\item $\dot\psi=2\dot\omega+\dot\Omega \approx 0$, 
 \item $\dot\psi=2\dot\omega+2\dot\Omega \approx 0$;
 \end{itemize}
 
\item $\dot\psi=\alpha\dot\Omega\pm\dot\omega\pm n_S \approx 0$,
  associated with SRP ($\alpha={0,1}$), \cite{gC62}-\cite{H77}.
\end{itemize}

 For $A/m=1$ m$^2/$kg, the only resonances which matter are the six resonances associated
with SRP, namely those in the last bullet, which can be rewritten, for the sake of clarity as, 
\begin{eqnarray*}
\dot\psi_1&=&\dot\Omega+\dot\omega- n_S \approx 0,\\
\dot\psi_2&=&\dot\Omega-\dot\omega- n_S \approx 0,\\
\dot\psi_3&=&\dot\omega- n_S \approx 0,\\
\dot\psi_4&=&\dot\omega+ n_S \approx 0,\\
\dot\psi_5&=&\dot\Omega+\dot\omega+ n_S \approx 0,\\
\dot\psi_6&=&\dot\Omega-\dot\omega+ n_S \approx 0.\\
\end{eqnarray*}
From Figure  \ref{fig:ai_res_2} (top), it is clear that, in this case, the variation
in eccentricity that can be obtained at a resonance can definitely allow for a
reentry. The amount of eccentricity increase can be estimated analytically in
the following way. Let us assume that the instantaneous variation given by the
Lagrange planetary equations can be written as

 $$
 \frac{de}{dt}=A\sin{\psi},
 $$
 where $A$ is a coefficient which depends on $(a,e,i)$, according to the given perturbation. By integration, we find that the detected variation in $e$ is proportional to $A$ and to the inverse of $\dot\psi$  \cite{A2018}. 
In Figure
\ref{fig:ai_res_2} (bottom), we show the variation in eccentricity computed in
this way for the six SRP resonances, by assuming the same $(i,a)$ grid adopted for the numerical simulation. The color bar is bounded to the maximum
value computed by propagation with FOP. Notice in particular the correspondence
of the yellow islands between the two plots (top and bottom) in the figure. The
effect of SRP resonances detected, especially in higher LEO and for $A/m=1$
m$^2/$kg, is of paramount importance both from a theoretical and an applied
point of view. Detailed theoretical findings obtained so far by our team can be found in \cite{A2018} and \cite{S2017}.
 
A remark to  the long-term variation in eccentricity due to the
zonal harmonic is in order. For the third harmonic $J_3$ it is given by:
$$
\Delta e_{J_3}=-\frac{1}{2}\frac{J_3}{J_2}\frac{R_E}{a}\frac{e}{1-e^2}\sin i \sin \omega.
$$
This is a classical result \cite{R}, obtained by averaging the
corresponding Lagrange planetary equation over the orbital period of the
spacecraft, as well as over the period of the argument of pericenter. In an
analogous way, it is possible to derive the long-term effect due to the
fifth zonal harmonic. We have:
\begin{equation}
\Delta e_{J_5}=\frac{5}{128}\frac{J_5}{J_2}\left(\frac{R_E}{a}\right)^3\frac{\sin i}{(1-e^2)^2(3+5\cos2i)}\left[f(e,i)\sin\omega+g(e,i)\sin3\omega\right].
\label{eq:j5}
\end{equation}
for some specific $f,g$ functions (we omit the variation caused by $J_4$,
because it varies with $ \sin 2\omega$ without any special remark). 
It is worth noting, in particular, the singularity in the expression for $\Delta e_{J_5}$ in Eq. (\ref{eq:j5}),
occurring at $i\approx 63.4^{\circ}$ (and $i\approx 116.6$), which corresponds to the critical inclination where also $\dot \omega=0$. Notice that this result can be confirmed also averaging the expression of the effect in eccentricity due to $J_5$, given in \cite{M1961}.

\begin{figure}[htbp!]
  \begin{center}
         \includegraphics[width=0.45\textwidth]{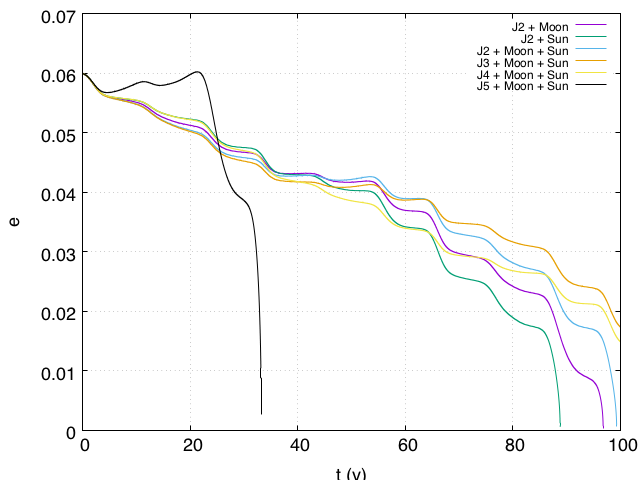}  \includegraphics[width=0.45\textwidth]{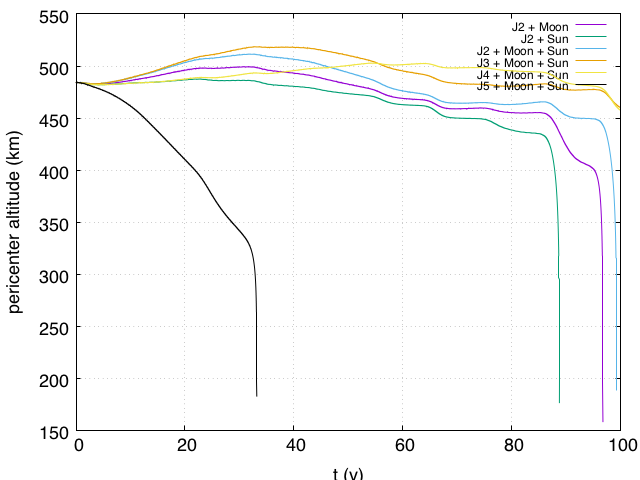}
\caption{Comparison of the lifetime computed starting from the critical inclination $i=63.4^{\circ}$, considering different contribution in the dynamical model. For all the curves, the effect of the drag was included. What changes from one curve to the other is the maximum degree of the expansion in the geopotential and the lunisolar perturbations. Left: eccentricity behavior in time. Right: pericenter altitude behavior in time. Initial condition: $a=7300.14$ km, $e=0.06$, $\Omega=0^{\circ}$, $\omega=90^{\circ}$,  epoch 2020, $C_R (A/m)=0.012$ m$^2/$kg.}  \label{fig:example_drag_lunisolar_j3j5}
\end{center}
\end{figure}

It can be shown that the additional effect of the $J_5$ harmonics
is significantly enhancing the perturbation in eccentricity with respect to the case where it is considered only $J_3$
 or the lunisolar perturbations. An example of this behavior is shown in Figure \ref{fig:example_drag_lunisolar_j3j5}, where different evolutions in  time of the eccentricity and the pericenter altitude are depicted, depending on the contributions considered in the dynamical model (always assuming the effect of the atmospheric drag). In particular, the sharp decrease in lifetime is clear when zonal harmonic of degree 5 is included.

The identification of the principal resonances are fundamental not only to design natural deorbiting highways, but also to understand the characteristic period in the neighborhood of the quasi-secular eccentricity
evolution. If an orbit is far
from any resonance, the eccentricity evolves on a time-scale $\sim
2\pi/\dot\omega$, which depends solely on the geopotential. Close to a resonance, the
eccentricity follows a longer secular time-scale, namely $\sim
2\pi/\dot\psi$, where $\psi$ is the critical angle of the dominant
resonance. A detailed analysis on the characteristic frequencies
detected for the eccentricity is under investigation.

\section{Conclusions}

We have shown the results obtained by mapping the LEO region from a
dynamical perspective. The instability of a given orbit and the
consequent lifetime change was evaluated as a function of the maximum
value of eccentricity computed in 120 years.  The whole mapping
revealed that dynamical resonances can modify significantly the
behavior in the long-term. For a typical value of area-to-mass ratio,
this effect is quite mild, though in higher LEO solar radiation
pressure effects can become important. The study paves the ways
towards different challenges to be faced. Considering the design of
affordable passive disposal solutions, we are now looking for the
optimal strategy to enter into one of the resonant corridors to
eventually achieve reentry. The focus will be put especially on the
most populated regions in terms of semi-major axis and inclination. On
the other hand, we are working in detail on the resonances found, in
particular on the role of the initial phasing and on the regularity of
the corresponding effects.

\begin{acknowledgements}
This work is funded through the European Commission Horizon 2020, Framework
Programme for Research and Innovation (2014-2020), under the ReDSHIFT project
(grant agreement n$^{\circ}$ 687500).

We are grateful to Bruno Carli, Camilla Colombo, Ioannis Gkolias, Kleomenis Tsiganis, Despoina Skoulidou and Volker Schaus for the useful
discussions. E.M.A. is also grateful to Florent Deleflie for the period
spent in Lille.  
\end{acknowledgements}


%
%

\end{document}